\documentclass[12pt]{article}

\pdfoutput=1

\usepackage{ifpdf}
\ifpdf
\usepackage{graphicx,color}
\usepackage{hyperref}
\else
\usepackage[dvipdfmx]{graphicx,color}
\usepackage[dvipdfmx]{hyperref}
\fi
\usepackage{amssymb,amsfonts,amsmath,cancel,cite,multirow}
\usepackage[capitalise]{cleveref}
\usepackage{slashed}
\usepackage{subcaption}
\usepackage[margin=1in]{geometry}

\newcommand{\eqs}[1]{\begin{equation}\begin{split} #1 \end{split}\end{equation}}

\begin{document}

\begin{titlepage}

\begin{flushright}
\end{flushright}

\vskip 1.35cm
\begin{center}

{\large
\textbf{
  LHC Lifetime Frontier and Visible Decay Searches in Composite Asymmetric Dark Matter Models
}}
\vskip 1.2cm

Ayuki Kamada$^{a,b,c}$
and Takumi Kuwahara$^{d,c}$

\vskip 0.4cm

\textit{$^a$
Institute of Theoretical Physics, Faculty of Physics, University of Warsaw, ul. Pasteura 5, PL–02–093 Warsaw, Poland
}

\textit{$^b$
Kavli Institute for the Physics and Mathematics of the Universe (WPI), The University of Tokyo Institutes for Advanced Study, The University of Tokyo, Kashiwa 277-8583, Japan
}

\textit{$^c$
Center for Theoretical Physics of the Universe,
Institute for Basic Science (IBS), Daejeon 34126, Korea
}

\textit{$^d$
Center for High Energy Physics, Peking University, Beijing 100871, China
}

\vskip 1.5cm

\begin{abstract}
  The LHC lifetime frontier will probe dark sector in near future, and the visible decay searches at fixed-target experiments have been exploring dark sector. 
  Composite asymmetric dark matter with dark photon portal is a promising framework explaining the coincidence problem between dark matter and visible matter.
  Dark strong dynamics provides rich structure in the dark sector: the lightest dark nucleon is the dark matter, while strong annihilation into dark pions depletes the symmetric components of the dark matter. 
  Dark photons alleviate cosmological problems.
  Meanwhile, dark photons make dark hadrons long-lived in terrestrial experiments.
  Moreover, the dark hadrons are produced through the very same dark photon. 
  In this study, we discuss the visible decay searches for composite asymmetric dark matter models. 
  For a few GeV dark nucleons, the LHC lifetime frontier, MATHUSLA and FASER, has a potential to discover their decay when kinetic mixing angle of dark photon is $\epsilon \gtrsim 10^{-4}$. 
  On the other hand, fixed-target experiments, in particular SeaQuest, will have a great sensitivity to dark pions with a mass below GeV and with kinetic mixing $\epsilon \gtrsim 10^{-4}$ in addition to the LHC lifetime frontier. 
  These projected sensitivities to dark hadrons in dark photon parameter space are comparable with the future sensitivities of dark photon searches, such as Belle-II and LHCb. 
\end{abstract}

\end{center}
\end{titlepage}

\section{Introduction} \label{sec:intro}

Lifetime is one of the most important properties of particles. 
In the Standard Model (SM), some particles are long-lived: for example, proton has a very long lifetime of $\gtrsim 10^{34}\,\mathrm{years}$, neutron decays into protons with a lifetime of $\mathcal{O}(10^3)\,\mathrm{s}$, and charged pions has a lifetime of $\mathcal{O}(10)\,\mathrm{ns}$.
As in the SM sector, a dark sector where dark matter (DM) particles reside has possibly a rich structure: it is very natural that dark particles have a variety of lifetime, and some dark particles may be long-lived.
At least, DM particles should have a lifetime longer than the age of the Universe: this is one of the known properties of DM particles.

The last decade has seen efforts in searching for the decay signal of long-lived particles in a dark sector by use of data collected at the past fixed-target experiments~\cite{Essig:2009nc,Reece:2009un,Aubert:2009cp,Bjorken:2009mm,Batell:2009di,Essig:2010gu,Andreas:2012mt,Lees:2014xha,Anastasi:2015qla}.
The main target particles in these experiments should have a lifetime of $\mathcal{O}(1)\,\mu \mathrm{s}$. 
The dark-sector particles may scatter with the SM particles via mediator particles that connect the dark sector with the SM sector. 
The scattering signals have been also explored at fixed-target experiments with downstream detectors such as LSND~\cite{Auerbach:2001wg,deNiverville:2011it}, E137~\cite{Bjorken:1988as,Batell:2014mga}, and MiniBooNE~\cite{Aguilar-Arevalo:2017mqx}.
The produced dark-sector particles may decay into lighter dark states, which is referred to as invisible decay search: for instance, the mono-photon search at BaBar~\cite{Lees:2017lec} puts a constraint on invisible decay of dark photon.
Besides reanalyzing data from the existing experiments, dark sector models will be tested at the lifetime frontier by currently running experiments, upgrades of the existing experiments, and newly proposed experiments:
displaced vertex searches at HPS~\cite{Celentano:2014wya}, SHiP~\cite{Anelli:2015pba,Alekhin:2015byh}, and SeaQuest~\cite{Gardner:2015wea,Aidala:2017ofy,Berlin:2018pwi}, and invisible decay at LDMX~\cite{Izaguirre:2014bca,Akesson:2018vlm} and Belle-II~\cite{Essig:2013vha,Kou:2018nap}.
These studies have been limited to the searches for the long-lived particle with mass of GeV or below since fixed-target experiments utilize low-energy beams.

The LHC experiment utilizes more energetic beams, and hence it is possible to produce long-lived particles with the mass of GeV or above.
Various strategies have been proposed to search for the long-lived particles at the existing LHC experiments near ATLAS, CMS, and LHCb detectors~\cite{Izaguirre:2015zva,Ilten:2016tkc,Liu:2018wte}: they are the displaced vertex searches, and the future sensitivity of the long-lived particles is limited by the detector sizes and the tracking system of the LHC experiments.
In particular, since there is no shield to reduce background events in the experiments, it is important to reconstruct the decay vertices in the detector.
Therefore, lifetime of interest is $\mathcal{O}(10\text{--}10^2)\,\mathrm{ps}$ in the LHC experiments with precise vertex reconstruction.
Meanwhile, a variety of experiments at the LHC has been proposed to search for long-lived particles recently by locating additional detectors far from the interaction points of the LHC: MATHUSLA~\cite{Chou:2016lxi}, FASER~\cite{Feng:2017uoz}, and CODEX-b~\cite{Gligorov:2017nwh}, collectively called LHC lifetime frontier.
The detector locations are $\mathcal{O}(10\text{--}10^2) \,\mathrm{m}$ away from the interaction point, and thus they would have a potential to explore the long-lived particles with a lifetime of $\mathcal{O}(10\text{--}10^2) \,\mathrm{ns}$.

We suggest that the composite asymmetric dark matter (ADM) framework, where DM particle masses are {\it to be} GeV and above, are very good target of the LHC lifetime frontier.
The basic concept of the composite ADM framework is the following.
Once the particle-antiparticle asymmetry is generated in either the SM or the dark sectors, it is shared among the two sectors via some portal interactions.
Consequently asymmetries of baryonic matter and DM have the same origin and close to each other.
Meanwhile, the observed energy densities of baryonic matter and DM are close to each other, $\Omega_\mathrm{DM} \simeq 5 \Omega_B$.
Thus, DM particle masses are to be GeV.
Compositeness plays three key roles: generation of the GeV mass, stability of the DM, and depletion of the symmetric DM component~\cite{Gudnason:2006yj, Dietrich:2006cm, Khlopov:2007ic, Khlopov:2008ty, Foadi:2008qv, Mardon:2009gw, Kribs:2009fy, Barbieri:2010mn, Blennow:2010qp, Lewis:2011zb, Appelquist:2013ms, Hietanen:2013fya, Cline:2013zca, Appelquist:2014jch, Hietanen:2014xca, Krnjaic:2014xza, Detmold:2014qqa, Detmold:2014kba, Asano:2014wra, Brod:2014loa, Antipin:2014qva, Hardy:2014mqa, Appelquist:2015yfa, Appelquist:2015zfa, Antipin:2015xia, Hardy:2015boa, Co:2016akw, Dienes:2016vei, Ishida:2016fbp, Lonsdale:2017mzg, Berryman:2017twh,  Gresham:2017zqi, Gresham:2017cvl, Mitridate:2017oky, Gresham:2018anj, Ibe:2018juk, Braaten:2018xuw, Francis:2018xjd, Bai:2018dxf, Chu:2018faw, Hall:2019rld, Tsai:2020vpi, Asadi:2021yml, Zhang:2021orr,Bottaro:2021aal, Hall:2021zsk} (see Ref.~\cite{Kribs:2016cew} for a review).
The strong dynamics in the dark sector dynamically explains the DM mass in the GeV range through the dimensional transmutation.
The accidental dark baryon number conservation ensures the longevity of the lightest dark nucleon, \textit{i.e.}, DM.
Annihilation of the dark nucleon and anti-nucleon into the dark pions leaves only the asymmetric component, so that the DM asymmetry determines the DM relic density today as in the SM nucleons.

In this article, we specifically consider a composite ADM model with dark photon~\cite{Ibe:2018juk}.
Dark photon plays a significant role in the composite ADM framework.
Without dark photon, the dark sector contains a large entropy density since in the early Universe two sectors are in thermal contact via portal interaction sharing the generated asymmetry between two sectors, and cause cosmological problems.
The dark photon decays into $e^+ e^-$ via the kinetic mixing with the SM photon~\cite{Holdom:1985ag}, and it releases the entropy of the dark sector into the SM sector.
We consider dark photons are the lightest particle in the dark sector in this study, and therefore the kinetic mixing of interest is below $10^{-3}$, evading constraints from searches for the prompt decay of dark photons at electron-positron collider experiments~\cite{Aubert:2009cp,Lees:2014xha,Archilli:2011zc,Babusci:2012cr,Anastasi:2015qla,Anastasi:2016ktq} (see also review~\cite{Bauer:2018onh}).

Through the very same dark photon, dark protons that are charged under dark quantum electrodynamics (QED) are produced in collider experiments.
Besides, through the dark photon, dark protons decay into dark neutrons that are neutral dark nucleons plus SM particles when dark protons are heavier than dark neutrons, and this transition can have a long lifetime due to the smallness of the kinetic mixing.
The lightest dark nucleon must have the multi-GeV mass to be the ADM,
and hence it is worthwhile for dark nucleon searches to explore the LHC lifetime frontier.
In addition to the dark nucleon searches at the LHC frontier, a rich structure of the dark sector allows us to take advantage of various search strategies for the dark sector.
When dark pions are heavier than dark photon but lighter than twice of the dark photon mass, dark pions decay into a dark photon emitting the SM particles via the anomaly-induced interaction. 
Dark pions are lighter than dark nucleons, and thus fixed-target experiments are also suitable for the dark pion visible decay signals in addition to the LHC lifetime frontier.
At the same time, as we mentioned, dark photon has been searched for by visible decay searches at fixed-target experiments and prompt decay searches at collider experiments. 

In this study, we investigate the sensitivity to the composite ADM models at the visible decay searches near future. 
We have found that the visible decay searches for the dark hadrons will explore the dark photon parameter space that the prompt decay searches of dark photon will also investigate.
In particular, the LHC lifetime frontier will have a sensitivity to the dark nucleons with multi-GeV mass, and the sensitivity area corresponds to the dark photon parameter with a few GeV mass and with the kinetic mixing $\epsilon \gtrsim 10^{-4}$.
As for dark pions, they are produced at both LHC and the fixed-target experiments.
SeaQuest, one of the on-going fixed-target experiments, will particularly have a great sensitivity to the dark pion with sub-GeV and the kinetic mixing $\epsilon \gtrsim 10^{-5}$.
The sensitivity area of the LHC lifetime frontier corresponds to the dark photon parameter with the sub-GeV mass and with the kinetic mixing $\epsilon \gtrsim 10^{-4}$.

This paper is organized as follows. 
In \cref{sec:model}, we briefly review a composite model of ADM~\cite{Ibe:2018juk}. 
We discuss mass spectrum and lifetime of dark hadrons, and existing constraints except for the visible decay searches in \cref{sec:existing}. 
In \cref{sec:darkhadron}, we discuss the visible decay searches of dark hadrons, and then we summarize sensitivities of visible decay searches in our composite ADM model in \cref{sec:results}.
\cref{sec:conclusion} is devoted to conclusions of our work.

\section{Composite Asymmetric Dark Matter Model \label{sec:model}}

\begin{table}
	\centering
	\caption{Charge assignment of the dark quarks and the dark Higgs for a composite ADM model.
	$SU(3)_D$ and $U(1)_D$ are gauge symmetries of the dark sector, while $U(1)_{B-L}$ is the global symmetry shared with the visible sector.
	}
	\label{tab:Charge_dark}
	\begin{tabular}{|c|c|c||c|}
		\hline
		& $SU(3)_D$ & $U(1)_D$ & $U(1)_{B-L}$ \\ \hline
		$U'$ & $\mathbf{3}$ & $2/3$ & $1/3$ \\
		$\overline U'$ & $\overline{\mathbf{3}}$ & $-2/3$ & $- 1/3$ \\
		$D'$ & $\mathbf{3}$ & $-1/3$ & $1/3$ \\
		$\overline D'$ & $\overline{\mathbf{3}}$ & $1/3$ & $- 1/3$ \\ \hline
		$\phi_D$ & $\mathbf{1}$ & $1$ & $0$ \\ \hline
	\end{tabular}
\end{table}

We review the composite ADM scenario with a dark photon portal and fix our notation in this section. 
We consider the vector-like two-flavor $SU(3)_D \times U(1)_D$ dynamics proposed by Ref.~\cite{Ibe:2018juk}.
We list the minimal particle contents in the dark sector in \cref{tab:Charge_dark}. 
The dark quarks are in (anti-)fundamental representation of $SU(3)_D$, are charged under the dark QED $U(1)_D$, and carry the same $B-L$ number as the SM quarks.
We refer to $U'$ and $D'$ as up and down dark quarks, respectively, since they have similar properties as the SM up and down quarks.
The dark quarks are confined into dark hadrons below the dynamical scale $\Lambda_\mathrm{QCD'}$ of the dark QCD, $SU(3)_D$.
DM consists of the dark nucleons, whose stability is ensured by the $B-L$ number conservation. 
The DM asymmetry has the same origin as the SM baryon asymmetry, and hence the mass of the dark nucleons is close to that of the SM nucleons as indicated from the coincidence of the mass densities of the baryonic matters and the DM, $\Omega_\mathrm{DM} \sim 5 \Omega_B$.

Dark pions also play a significant role in the composite ADM models; the annihilation into the dark pions depletes the symmetric part of the DM.
If the lightest pion is stable, it could lead to the overclosure of the Universe or a too large effective number of neutrino species.
The dark QED, $U(1)_D$, is therefore introduced to release the entropy of the dark sector to the SM sector.
The dark photon gets its mass via spontaneous symmetry breaking of $U(1)_D$ by the dark Higgs $\phi_D$.

The Lagrangian density of the dark sector is given by
\eqs{
  \mathcal{L} = \mathcal{L}_{Q'} + \mathcal{L}_{\phi_D} + \mathcal{L}_\mathrm{gauge} + \mathcal{L}_\mathrm{portal} \,. 
}
Here, $\mathcal{L}_\mathrm{gauge}$ denotes the kinetic terms of the dark photon and the dark gluons, and $\mathcal{L}_{\phi_D}$ contains the kinetic term and the potential term of the dark Higgs.
The dark photon mass arises from the vacuum expectation value (VEV) $v_D$ of the dark Higgs, $m_{A'}^2 = 2 e'^2 v_D^2$ with the $U(1)_D$ coupling $e'$. 

$\mathcal{L}_\mathrm{portal}$ connects the dark sector to the SM sector. 
This includes two kinds of portal interactions: one is called as the intermediate-scale portal interaction sharing the generated asymmetry between the SM sector and the dark sector, another is the dark photon portal interaction that releases the entropy in the dark sector to the SM sector.
As for the intermediate-scale portal interaction, the charge assignment of the dark quarks in \cref{tab:Charge_dark} allows us to write the following interactions between our sector and the dark sector~\cite{Ibe:2018juk}:
\eqs{
  \mathcal{L}_\mathrm{portal} \supset \frac{1}{\Lambda_\ast^3} (\overline U' \overline D' \overline D') (LH) + \frac{1}{\Lambda_\ast^3} (U'^\dag D'^\dag \overline D') (LH) + \mathrm{h.c.} \,,
  \label{eq:Intermediate_Portal}
}
where $\Lambda_\ast$ collectively denotes mass-dimension one coefficients.
These operators are non-renormalizable and violate the global $B-L$ symmetry.
As for ultraviolet (UV) origin of the operators, we can consider right-handed neutrinos and a dark-colored scalar~\cite{Ibe:2018juk,Ibe:2018tex}: the dark-colored scalar has Yukawa couplings to $U' D'$ and $\overline U' \overline D'$, while the right-handed neutrinos with the Majorana mass term have Yukawa couplings to $D'$ and the dark-colored scalar.
The effective operators in \cref{eq:Intermediate_Portal} appears as the low-energy effective theory below the mass thresholds of the right-handed neutrinos and the dark-colored scalar.

The intermediate-scale portal interaction leads to the DM decay into anti-neutrino~\cite{Fukuda:2014xqa}, but the lifetime is severely constrained by the neutrino flux measurement at the Super-Kamiokande~\cite{Desai:2004pq,Covi:2009xn}.
The lower limit on the DM lifetime is about $10^{23}\,\mathrm{s}$, and hence it is too long-lived to find the DM at the lifetime frontier.
Furthermore, once we consider a UV completion of the composite ADM~\cite{Ibe:2018tex}, the same origin of the intermediate-scale portal interaction provides an effective operator causing oscillation of DM and anti-DM, and it leads to a late-time annihilation of DM into the SM particles, and it can be tested by $\gamma$-ray and electron-positron flux observations~\cite{Ibe:2019yra}.
We ignore the intermediate-scale portal interaction that is irrelevant in the rest of this study.

The dark photon portal interaction is given by
\eqs{
  \mathcal{L}_\mathrm{portal} \supset - \frac{\epsilon}{2} F^{\mu \nu} F'_{\mu\nu} \,,
}
where $F_{\mu\nu}$ and $F'_{\mu\nu}$ are field strength tensors of the SM photon and the dark photon, respectively.
After diagonalizing the kinetic term, the dark photon $A'$ couples to the SM particles, and the coupling constant is proportional to their charge, $\epsilon e Q_f$, while the dark-sector particles remain electrically neutral.
The dark photon decays into the SM particles through the coupling,
\eqs{
  \Gamma(A' \to \mu^+ \mu^-) & = \frac13 \alpha \epsilon^2 m_{A'} \sqrt{1 - \frac{4 m_\mu^2}{m_{A'}^2}} \left( 1 + \frac{2 m_\mu^2}{m_{A'}^2} \right) \,, \\
  \Gamma(A' \to \text{hadrons}) & = \Gamma(A' \to \mu^+ \mu^-) R(\sqrt{s} = m_{A'}) \,,
  \label{eq:darkphotondecay}
}
where $R(\sqrt{s}) = \sigma(e^+e^- \to \text{hadrons})/\sigma(e^+e^- \to \mu^+\mu^-)$ takes into account the hadronic resonances \cite{Ilten:2018crw,Zyla:2020zbs}.
The darkly charged particles interact with the SM particles through dark photon exchange, and hence DM interact with ordinary matter when the DM is darkly charged.

The Lagrangian density $\mathcal{L}_{Q'}$ contains the dark quarks:
\eqs{
  \mathcal{L}_{Q'} = \sum_{\hat Q'} \hat Q'^\dag i \bar\sigma^\mu D_\mu \hat Q' 
  - (m_{U'} \hat{\overline U'} \hat U' + m_{D'} \hat{\overline D'} \hat D' + y_1 \phi_D \hat{\overline U'} \hat D' + y_2^\ast \phi_D^\ast \hat{\overline D'} \hat U' + \mathrm{h.c.}) \,.
  \label{eq:QuarkLagrangian}
}
Here, we use the hatted-notation for fields in the charge basis. 
The vector-like masses, $m_{U'}$ and $m_{D'}$, are assumed to be smaller than the dynamical scale $\Lambda_\mathrm{QCD'}$.
The dark QCD interactions preserve the flavors of dark quarks, and hence there would be approximate global symmetry interchanging up and down dark quarks, which we refer to as isospin symmetry of the dark quarks.
We introduce the isospin doublet of the dark quarks as $\hat Q' \equiv (\hat U' \, \hat D')^T$ and its vector-like counterpart as $\hat{\overline Q'} \equiv (\hat{\overline U'} \, \hat{\overline D'})^T$.
The kinetic term of the dark quarks preserves a flavor symmetry $SU(2)_L \times SU(2)_R \times U(1)_A \times U(1)_B$.
Under the flavor symmetry, the dark quarks are transformed as follows.
\eqs{
  \hat Q' \to e^{i \alpha/2} e^{i \beta/2} L \hat Q' \,, \qquad
  \hat{\overline Q'} \to e^{i \alpha/2} e^{- i \beta/2} \hat{\overline Q'} R^\dag \,, 
  \label{eq:Quark_Flavor_Rot}
}
where $\alpha$ and $\beta$ are phases of $U(1)_A$ and $U(1)_B$, while $L$ and $R$ denote $SU(2)_L$ and $SU(2)_R$ rotations, respectively.
The chiral symmetry in the dark sector, $SU(2)_L \times SU(2)_R$, is explicitly broken by the quark mass terms and $U(1)_D$ interactions.

We systematically include the explicit breaking of chiral symmetry in the dark hadron spectrum discussed later by treating the breaking terms as spurion fields.
The mass matrix of the dark quarks, $M_{Q'}$, has off-diagonal entries when the dark Higgs obtains the VEV $v_D$%
\footnote{
  The VEV of the dark Higgs $\phi_D$, $v_D$, is determined by whole potential. 
  As we will see later, we consider the Linear Sigma Model as the low-energy effective theory. 
  In \cref{app:LSM}, we discuss the determination of the dark Higgs VEV including the Linear Sigma Model field.
}:
\eqs{
  M_{Q'} = 
  \begin{pmatrix}
    m_{U'} & y_1 v_D \\
    y_2^\ast v_D & m_{D'} \\
  \end{pmatrix} \,.
}
The mass matrix transforms as a spurion under the global flavor symmetry as $M_{Q'} \to R M_{Q'} L^\dag$.
This matrix is diagonalized by unitary matrices $U_L$ and $U_R$ as $M^\mathrm{diag}_{Q'} = U_R^\dag M_{Q'} U_L = \mathrm{diag}(M_{1},M_{2})$.
In the following, we assume that $M_{Q'}$ is hermitian (parity-conserving), \textit{i.e.}, $y_1 = y_2$, and thus the diagonalizing matrices are the same, $U_L = U_R \equiv U_V$;
we also discuss the case that $M_{Q'}$ is non-hermitian in \cref{app:LSM}.
The covariant derivative and the charge matrix for dark quarks is defined by 
\begin{align}
  D_\mu \hat Q' & =
  \partial_\mu \hat Q' - i e' \hat q_L A'_\mu \hat Q' \,, 
  &
  D_\mu \hat{\overline Q'} & =
  \partial_\mu \hat{\overline Q'} + i e' A'_\mu \hat{\overline Q'} \hat q_R \,, 
  \\
  \hat q & \equiv \hat q_L = \hat q_R = \frac13
  \begin{pmatrix}
    2 & 0 \\
    0 & -1
  \end{pmatrix}\,.
\end{align}
Here, we omit the dark QCD interaction. 
The charge matrices transform as spurions under the global flavor symmetry as $\hat q_L \to L \hat q_L L^\dag$ and $\hat q_R \to R \hat q_R R^\dag$.

The dark quarks in the charge basis and in the mass basis are related with each other by the unitary transformation:
\begin{align}
  \hat Q' & = U_V Q' \,, &
  \hat{\overline Q'} & = \overline Q' U_V^\dag \,, & 
  U_V & = 
  \begin{pmatrix}
    \cos \theta_V & - e ^{-i \alpha_V} \sin \theta_V \\
    e ^{i \alpha_V} \sin \theta_V & \cos \theta_V \\
  \end{pmatrix} \,,
\end{align}
where $\hat Q'$ and $\hat{\overline Q'}$ indicate dark quarks in the charge basis, again, while $Q'$ and $\overline Q'$ are the quarks in the mass basis.
The charge matrices in the mass basis are given by
\eqs{
  q_L = q_R = U_V^\dag \hat q U_V \,,
}
and are different as $U_L \neq U_R$. 

We discuss the long-lived dark hadrons in the composite ADM scenario in this article. 
The isospin symmetry is broken by the dark quark masses and the $U(1)_D$ interaction.
This leads to the mass difference among states in an isospin multiplet even in hadronic picture below the dynamical scale, and thus the heavier states can decay into the lighter state.
The heavier state can be a long-lived particle when the isospin symmetry is slightly broken.

We use the linear sigma model (LSM) as a low-energy effective theory to describe the mass spectrum of the dark hadrons and the interactions among the dark hadrons below $\Lambda_\mathrm{QCD'}$.
We discuss the LSM in detail in \cref{app:LSM}.
The LSM field $\Phi$ is a matrix-form scalar field, and corresponds to the quark bilinear $\Phi_{ij} \sim \overline Q'_j Q'_i$ where the dark quarks $Q'_i$ and $\overline Q'_i$ are in mass basis.
On the other hand, the isospin-doublet dark nucleons are denoted by the Weyl fermions $N$ and $\overline N$.
The transformation of $\Phi \,, N$\,, and $\overline N$ under the flavor symmetry is given by
\eqs{
  \Phi \to e^{i \alpha} L \Phi R^\dag \,, \qquad 
  N \to e^{i \alpha/2} e^{3 i \beta/2} L N \,, \qquad 
  \overline N \to e^{i \alpha/2} e^{- 3 i \beta/2} \overline N R^\dag \,.
}
Here, $L\,,R$ and $\alpha\,,\beta$ characterize the same transformation under the flavor symmetry as dark quarks. 

The low-energy effective Lagrangian consists of the LSM field $\Phi$, the dark nucleons, and the dark photon.
The LSM Lagrangian has the following form:
\eqs{
  \mathcal{L}_\mathrm{LSM} & = \mathrm{tr}(D_\mu \Phi^\dag D^\mu \Phi) - V_\mathrm{LSM}(\Phi) \,, \\
  V_\mathrm{LSM}(\Phi) & = - \mu^2 \mathrm{tr}(\Phi^\dag \Phi) + \lambda_1 [\mathrm{tr}(\Phi^\dag \Phi) ]^2 + \lambda_2 \mathrm{tr}(\Phi^\dag \Phi)^2 \\
  & \qquad - c \left( \det \Phi +\det \Phi^\dag \right) - \mathrm{tr} H (\Phi + \Phi^\dag) \,.
}
Here, a VEV of $\Phi$ develops since $\mu^2$ is positive, and hence the VEV breaks the global flavor symmetry $SU(2)_L \times SU(2)_R$.

The covariant derivative of the LSM field is defined by 
\eqs{
  D_\mu \Phi = \partial_\mu \Phi - i e' A'_\mu (q_L \Phi - \Phi q_R) \,,
  \label{eq:covariantD_Phi}
}
where $q_L$ and $q_R$ are the charge matrices.
We note that the charge matrices are identical ($q_L = q_R$) when $\Phi$ is in the charge basis.
The na\"ive dimensional analysis (NDA) with large-$N_C$ scaling of the low-energy parameters in the LSM is given by (see Ref.~\cite{Manohar:1983md,Georgi:1992dw} for NDA and Refs.~\cite{tHooft:1973alw,tHooft:1974pnl,Witten:1979vv,Witten:1979kh,Coleman:1980mx,Witten:1980sp} for large-$N_C$ scaling)
\begin{align}
  \lambda_1 \,, \lambda_2 & \simeq \frac{(4\pi)^2}{N_C} \,, & 
  \mu^2 & \simeq \Lambda_{\chi\mathrm{SB}'}^2 \, &
  c & \simeq \frac{1}{N_C} \Lambda_{\chi\mathrm{SB}'}^2 \, &
\end{align}
with the number of dark colors, $N_C = 3$, and the chiral symmetry breaking scale in the dark sector, $\Lambda_{\chi\mathrm{SB}'}$.
The fourth term of $V_\mathrm{LSM}$ comes from the instanton-induced quark interaction and breaks $U(1)_A$~\cite{Kobayashi:1970ji,tHooft:1976rip}, which we refer to as the anomaly term.
We set the coefficient of the term $c$ to be positive so that the term is parity symmetric, which directly gives the mass difference between the dark $\eta$ meson and the dark pions in the two-flavor case.
The last term originates from the quark mass term, and thus $H$ is hermition, and $H \equiv j^a T^a = j^0 T^0 + j^3 T^3$ is proportional to the diagonalized mass matrix of the dark quarks,
\eqs{
  H \simeq \frac{\sqrt{N_C}}{4\pi} \Lambda_{\chi\mathrm{SB}'}^2 M_{Q'}^\mathrm{diag} \,.
}
We define scalar and pseudoscalar components of $\Phi$ as follows.
\eqs{
  \Phi = (\sigma^a + i \pi^a) T^a \,, \qquad
  T^a = \frac{\hat\lambda^a}{2} \,, \qquad
  (a = 0\,, 1 \,, 2\,, 3)\,,
}
where the matrix $\hat\lambda^0$ is a unit matrix and $\hat \lambda^a ~ (a=1,2,3)$ are the Pauli matrices.

First, we consider the mass spectrum of dark pions by use of the LSM. 
In the absence of the source term $H$, dark pions are massless Nambu--Goldstone bosons.
$H$ tilts the LSM potential and gives the masses of dark pions.
In this article, we consider the case $j^0$ dominates the source term with a small perturbation $j^3$ that describes isospin violation in dark quarks, while we also discuss other cases in \cref{app:LSM}. 
In this case, $\sigma^0$ obtains its VEV of order of the dynamical scale $\sqrt{N_C} \Lambda_{\chi\mathrm{SB}'}/4\pi$, while $\sigma^3$ has its VEV due to a tilt by a source term $j^3 \sigma^3$, and thus the VEV $\langle \sigma^3 \rangle$ is proportional to the source $j^3 \Lambda_{\chi\mathrm{SB}'}^{-2}$.
The spectrum of pseudoscalars is as follows.
\begin{align}
  m_{\pi^0}^2 & \simeq 2 c \,, &
  m_{\pi^{1,2}}^2 & = \frac{j^0}{f_{\pi'}} \,, &
  m_{\pi^{3}}^2 & = \frac{j^0}{f_{\pi'}} - \frac{\lambda_2 j^3 \delta}{2c} \,.
  \label{eq:pion_mass}
\end{align}
Here, $f_{\pi'} \equiv \langle \sigma^0 \rangle \simeq (4\pi)^{-1} \sqrt{N_C} \Lambda_{\chi\mathrm{SB}'}$ corresponds to the pion decay constant, and $\delta \equiv \langle \sigma^3 \rangle$.
$\pi^0$ mass is dominated by the anomaly term, which corresponds to the $\eta'$ meson in the SM, while the mass of other pseudoscalars is proportional to $j^0$ at the leading order. 
The mass difference within the isospin multiplet is proportional to square of the dark quark mass difference, $(M_{1}-M_{2})^2$ (see \textit{e.g.},~\cite{Gasser:1982ap,Donoghue:1996zn}).

Besides the quark mass difference, the $U(1)_D$ interaction explicitly violates the isospin symmetry and provides the pion mass difference.
The effective Lagrangian up to $\mathcal{O}(e'^2)$ is given by
\eqs{
  \mathcal{L} & \supset 
  - c_q^{\pi'} \frac{\alpha'}{4\pi} \Lambda_{\chi\mathrm{SB}'}^2 \left[ 
  \mathrm{tr} (\Phi^\dag q_L \Phi q_R)
  - \frac{1}{2} \mathrm{tr} (\Phi^\dag q_L^2 \Phi) 
  - \frac{1}{2} \mathrm{tr} (\Phi^\dag \Phi q_R^2) 
  +  \mathrm{h.c.} \right] \,,
}
where $c_q^{\pi'}$ is an $\mathcal{O}(1)$ dimensionless parameter, and $\alpha' = e'^2/4\pi$.
This Lagrangian is invariant under the global flavor symmetry as the charge matrices $q_L$ and $q_R$ are spurions. 
We determine the coefficients to give a contribution only to the charged pion mass when the charge matrices are in charge basis ($q_L = q_R = \hat q$).
\begin{align}
  \mathcal{L} & \supset \frac{c_q^{\pi'}}{2} \frac{\alpha'}{4\pi} \Lambda_{\chi\mathrm{SB}'}^2 \left[(\pi^1)^2 + (\pi^2)^2 \right] \,.
\end{align}
Here, the charges of dark quarks determine the charge matrix in the charge basis, $\hat q$.%
\footnote{
  The charge matrix of dark quarks is determined by $T^3 + B/2$ with isospin $T^3$ and baryon number $B$ of dark quarks.
  The baryon number is zero for $\Phi$, and the charge matrix of $\Phi$ can be determined only by the isospin.
  However, since $\Phi$ is identified with $Q'\overline Q'$, the charge matrix of $\Phi$ can also be determined by $\hat q$, the charge matrix of $Q'$.
  Since we can add any matrix proportional to unit matrix to the charge matrix in charge basis, both definitions of the charge matrix do not change the results (except for anomaly induced interactions).
  It is easy to compute the spectrum by use of the charge defined by $T^3 + 1/2 = \mathrm{diag}(1\,,0)$. 
}
The mass differences among dark pions originate from $U(1)_D$ interaction and dark quark mass difference, and each of them is numerically given by
\eqs{
  \Delta_{\pi'} & \equiv \frac{m_{\pi'^1} - m_{\pi'}}{m_{\pi'}} 
  = \Delta_{\pi'}^\mathrm{QED'} + \Delta_{\pi'}^\mathrm{iso} \,, \\
  \Delta_{\pi'}^\mathrm{QED'} & = \frac{c_q^{\pi'} \alpha' \Lambda_{\chi\mathrm{SB}'}^2}{16 \pi m_{\pi'}^2} \simeq 0.13 \left( \frac{\alpha'}{0.1} \right) \left( \frac{\Lambda_{\chi\mathrm{SB}'}}{10\,\mathrm{GeV}}\right)^2 \left( \frac{1\,\mathrm{GeV}}{m_{\pi'}} \right)^2 \,, \\
  \Delta_{\pi'}^\mathrm{iso} & = \frac{\lambda_2 j^3 \delta}{4 c m_{\pi'}^2} 
  \simeq \frac{N_C}{2 m_{\pi'}^2} (M_1-M_2)^2 
  \simeq 0.02 \left( \frac{1\,\mathrm{GeV}}{m_{\pi'}^2} \right)^2 \left( \frac{M_1-M_2}{0.1\,\mathrm{GeV}} \right)^2\,.
}
Here, we use $\delta \simeq j^3 \Lambda_{\chi\mathrm{SB}'}^{-2}$, $j^3 \simeq (4\pi)^{-1} \sqrt{N_C} \Lambda_{\chi\mathrm{SB}'}^2 (M_1-M_2)$, and large-$N_C$ scaling of other parameters.
$m_{\pi'}$ denotes the mass of $\pi^3$. 
As with the pion mass difference in the SM (see Ref.~\cite{Colangelo:2010et}), the $U(1)_D$ correction dominates the dark pion mass difference unless dark quark mass difference is quite large. 

After diagonalizing the dark pion mass terms including the $U(1)_D$ correction, the dark photon couples to dark pions. 
\eqs{
  \mathcal{L}_\mathrm{LSM} \supset 
  e' A'^\mu \left[  
    \cos(2 \theta_V + \beta_V) (\pi'^1 \partial_\mu \pi'^2 - \pi'^2 \partial_\mu \pi'^1)
    + \sin(2 \theta_V + \beta_V) (\pi'^2 \partial_\mu \pi'^3 - \pi'^3 \partial_\mu \pi'^2)
  \right]
}
Here, we use primed notation for the hadronic field in mass basis after taking into account the $U(1)_D$ correction.
An additional angle $\beta_V$ arises from diagonalizing the mass terms including the $U(1)_D$ correction in addition to $\theta_V$.
We discuss the dark pion mass spectrum in detail in \cref{sec:spectrum}.

Next, we consider the spectrum of light dark nucleons. 
The Lagrangian density of the dark nucleons is
\eqs{
  \mathcal{L}_B & = \overline N^\dag i\cancel{D} \overline N + N^\dag i\cancel{D} N - (\xi \overline N M_{Q'}^\mathrm{diag} N + g \overline N \Phi^\dag N + \mathrm{h.c.}) \,.
}
A typical scale of the dark nucleon mass is determined by the dynamical scale, $g f_{\pi'}/2$, and the NDA with the large-$N_C$ scaling of the coupling $g$ is $g \simeq 4 \pi \sqrt{N_C}$.
We take into account two mass terms when we consider the mass difference of the dark nucleons: one comes from the VEV of the LSM field, and another comes from the dark-quark mass matrix itself which is proportional to an $\mathcal{O}(1)$ dimensionless parameter $\xi$.
The mass difference that comes from the VEV of the LSM field is proportional to $j^3$, which originates from the mass difference of the dark quarks.
When $j^0$ is dominated, the dark nucleon masses are given by
\eqs{
  m_{N_1} = \frac{g}{2}(f_{\pi'}+\delta) + \xi M_{1} \,, \qquad
  m_{N_2} = \frac{g}{2}(f_{\pi'}-\delta) + \xi M_{2} \,.
  \label{eq:nucleonmass}
}
The lightest dark nucleon is the candidate of the composite ADM. 
As shown in Refs.~\cite{Ibe:2011hq,Fukuda:2014xqa,Ibe:2018juk}, if the asymmetry is fully shared between the dark and SM sectors, the DM mass is to be $8.5/n_{g'}\,\mathrm{GeV}$ with $n_{g'}$ being the number of generations of dark quarks.
When we consider mirror-sector scenario to explain the coincidence of the dynamical scales in the visible and the dark sectors~\cite{Ibe:2019ena}, $n_{g'}$ should be integer. 
As long as we consider vector-like QCD in the dark sector, $n_{g'}$ can be half-integer.
The one-loop beta function of the dark QCD coupling is proportional to $11 - 4 n_{g'}/3$. 
The dark QCD is asymptotically free as long as $n_{g'} \leq 8$, and this gives a rough lower bound on the ADM mass to be larger than $1.1\,\mathrm{GeV}$. 
On the other hand, the large number of generations causes the Landau pole of the dark QED coupling.
The one-loop beta function of the dark QED coupling is proportional to $20 n_{g'}/9+1/3$.
The dark QED coupling does not diverge up to the Planck scale when we take the dark QED coupling to be $\alpha' \simeq 0.06/(n_{g'}+0.15)$ at $1\,\mathrm{GeV}$ for $n_{g'}$ generations.
In the following, we assume $N_2$ is the lightest nucleon, and we set $m_{N_2} = 8.5/n_{g'}\,\mathrm{GeV}$.

In addition to the mass difference from $\delta$ and dark quark mass difference, we can add the $U(1)_D$ correction as with the dark pions. 
The $U(1)_D$ correction is typically $\alpha' f_{\pi'}/4\pi$ in the limit of small $\theta_V$ and $\delta$ . 
\eqs{
  \Delta_N & \equiv \frac{m_{N_1} - m_{N_2}}{m_{N_2}} 
  = \Delta_N^\mathrm{QED'} + \Delta_N^\mathrm{iso} \,, \\
  \Delta_N^\mathrm{QED'} & \simeq \frac{\alpha' f_{\pi'}}{2 \pi g f_{\pi'}} \simeq 7 \times 10^{-4} \left( \frac{\alpha'}{0.1} \right) \,, \\
  \Delta_N^\mathrm{iso} & = \frac{2 g \delta}{g f_{\pi'}} 
  \simeq \frac{2 (M_1-M_2)}{\Lambda_{\chi\mathrm{SB}'}}
  \simeq 0.02 \left( \frac{10\,\mathrm{GeV}}{\Lambda_{\chi\mathrm{SB}'}}\right) \left( \frac{M_1-M_2}{0.1\,\mathrm{GeV}} \right)\,.
}
As for the dark nucleons, we ignore the $U(1)_D$ correction in the following since $\Delta_N^\mathrm{QED'}$ is smaller than $\Delta_N^\mathrm{iso}$.

The dark hadrons with different $U(1)_D$ charges can mix with each other since $U(1)_D$ is broken in our model.
In this case, the heavier dark hadrons mainly decay into the lighter ones and (off-shell) dark photon.
The charge matrices in the mass basis are important for determining the decay rate.
As we assume the quark mass matrix $M_{Q'}$ to be hermitian in this article, the diagonalizing matrices are identified, $U_L = U_R = U_V$.
The dark hadrons in the charge basis are obtained by the same unitary matrix $U_V$:
\begin{align}
  \hat N & = U_V N \,, &
  \hat{\overline N} & = \overline N U_V^\dag\,, &
  \hat \Phi & = U_V \Phi U_V^\dag \,.
\end{align}
This rotation also gives the relation between the charge matrices in the mass basis and in the charge basis as follows:
\eqs{
  q^N_L = q^N_R & = U_V^\dag \hat q_N U_V 
  = 
  \begin{pmatrix}
    \cos^2 \theta_V & - e ^{-i \alpha_V} \sin \theta_V \cos\theta_V \\
    - e ^{-i \alpha_V} \sin \theta_V \cos\theta_V & \sin^2 \theta_V
  \end{pmatrix}
  \,,
  \label{eq:charge_mat}
}
where $\hat q_N = \mathrm{diag}(1,0)$ is the charge matrix of dark nucleons in charge basis.
The dark nucleons in mass basis have an inelastic gauge interaction with dark photons,
\eqs{
  \mathcal{L} & \supset 
  - \sin \theta_V \cos \theta_V e' A_\mu \left[ 
    e^{-i \alpha_V} \left(N_2^\dag \bar\sigma^\mu N_1 
    + \overline N_2 \sigma^\mu \overline N_1^\dag\right) 
    + \mathrm{h.c.}
  \right] \\
  & + \cos^2 \theta_V e' A_\mu \left( 
    N_1^\dag \bar\sigma^\mu N_1
    + \overline N_1 \sigma^\mu \overline N_1^\dag
  \right) 
  + \sin^2 \theta_V e' A_\mu \left( 
    N_2^\dag \bar\sigma^\mu N_2 
    + \overline N_2 \sigma^\mu \overline N_2^\dag 
  \right) \,.
  \label{eq:interaction_Nucleons}
}
We ignore the phase $\alpha_V$ in the following.

Since we assume that $U(1)_D$ is spontaneously broken by the dark Higgs $\phi_D$, the dark photon mass is not related to the mass spectrum of dark hadrons. 
We take the mass of dark photons as a free parameter in this study. 
The dark photon mass is related with the dark pion mass, for instance in a chiral setup of the dark confining sector~\cite{Harigaya:2016rwr,Co:2016akw,Ibe:2021gil}.

\section{Dark Hadron Mass Spectrum and Lifetime \label{sec:existing}}

We have discussed the generic mass spectrum and interactions in the composite ADM model with a dark photon in the previous section.
The search strategies for the long-lived particles that are produced at the collider experiments depend on their lifetime.

When the produced particles are quite long-lived or are stable in the dark sector, they leave missing signals or signals from scattering with the detector material. 
In the composite ADM models, the dark proton which has a $U(1)_D$ charge scatters with the SM particle via dark photon, and then the scattering gives the recoil energy of nuclei in the detector material~\cite{Izaguirre:2013uxa}. 
When the dark neutron that does not have a $U(1)_D$ charge is stable, it escapes from the detectors, and then this process has been and will continue to be explored by searches for invisible decay at NA64~\cite{Banerjee:2016tad,Banerjee:2017hhz}, LDMX~\cite{Izaguirre:2014bca,Akesson:2018vlm}, and Belle-II~\cite{Essig:2013vha,Kou:2018nap}.

The dark pions are lightest among the dark hadrons. 
When the dark pions are lighter than dark photons, the dark pions decay into visible particles with off-shell dark photons. 
In this case, the dark pions are quite long-lived, and hence their decay leads to problematic energy injection in electromagnetic channels at the late time, in particular, when their lifetime is longer than $10^{4}\,\mathrm{s}$~\cite{Poulin:2016anj}. 
As for the dark photon decay, we have different phenomenology when a decay mode $A' \to \pi' \pi'$ opens.
The decay has been investigated by the invisible dark photon decay searches.

On the other hand, in this study, we focus only on the visible decay signals from dark hadrons. 
The dark hadrons should have decay length of $\mathcal{O}(1)\,\mathrm{m}$ in order to leave visible signals at the detector. 
To begin with, we discuss the lifetime of the dark hadrons, and determine the mass spectrum that is the most relevant to the visible decay searches.
Then, we fix the nucleon mass spectrum from the direct detection experiment for the dark matter. 
Dark photons are lightest in the dark sector, and thus we discuss the dark photon searches in the end of this section.
We will discuss the production of dark hadrons at the collider experiments and the fixed-target experiments in \cref{sec:darkhadron}.

\subsection{Decay and Transition of Dark Hadrons \label{sec:decay_hadrons}}

When the mass difference of the dark nucleons is smaller than the dark photon mass $m_{A'}$, the heavier dark nucleon $N_1$ mainly decays to the lighter nucleon $N_2$ with the SM fermions.
The differential decay rate is given by
\eqs{
  \frac{d\Gamma(N_1 \to N_2 + f \bar f)}{d s_{f\bar f}} = 
  \frac{1}{\pi} \Gamma_{N_1 \to A' N_2}(s_{f\bar f}) \frac{\sqrt{s_{f\bar f}} \Gamma_{A'}(m_{A'}=\sqrt{s_{f\bar f}})}{(s_{f\bar f}-m_{A'}^2)^2 + (\sqrt{s_{f\bar f}} \Gamma_{A'})^2 }
  \,,
}
Here, $s_{f\bar f} \equiv (p_f + p_{\bar f})^2$.
$\Gamma_{N_1 \to A' N_2}$ is the decay width into on-shell dark photon with the dark photon mass replaced with $m_{A'} \to \sqrt{s_{f\bar f}}$.
$\Gamma_{A'}$ denotes the decay width of the dark photon with the mass of $\sqrt{s_{f\bar f} }$, which is given by \cref{eq:darkphotondecay}. 
We include the hadronic final state when the fictitious mass $\sqrt{s_{f\bar f}}$ exceeds the threshold.
The momentum transfer is much smaller than the dark photon mass, $s_{f\bar f} \ll m_{A'}^2$, when the decay mode into on-shell dark photon does not open.
In this case, in the small limit of $s_{f\bar f}$ and the mass difference, denoted by $\Delta_N = (m_{N_1}-m_{N_2})/m_{N_2}$, we get
\eqs{
  \frac{d\Gamma(N_1 \to N_2 + f \bar f)}{d s_{f\bar f}} \simeq 
  \frac{1}{\pi} \frac{\alpha'}{2 s_{f\bar f}} \sin^2 2 \theta_V (m_{N_2}^2 \Delta_N^2 - s_{f\bar f})^{3/2} \frac{\sqrt{s_{f\bar f}} \Gamma_{A'}(m_{A'}=\sqrt{s_{f\bar f}})}{m_{A'}^4}
  \,,
  \label{eq:DecayRate_Nucleon}
}
where $\alpha' = e'^2/4\pi $ is the fine structure constants of $U(1)_D$.
Once the mass of the SM fermions is neglected, the decay rate is approximately given by
\eqs{
  \Gamma(N_1 \to N_2 + f \bar f) = \frac{\epsilon^2 \alpha' \alpha Q_f^2}{15 \pi} \sin^2 2 \theta_V \frac{m_{N_2}^5 \Delta_N^5}{m_{A'}^4} \,.
}
Here, $\alpha = e^2/4\pi$ is the fine structure constants of the electromagnetism and $Q_f$ is the electromagnetic charge of $f$.
The approximate decay length is given by
\eqs{
  c \tau(N_1 \to N_2 + f \bar f) \simeq 3 \,\mathrm{m}
  \left( \frac{0.1}{\Delta_N} \right)^5
  \left( \frac{8.5~\mathrm{GeV}}{m_{N_2}} \right)^5
  \left( \frac{m_{A'}}{3~\mathrm{GeV}} \right)^4
  \left( \frac{10^{-2}}{\sin2\theta_V} \right)^2
  \left( \frac{5\times 10^{-3}}{\epsilon} \right)^2
  \left( \frac{\alpha}{\alpha'} \right) \,. 
}
The lifetime of the heavier state is much smaller than 1\,s, and then the heavier states that are produced in the early universe do not lead to the problematic energy injection to the electromagnetic channels at the late time.
Therefore, their decay is free from the cosmological constraints, and we focus on the visible decay searches and will discuss future sensitivities to the decay in \cref{sec:LHCNucleon}.

Similarly, the heavier dark pion (with mass $m_{\pi'^1}$) can decay into the lighter one (with mass $m_{\pi'}$) by emitting the off-shell dark photon.
Once we neglect mass of final state fermions, the decay rate through the dark pion transition is given by
\eqs{
  \Gamma(\pi'^1 \to \pi'^3 f \bar f) =
  \frac{\epsilon^2  \sin^2 (2 \theta_V + \beta_V) Q^2_f \alpha' \alpha}{15 \pi} \frac{m_{\pi'}^5}{m_{A'}^4} \Delta_{\pi'}^5 \,.
}
Here, $\Delta_{\pi'} = (m_{\pi'^1} - m_{\pi'})/m_{\pi'}$ and $2 \theta_V + \beta_V$ denotes the mixing angle between charged and neutral dark pions, which we discuss in \cref{sec:spectrum}.
However, since dark pions are lighter than dark nucleons, decay length of pions through transition tends to be much longer than that of nucleons. 
Unless pion mass is close to nucleon mass, the dark pion decay via the pion transition is expected to be beyond the sensitivity of the visible decay searches. 
Therefore, we do not include the dark pion transition in this study.

\begin{figure}
  \centering
  \includegraphics[width=\textwidth]{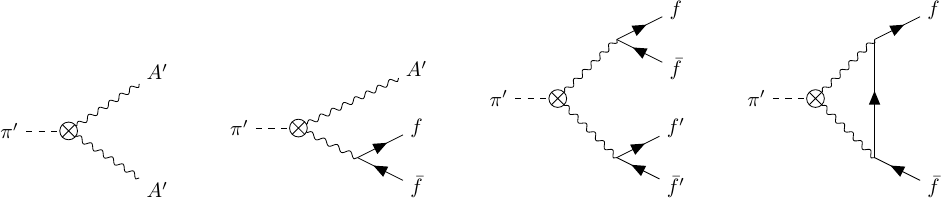}
  \caption{
    Feynman diagrams for the dark pion decay to (off-shell) dark photon. 
  }
  \label{fig:piondecay}
\end{figure}

A dark-neutral pion can couple to dark photons through the anomaly-induced interaction:
\eqs{
  \mathcal{L}_\mathrm{eff} & = - \frac{\alpha'}{4 \pi f_{\pi'}} \frac{1}{2i} \mathrm{tr}[\hat q^2 (\hat\Phi - \hat \Phi^\dag)] F'^{\mu \nu} \tilde F'_{\mu\nu} \\
  & = - \frac{\alpha'}{8 \pi f_{\pi'}} \left[ \cos(2 \theta_V - \beta_V) \pi'^3 - \sin(2 \theta_V - \beta_V) \pi'^1 \right]F'^{\mu \nu} \tilde F'_{\mu\nu} \,.
}
Here, $f_{\pi'}$ is a decay constant of the dark pion.
The dark pion mainly decays into (off-shell) dark photons via this interaction: as illustrated in \cref{fig:piondecay}, decay into two on-shell dark photons as $2 m_{A'} < m_{\pi'}$, decay into a dark photon and a pair of the SM fermions as $m_{A'} < m_{\pi'} < 2 m_{A'}$, and decay only into the SM fermions as $m_{\pi'} < m_{A'}$.
The decay rates to two dark photons and to a dark photon and a pair of the SM fermions are, respectively, given by
\begin{align}
  \Gamma(\pi'^3 \to A' A') & = \frac{1}{16 \pi} \left( \frac{\alpha' \cos(2 \theta_V - \beta_V)}{2\pi f_{\pi'}}\right)^2 (m_{\pi'}^2 - 4 m_{A'}^2)^{3/2} \,, \\
  \Gamma(\pi'^3 \to A' f \bar f) &= \frac{\epsilon^2 Q_f^2 \alpha}{64 \pi^2} \left( \frac{\alpha' \cos(2 \theta_V - \beta_V)}{2\pi f_{\pi'}}\right)^2 m_{\pi'}^3 F\left( \frac{m_{A'}}{m_{\pi'}} \right) \,.
\end{align}
Here, the decay rate for $\pi'^3 \to A' f \bar f$ is computed neglecting the fermion mass, and $F(x)$ is a function originated from the three-body phase space integral:
\eqs{
  F(x)
  & = - \frac19 (x^6 + 63 x^4 -81x^2+17) + \frac23 (6x^4 - 9 x^2 + 1) \ln x \\ 
  & \qquad - \frac23 \frac{1- 11x^2+28x^4}{\sqrt{4x^2-1}} \left[ \tan^{-1}\left( \frac{1-3x^2}{(1-x^2)\sqrt{4x^2-1}} \right) + \frac{\pi}{2}\right]\,.
}
This function vanishes as $x \to 1$.
The dark pions promptly decay into to dark photons when $2 m_{A'} < m_{\pi'}$.
On the other hand, the dark pions are long-lived when $m_{A'} < m_{\pi'} < 2 m_{A'}$ due to the three-body phase space and the kinetic mixing $\epsilon$.
When we set $m_{\pi'} = 1.8 m_{A'}$, a proper decay length of the dark-neutral pions is given by
\eqs{
  c \tau(\pi'^3 \to A' + f \bar f) \simeq 0.3 \,\mathrm{m}
  \left( \frac{1.0~\mathrm{GeV}}{m_{A'}} \right)^3
  \left( \frac{f_{\pi'}}{0.8~\mathrm{GeV}}\right)^2
  \left( \frac{10^{-3}}{\epsilon} \right)^2
  \left( \frac{0.1}{\alpha'} \right)^2 \,. 
}

When the dark photon is heavier than the dark pion, all final states of the decay process are the SM fermions.
As shown in \cref{fig:piondecay}, we have two processes only with the SM fermion final states: one is the four-body tree-level decay, and another is the loop-induced two-body decay.
We use the four-body decay rate in the massless fermion limit and the two-body decay rate that are computed in Ref.~\cite{Katz:2020ywn} as dark $\eta$ meson decay.
We take into account the difference of the anomaly coefficients as the dark pion decay.
\eqs{
  \Gamma(\pi'^3 \to f \bar f f' \bar f') & = \frac{16 \epsilon^4 \alpha'^2 \alpha^2 \cos^2(2 \theta_V - \beta_V) Q^2_f Q^2_{f'} m_{\pi'}}{6301 \times 8 \pi (4 \pi)^4} \left(\frac{m_{\pi'}^5}{f_{\pi'} m_{A'}^4} \right)^2 \,, \\
  \Gamma(\pi'^3 \to f \bar f) & = \frac{21}{4} \frac{\epsilon^4 \alpha'^2 \alpha^2 \cos^2(2 \theta_V - \beta_V) Q_f^4}{8 \pi (4 \pi)^4} \frac{(m_{\pi'}/2)^5 m_f^2}{f_{\pi'}^2 m_{A'}^4} \sqrt{1-\frac{4 m_f^2}{m_{A'}^2}} \,.
}
Due to the extra suppression from kinetic mixing $\epsilon^4$ and the four-body phase space or the loop suppression factor, the decay rate without an on-shell dark photon final state is very tiny.
Hence, the visible decay searches are not promising when the dark pion is lightest in the dark sector.

\begin{figure}
  \centering
  \includegraphics[width=8cm]{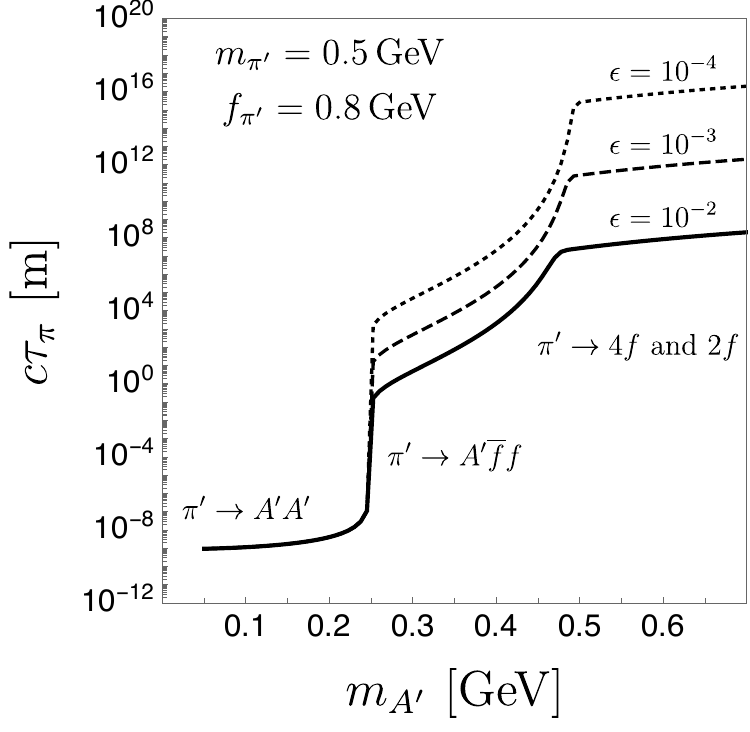}
  \caption{
    Decay lengths of dark-neutral pions: 
    The kinetic mixing is assumed to be $\epsilon = 10^{-2}$ (solid), $\epsilon = 10^{-3}$ (dashed), and $\epsilon = 10^{-4}$ (dotted).
    We take $\alpha' = 0.05 \,, m_{\pi'} = 0.5\,\mathrm{GeV} \,, f_{\pi'} = 0.8\,\mathrm{GeV}$.
    We include all possible final state for $\pi'^3 \to A' f \bar f$, but include only leptonic final states for $\pi'^3 \to f \bar f$ and $\pi'^3 \to f \bar f f' \bar f'$.
  }
  \label{fig:piondecaylength}
\end{figure}

We show the decay length of the $U(1)_D$-neutral dark pion in \cref{fig:piondecaylength} with the dark pion mass of $0.5\,\mathrm{GeV}$ and the decay constant $f_{\pi'} = 0.8\,\mathrm{GeV}$.
The different line-types in the figure correspond to different values of kinetic mixing $\epsilon$: $\epsilon = 10^{-2}$ (solid), $\epsilon = 10^{-3}$ (dashed), and $\epsilon = 10^{-4}$ (dotted).
Each lines is composed of three parts: prompt decay via $\pi'^3 \to A' A'$ for $2 m_{A'} < m_{\pi'}$, long-lived decay via $\pi'^3 \to A' f \bar f$ for $m_{A'} < m_{\pi'} < 2 m_{A'}$, and very long-lived decay via four-body/loop-induced decay.
As shown in \cref{fig:piondecaylength}, dark pions would have the typical decay length of $\mathcal{O}(10^{-2})\text{--}\mathcal{O}(10^2)\,\mathrm{m}$ via three-body decay when $m_{\pi'}$ is close to $2 m_{A'}$.
Therefore, the visible decay searches have a potential to explore the dark pion decay in this mass range.

The darkly charged pions can also decay through the pion mixing and chiral anomaly. 
When the darkly charged pions decay into the SM particles, their lifetime is longer than dark-neutral pions due to the extra factor of the pion mixing. 
When the kinetic mixing is of order of $\mathcal{O}(10^{-3})$, $\pi'^\pm \to A' A'$ will give the visible signal since the dark photon promptly decays into the visible particles in this range of the kinetic mixing.
To get the decay length of $\mathcal{O}(1)\,\mathrm{m}$, the pion mixing angle should be or $\mathcal{O}(10^{-4})$ or below. 
Meanwhile, we assume the dark nucleon mixing $\theta_V$ to be $\mathcal{O}(10^{-2})$ in order that the decay length of the dark nucleon transition is of $\mathcal{O}(1)\,\mathrm{m}$.
The dark pion mixing is expected to be the same order of magnitude as the dark nucleon mixing, and hence we focus only on the decay of dark-neutral pions in this study.

In this article, we focus on the visible decay searches, namely we focus on $ m_{A'} \leq m_{\pi'} \leq 2 m_{A'}$.
We will discuss the sensitivity of the searches for the dark pion decay in \cref{sec:pion_seaquest}.

\subsection{Direct Detection \label{sec:directdetection}}

The DM has a coupling to the dark photon through the nucleon mixing as shown in \cref{eq:interaction_Nucleons}, and hence the DM direct detection experiments put constraint on the kinetic mixing $\epsilon$.
The $U(1)_D$ charge of the lightest dark nucleon is $\sin^2 \theta_V$, where $\theta_V$ is the nucleon mixing angle.
The predicted ADM mass is $m_{N_2} = 8.5/n_{g'} \, \mathrm{GeV}$, and in particular we consider $m_{N_2} = 8.5 \,, 4.3 \,, 2.1 $, and $1.1 \, \mathrm{GeV}$ (corresponding to $n_{g'} = 1\,, 2\,, 4 \,,$ and 8\,, respectively) in the following.
The scattering of the ADM off nuclei leaves recoil energy that is close to the energy threshold of the direct detection experiments with liquid noble gas.
For this reason, we use the constraints given by the direct detection experiments with unconventional methods, in particular for the light ADM.

The current bounds on the DM-nucleon cross section is the following.
The direct detection experiments using liquid xenon constraint on the DM-nucleon cross section get weaker for the DM lighter than $10\,\mathrm{GeV}$.
PandaX-II experiment puts constraints on the DM-nucleon cross section both via a heavy mediator (contact interaction of DM and nucleons) and via a light mediator~\cite{Ren:2018gyx,PandaX-II:2021lap}.
When the DM mass is $8.5\,\mathrm{GeV}$ and the dark photon mass is $1.0\,\mathrm{GeV}$, the upper bound on the cross section is given by $\sigma \leq 6 \times 10^{-45} \, \mathrm{cm}^2$.
Comparable constraints are obtained from XENON1T (S2-only technique, using ionization signals)~\cite{XENON:2019gfn} and LUX~\cite{LUX:2016ggv}, and one order of magnitude severer constraints are obtained from XENON1T (with the conventional method)~\cite{XENON:2018voc}.
We use the constraints on the spin-independent cross section for the DM mass of 4.3\,GeV and 2.1\,GeV by DarkSide-50~\cite{DarkSide:2018bpj}, which uses the liquid argon.
The corresponding bounds on the DM-nucleon cross section are given by $\sigma \leq 2 \times 10^{-42} \, \mathrm{cm}^2$ for the DM mass 4.3\,GeV and $\sigma \leq 10^{-41} \, \mathrm{cm}^2$ for the DM mass 2.1\,GeV. 
Even though we use the constraints by DarkSide-50, as for the DM mass of 4.3\,GeV, XENON1T has put a constraint one order of magnitude severer than that of DarkSide-50~\cite{XENON:2019gfn}.
Direct detection experiments with solid-state detectors have placed constraints on the DM-nucleon cross section even for the DM mass of GeV.
The CRESST-III experiment has already set an upper bound on the cross section to be less than $5 \times 10^{-39}\,\mathrm{cm}^2$ for $1\,\mathrm{GeV}$ DM mass~\cite{CRESST:2019jnq}, which is about six orders of magnitude larger than the constraint for the ADM mass of $8.5\,\mathrm{GeV}$. 
A light dark matter search at XENON1T (with the use of Migdal effect~\cite{Migdal,Baur:1983,Ibe:2017yqa}) has reported, and the constraint is comparable to the CRESST-III experiment~\cite{XENON:2019zpr}.

When the dark protons are the lightest dark nucleon, the bound on the kinetic mixing is given by $\epsilon \leq 1.4 \times 10^{-7} (m_{A'}/1\,\mathrm{GeV})^2$ as the DM mass $m_{N_2} = 8.5 \,\mathrm{GeV}$ and the fine structure constant $\alpha' = \alpha$~\cite{Ibe:2018juk}.
As we will see in the next subsection, this bound is far below the future sensitivity to the visible decay searches of dark photons.
In contrast, direct detection bounds get much weaker when dark-neutral baryons, \textit{i.e.}, dark neutrons, are the lightest nucleons since dark neutrons do not have an electric coupling to dark photons.%
\footnote{
  The dark neutrons can have magnetic dipole interactions to the dark photon that originates from magnetic dipoles of the dark quarks (see an Appendix in Ref.~\cite{Kamada:2020buc}). 
  Due to the magnetic coupling, the scattering cross section is suppressed by the DM velocity, and then the bound on $\epsilon$ is weakened by $1/v^2 \sim 10^6$ from the bound in the case of dark protons. 
  When the dark QCD has a sizable parity violation, dark neutrons would have the electric dipole moment via the dark-QCD $\theta$ term. 
  In this case, the velocity suppression of the cross section via the electric dipole interaction is milder than that via the magnetic dipole since electric field couples to the nucleon current without any velocity suppression. 
  When the dark sector violates parity, the direct detection may place a severe constraint on the kinetic mixing and the dark-QCD $\theta$.
} 
Indeed, a dark-neutral baryon is essential in composite ADM models, in order to share the generated asymmetry between two sectors via the portal interactions \cref{eq:Intermediate_Portal}.

In our study, the lightest dark nucleon is mainly composed of the dark neutron, and mixes with the dark proton.
The bounds from the direct detection experiments are replaced by the bound on the kinetic mixing and the nucleon mixing as follows.
\eqs{
  \epsilon \sin^2 \theta_V \leq 1.4 \times 10^{-7} \left( \frac{m_{A'}}{1\,\mathrm{GeV}} \right)^{2}\left( \frac{\alpha'}{1/137} \right)^{-1/2} \left( \frac{\sigma^\mathrm{bound}}{6 \times 10^{-45} \, \mathrm{cm}^2} \right)^{1/2} \,.
  \label{eq:DDBound}
}
Here, $\sigma^\mathrm{bound}$ denotes the upper bound on the nucleon-DM scattering cross section given by PandaX-II, DarkSide-50, and CRESST-III as shown above.
The upper bound on the kinetic mixing depends also on the detector material as $A/Z$, which is taken to be that in PandaX-II, namely xenon.
The upper bound on the kinetic mixing is relaxed thanks to the small mixing $\theta_V$.
As for the ADM with the mass of $8.5\,\mathrm{GeV}$ and small $m_{A'}$, we take into account the constraints on the scattering cross section with a light mediator given by PandaX-II~\cite{Ren:2018gyx,PandaX-II:2021lap}.

\subsection{Dark Photon Searches \label{sec:darkphotonsearch}}

The search strategy of the dark photon visible decay changes depending on lifetime.
Various prompt decay searches have constrained the dark photon parameters. 
For the dark photon mass above GeV, dark photons can be produced via Drell-Yan production and heavy rare meson decay in collider experiments, and hence collider experiments put constraints on the kinetic mixing from the absence of the signals.
LHCb has performed the inclusive searches for the dark photon (\textit{e.g.}, $pp \to X A' \to X \mu^+ \mu^-$ with any final states $X$), and put the upper bound on the kinetic mixing of $\epsilon \lesssim 10^{-3}$ from prompt-like decay search~\cite{Ilten:2015hya,Ilten:2016tkc,Aaij:2017rft}.
At electron colliders, dark photons are produced by meson decay (\textit{e.g.}, $\phi \to \eta A'$) and radiative return process ($e^+e^- \to \gamma A'$).
The absence of a deviation from the SM background has been already reported at BaBar~\cite{Lees:2014xha} and KLOE~\cite{Archilli:2011zc,Babusci:2012cr,Anastasi:2015qla,Anastasi:2016ktq} and has set upper bounds on the kinetic mixing in a broad dark photon mass range from MeV to GeV. 

There have been several bounds on dark photon parameters from astrophysical observations.
New light particles with weak coupling to the SM particles can be produced in the hot supernova core via mainly bremsstrahlung, and can be radiated. 
This process provides a new cooling process of supernova, and leads to inconsistency with the observation of neutrinos from supernova, SN1987A~\cite{Rrapaj:2015wgs,Chang:2016ntp,Hardy:2016kme,Mahoney:2017jqk,Chang:2018rso}.
A typical decay length of the dark photon that is constrained by SN1987A is about 10\,km corresponding to the size of the supernova core.
The dominant bound arises from the supernova cooling for dark photons with the mass below $100~\mathrm{MeV}$ and with the kinetic mixing of $\mathcal{O}(10^{-9}\text{--}10^{-7})$, and therefore the parameter space constrained by SN1987A is beyond our interests.

Fixed-target experiments have the sensitivity to the parameter space between constraints by the prompt decay search in collider experiments and the SN1987A observation. 
Electron beam-dump experiments, which have the beam dump area without any tracking system of daughter particles, put constraints on the parameter space above $\epsilon \gtrsim 10^{-7}$ and below a few hundred MeV. 
The dark photons are produced via Bremsstrahlung at the electron beam-dump experiments.
The strong constraints have mainly come from the E137~\cite{Bjorken:1988as} and E141~\cite{Riordan:1987aw} experiments at SLAC, the Fermilab E774 experiment~\cite{Bross:1989mp}, Orsay~\cite{Davier:1989wz}, and KEK~\cite{Konaka:1986cb}.
Though they have mainly searched for axion-like particles and light Higgs bosons, the constraints on the dark photon models are obtained by reanalyzing these constraints~\cite{Bjorken:2009mm,Andreas:2012mt}.
The electron fixed-target (non beam-dump) experiments have a sensitivity between those of prompt decay experiments and beam-dump experiments since their detectors are not located very far from targets.
There have been the electron-beam fixed target experiments searching for visible final states: A1/MAMI~\cite{Merkel:2014avp}, APEX~\cite{Essig:2010xa,APEX:2011dww}, and HPS~\cite{Battaglieri:2014hga,Celentano:2014wya,HPS:2018xkw}.
Further running of APEX and HPS is planned in the near future.

Similarly to the electron beam-dump experiments, data from previous proton beam-dump experiments have been also reanalyzed to constrain the dark photon visible decay: 
CHARM~\cite{Bergsma:1985is,Gninenko:2012eq}, LSND~\cite{Athanassopoulos:1997er,Batell:2009di}, and U70/$\nu$Cal~\cite{Blumlein:2011mv,Blumlein:2013cua}. 
The use of proton beam provides the enhancement of dark photon production, due to high penetration rate of protons on target materials and enhanced production mechanism from meson decays besides Bremsstrahlung processes.
The on-going and future fixed-target experiments with proton beam, SeaQuest~\cite{Aidala:2017ofy,Berlin:2018pwi} and SHiP~\cite{Anelli:2015pba,Alekhin:2015byh}, have been proposed. 
Since the decay volumes of the proposed experiments are shorter than those of the past proton beam-dump experiments, the proposed experiments have the sensitivity to dark photons with short lifetime.

\section{Dark Hadron Searches \label{sec:darkhadron}}

In this section, we discuss visible decay searches of dark hadrons at the lifetime frontier. 
In particular, we consider the dark nucleon transition and the three-body decay process of the dark-neutral pion.
We assume that the dark hadrons made of one of $n_{g'}$ generations dark quarks are predominantly produced at the LHC and the fixed-target experiments.
The dark hadrons are produced via off-shell dark photons at the LHC and the fixed-target experiments since the dark photons are lightest in the dark sector. 
The differential cross section for the production of dark hadrons is written as~\cite{Izaguirre:2013uxa}
\eqs{
  \frac{d \sigma}{dm_{A'}^{\ast2} dx}
  = \frac{d \sigma_{A'}(m_{A'}^{\ast})}{dx}
  \times \frac{1}{\pi} \frac{m_{A'}^{\ast} \Gamma_{A'}(A' \to \text{dark hadrons})}{(m_{A'}^{\ast2}-m_{A'}^2)^2 + (m_{A'}^{\ast} \Gamma_{A'})^2}\,.
}
Here, $x$ collectively denotes the variables for the dark photon production, and $m_{A'}^{\ast}$ is the fictitious dark photon mass corresponding to the invariant mass of final states.
$\sigma_{A'}$ denotes the cross section of on-shell dark photon with the dark photon mass $m_{A'}^{\ast}$ being the fictitious mass, and $\Gamma_{A'}$ is the partial decay width of $A'$ with the mass of $m_{A'}^{\ast}$. 
From this formula, we evaluate the produced number of dark hadrons via the off-shell dark photons when the invariant mass is above the dark photon mass as follows.
\eqs{
  N \simeq 
  \int d m_{A'}^{\ast2} \frac{1}{\pi} \frac{m_{A'}^{\ast}\Gamma_{A'}(m_{A'} = m_{A'}^{\ast})}{m_{A'}^{\ast 4}} \left. N_{A'} \right|_{m_{A'} = m_{A'}^{\ast}} \,.
  \label{eq:hadron_producednumber}
}
Here, $N_{A'}$ denotes the number of produced dark photons.

The visible signal would be detected when the long-lived particles decays into the SM particles inside the detector volume or generically in a specific region.
The number of signals is roughly evaluated as follows:
\begin{align}
  N_\mathrm{signal} & \simeq 
  \int d m_{A'}^{\ast2} (\mathrm{eff}) \frac{1}{\pi} \frac{m_{A'}^{\ast}\Gamma_{A'}(m_{A'} = m_{A'}^{\ast})}{m_{A'}^{\ast 4}} \left. N_{A'} \right|_{m_{A'} = m_{A'}^{\ast}}
  \simeq N \times (\mathrm{eff}) \,, 
  \label{eq:Signal_Number} \\
  (\mathrm{eff}) & \simeq \langle (e^{-r_\mathrm{min}/d}-e^{-r_\mathrm{max}/d}) A_\mathrm{eff} \rangle \,.
\end{align}
Here, we assume the branching fraction of the long-lived particles to the visible particles to be unity.
$(\mathrm{eff})$ denotes the total efficiency when the geometric detector acceptance has weak dependence on the location of the decay: $A_\mathrm{eff}$ is the acceptance of detector, $r_\mathrm{min}$ and $r_\mathrm{max}$ are, respectively, distances of the detector starting and end points from the interaction point, and $d \equiv c \tau p m^{-1}$ is the boosted decay length of the long-lived particle with the boost factor $p m^{-1}$, and $\langle \dots \rangle$ denotes the ensemble average.
Though the efficiency depends on $m_{A'}^\ast$, the fictitious mass of off-shell dark photon, we factorize the efficiency $(\mathrm{eff})$ at $m_{A'}^\ast$ where the signal number $N_\mathrm{signal}$ is dominated. 
The acceptance $A_\mathrm{eff}$ includes the geometric acceptance $A^\mathrm{geo}$ and the depletion of the produced numbers at a certain momentum differs from typical momentum $A^\mathrm{prod}$: $A_\mathrm{eff} = A^\mathrm{geo} A^\mathrm{prod}$.
Since the detectors are located away from the interaction points at collider or the beam-dump areas, the angle of the emitted particle with respect to the beam-axis is essential to detect the visible signal.
The angle of the emitted particle is determined by the decay of off-shell dark photon into the long-lived particles since the emission of the off-shell dark photon is collinear to the beam axis in the following.
Meanwhile, the energy threshold of detectors is imposed for some reasons discussed later, for instance to remove low-energy backgrounds. 
When the long-lived particles predominantly coming to a detector have the energy below the threshold, most of long-lived particles is vetoed, and the number of long-lived particles leaving visible signal decreases.
We will discuss the acceptance that we use in this study later in detail.

In the following, we consider the LHC experiment and the fixed-target experiments. 
The dark photons are mainly produced through the Drell-Yan process in the former and the Bremsstrahlung process in the latter. 
We use the produced number of dark photons through the Drell-Yan process at the LHC~\cite{Berlin:2018jbm} and through the Bremsstrahlung process at the electron beam fixed-target experiment, E137~\cite{Izaguirre:2013uxa}, and at the proton beam fixed-target experiment, SeaQuest~\cite{Berlin:2018pwi}.
The produced number $N_{A'}$ via the Drell-Yan process and the electron Bremsstrahlung process is proportional to $\epsilon^2 m_{A'}^{-2}$, while $N_{A'}$ via the proton Bremsstrahlung does not depend on the dark photon mass for $m_{A'} \lesssim 500\,\mathrm{MeV}$.
As for the proton Bremsstrahlung, $N_{A'}$ is enhanced due to the mixing of $A'$ and the SM $\rho$-meson near $m_{A'} \simeq m_\rho$~\cite{deNiverville:2016rqh}, and $N_{A'}$ is sharply suppressed due to the form-factor for $m_{A'}$ above $1\,\mathrm{GeV}$, where the internal structure of nucleons is essential.%
\footnote{
  The forward production at the proton Bremsstrahlung has an uncertainty due to non-perturbative effect of QCD \cite{Foroughi-Abari:2021zbm}.
}

When the long-lived particle is elementary, the decay rate $\Gamma_{A'}$ is proportional to $m_{A'}$, and therefore the integral given by \cref{eq:hadron_producednumber} is dominated by the infrared (IR) contribution for the electron Bremsstrahlung.
As discussed in Ref.~\cite{Izaguirre:2013uxa} for the electron Bremsstrahlung, through the off-shell dark photon, the dark-sector particles are dominantly produced at $m_{A'}^\ast$ near the kinematical threshold $2 m_\chi$ with $m_\chi$ being the dark-sector particle mass: $(\alpha'/\pi) N_{A'}|_{m_{A'}^{\ast} = \sqrt{10} m_\chi}$.
On the other hand, when we consider composite particles, the decay rate can have a higher power of $m_{A'}$, and then the UV contribution (\textit{e.g.}, dark dynamical scale) can dominate the produced number $N$.
We will discuss the UV dominance in the dark hadron production later. 
As for the proton Bremsstrahlung, $N$ is predominantly determined by the SM $\rho$-meson resonance even when we consider composite particles due to the enhancement of $N_{A'}$ at the $\rho$-meson resonance.

\subsection{Dark Nucleon Searches at the LHC \label{sec:LHCNucleon}}

We discuss the visible searches at the LHC lifetime frontier for the transition of the dark nucleon. 
The off-shell dark photon is produced via the Drell-Yan process, and then $N_1$, the heavier dark nucleon, is produced via the dark hadronization when a sufficient energy through the off-shell dark photon is injected to the dark sector.
Various experiments have been proposed to explore long-lived particles with the decay length of $\mathcal{O}(1) \text{-} \mathcal{O}(10^2)\,\mathrm{m}$ at the LHC: MATHUSLA~\cite{Chou:2016lxi}, FASER~\cite{Feng:2017uoz}, and CODEX-b~\cite{Gligorov:2017nwh}.
First, we summarize the LHC lifetime frontier.

The MATHUSLA experiment has been proposed~\cite{Chou:2016lxi} to search for long-lived particles at the LHC. 
Its detector would be placed on the surface $\sim 100\,\mathrm{m}$ above the LHC beam line and $\sim 100\,\mathrm{m}$ downstream from the interaction point, and would take data during the HL-LHC. 
The MATHUSLA detector consists of on-top five tracking layers with timing resolution of $1\,\mathrm{ns}$, air-filled decay volume, and floor-detector reducing background muons from the CMS interactions point.
Although, different decay volumes, so called MATHUSLA50, MATHUSLA100, and MATHUSLA200, have been considered in the MATHUSLA proposal~\cite{Chou:2016lxi,Alpigiani:2018fgd,Lubatti:2019vkf}, we focus only on MATHUSLA200 with decay volume of $200\,\mathrm{m} \times 200\,\mathrm{m} \times 20\,\mathrm{m}$ in this article.
At the MATHUSLA experiment, various backgrounds are expected to be reduced by the following. 
The rock is present between the LHC ring and the detector location, good time resolution of the tracking layer differentiates upward going (long-lived particle) signals from downward going (cosmic ray) signals.
Therefore, the searches for the long-lived particles are assumed to be background free at the MATHUSLA.

The FASER experiment has been proposed~\cite{Feng:2017uoz} to search for long-lived particles in the very forward direction of the ATLAS interaction point.
Its detector is located in the very forward region along the beam line of the LHC, and is positioned $\sim 480\,\mathrm{m}$ away from the interaction point, where the LHC tunnel just starts to curve.
The detector has a cylindrical shape, and its longitudinal axis coincides with the LHC beam collision axis.
Various versions of the FASER experiment have been proposed:
FASER~\cite{Ariga:2018zuc,Ariga:2018pin} and FASER2~\cite{Ariga:2018uku}.
The former collects data during LHC Run 3 and its detector have the cylindrical shape with $1.5\,\mathrm{m}$ depth and $10\,\mathrm{cm}$ radius, while the latter takes data during HL-LHC with the cylindrical shape of $5\,\mathrm{m}$ depth and $1\,\mathrm{m}$ radius.
As with the MATHUSLA experiment, background events are expected to be reduced by the following. 
There are the $100\,\mathrm{m}$ of rock and concrete shield of the detector from the ATLAS interaction point, and a charged particle veto in front of the detector. 
We focus on the future sensitivity of the dark nucleon at the FASER2 experiment.

Similarly to the MATHUSLA experiment, it has been proposed to put a new detector near the LHCb interaction point, which is named as the CODEX-b experiment~\cite{Gligorov:2017nwh}.
The detector would be placed about $25\,\mathrm{m}$ away from the interaction point, with a $10 \times 10 \times 10\,\mathrm{m}^3$ volume.
The background events are expected to be reduced by the existing concrete wall of $\sim 3\,\mathrm{m}$, an additional lead shield of $\sim 4.5\,\mathrm{m}$, and a charged-particle veto inside the lead shield~\cite{Aielli:2019ivi}.
In comparison with the MATHUSLA detector, CODEX-b has a smaller detector volume, the detector location is closer to the interaction point, and the integrated luminosity at LHCb is smaller than that at ATLAS. 
Due to the geometry and the expected luminosity of CODEX-b, the future sensitivity of CODEX-b is comparable with but lower than that of MATHUSLA in the inelastic dark matter model~\cite{Berlin:2018jbm}. 
Thus, we do not include the future sensitivity of CODEX-b in the following analysis.


The searches for long-lived particles at the LHC have been studied in the context of the inelastic dark matter model in Refs.~\cite{Izaguirre:2015zva,Berlin:2018jbm}.
We map their results into the dark nucleon transition (the dark-neutral pion decay in the next subsection) by taking into account important differences.
To this end, we outline their setup to compare with our model. 
In these studies, $U(1)_D$ gauge invariance allows only a Dirac mass term of two Weyl fermions, and Majorana masses arise from the spontaneous $U(1)_D$ breaking. 
In the mass basis, the lightest state $\chi_1$ is the DM and inelastically couples to the heavier state $\chi_2$ with dark photons.
Their mass splitting is controlled by the nearly degenerated Majorana masses, and the mass splitting is small when the Majorana masses are smaller than the Dirac mass.
Dark photons are assumed to be heavier than DM particles; in particular they assume $m_{A'} = 3 m_\chi$ where $m_\chi$ is the DM mass. 
$\chi_2$ is produced via dark photon decay, and hence the produced number of $\chi_2$ denoted by $N_\chi^0$ is equal to that of dark photons: $N_\chi^0 = N_{A'}$.
Then, the produced $\chi_2$ decays into $\chi_1$ and the SM fermions via off-shell dark photons.

In our study, we consider that dark photons are lighter than dark nucleons as discussed before.
In this case, we have an additional factor for production rate of $N_1$ via the dark hadronization as follows:%
\footnote{
  We only consider dark hadrons made of one of $n_{g'}$ generations. 
  When we include the whole generations, the production rate would be multiplied by $n_{g'}$.
}
\eqs{
  N_{N_1} \simeq \left. \frac{2 n_{N_1} \alpha'}{\pi} \frac53 N_{A'}\right|_{m_{A'}^\ast = \sqrt{10} m_{N_1}} \,.
  \label{eq:ProducedNumberN}
}
Here, $N_{N_1}$ denotes the number of $N_1$ and $\overline N_1$ produced in collider experiments.
The factor $5/3$ originates from the sum of the charge squared of dark quarks.
We take $m_{A'}^\ast = \sqrt{10} m_{N_1}$ by analogy with the off-shell production of the elementary dark particles~\cite{Izaguirre:2013uxa}.
This may be a crude approximation since it is not clear if the contribution near the kinematic threshold dominates the production even for the composite dark particle.
$n_{N_1}$ denotes the multiplicity of $N_1$ for the production through the dark photon mediation. 
For simplicity, we take a similar value, $n_{N_1} = 0.04$, to the multiplicity of protons by the $e^+ e^-$ annihilation for the center of mass energy just above $J/\psi$ threshold~\cite{DASP:1978ftr}.
$n_{N_1}$ depends on the choice of $m_{A'}^\ast$ since the multiplicity depends on the injected energy.
We discuss how the change of the multiplicity affects the sensitivity of the visible decay searches at the LHC lifetime frontier in \cref{sec:production}. 

The sensitivity range of the kinetic mixing has a maximum value, $\epsilon_\mathrm{max}$, and a minimum value, $\epsilon_\mathrm{min}$, with other parameters being fixed.
Due to detector designs of MATHUSLA and FASER, the visible decay products are vetoed as background events if $N_1$ decays before the decay volume area.
This means that there is a lower limit on the decay length of $N_1$, and below this limit $N_1$ cannot leave any visible signal at the detector even when $N_1$ is maximally boosted. 
It leads to the maximum value $\epsilon_\mathrm{max}$. 
The maximum value $\epsilon_\mathrm{max}$ is evaluated by use of \cref{eq:Signal_Number} for the boosted decay length $d \lesssim r_\mathrm{min} < r_\mathrm{max}$:
\eqs{
  \frac{c \tau_{N_1} p^\ast_\mathrm{max} m_{N_1}^{-1}}{r_\mathrm{min}} \ln \left(\frac{N_{N_1}}{N_\mathrm{signal}} A_\mathrm{eff}(p^\ast_\mathrm{max}) \right)\simeq 1 \,.
  \label{eq:epsilon_max_relation}
}
Here, $p^\ast_\mathrm{max}$ (subscript corresponds to $\epsilon_\mathrm{max}$, and does not necessarily coincide with the maximum energy) is given by the maximum momentum, which we will discuss later. 
Instead of directly computing $N_{N_1}$ and $A_\mathrm{eff}$, we consider a formula similar to \cref{eq:epsilon_max_relation} in the inelastic DM, with $\epsilon_\mathrm{max}$ and other parameters given in Ref.~\cite{Berlin:2018jbm}.
By using the similar formula in the inelastic DM, we can rewrite \cref{eq:epsilon_max_relation} as follows:
\eqs{
  \ln \frac{A_\mathrm{eff}(p^\ast_\mathrm{max})}{A^0_\mathrm{eff}(p_\mathrm{max})} + \ln \frac{N_{N_1}}{N^0_{\chi}} + \frac{r_\mathrm{min}}{d_0(p_\mathrm{max})} = \frac{r_\mathrm{min}}{c \tau_{N_1} p^\ast_\mathrm{max} m_{N_1}^{-1}} \,.
  \label{eq:fitting_epsilon_max}
}
Here, $d_0(p_\mathrm{max})$ is the boosted decay length of $\chi_2$ computed with the reference parameter set in the inelastic DM that corresponds to a reference point on the upper boundary $\epsilon_\mathrm{max}$ of sensitivity area in a $m_1$-$\epsilon$ plot read out from Ref.~\cite{Berlin:2018jbm} and with the momentum $p_\mathrm{max}$ same as $p^\ast_\mathrm{max}$.
In this formula, $N_{N_1}$ appears as a ratio to $N^0_{\chi}$, and therefore we only have to care about the parameter dependence of $N_{A'}$ in \cref{eq:ProducedNumberN}.
In particular, $N_{A'}$ is proportional to $\epsilon^2 m_{A'}^{-4}$ for FASER ($A'$ only in the forward region) and $\epsilon^2 m_{A'}^{-2}$ for MATHUSLA.
$A_\mathrm{eff}$ and $A^0_\mathrm{eff}$ denote the efficiencies in the composite ADM model and in the inelastic DM model, respectively. 
The efficiency also appears as a ratio to $A^0_\mathrm{eff}$, and hence we only have to care about the model parameter dependence of the efficiency.
We will discuss the efficiency in the end of this subsection in detail.
We demonstrate the validity of our approximation in \cref{app:fitting} by applying the formula to the other parameter sets in the inelastic DM model.

Concerning $p^\ast_\mathrm{max}$, we take $p^\ast_\mathrm{max} = 7 \, \mathrm{TeV}$ for FASER and $p^\ast_\mathrm{max} = 100\,\mathrm{GeV}$ for MATHUSLA.
We expect that $p^\ast_\mathrm{max}$ hardly depends on the model parameters.
Dark nucleons are assumed to be mainly produced by the Drell-Yan process at the LHC lifetime frontier.
Due to the parton distribution function, the production cross section through the Drell-Yan process sharply decreases above about $100\,\mathrm{GeV}$ (\textit{e.g.}, see Ref.~\cite{Basso:2015lua}).
This implies that it would be hard to produce the long-lived particles with the transverse momentum above $100\,\mathrm{GeV}$ at LHC collision. 
Meanwhile, in the forward direction to the beam line, the produced particles can carry the momentum of the beam energy.
Therefore, we take $p^\ast_\mathrm{max} = 100\,\mathrm{GeV} \, (7 \, \mathrm{TeV})$ for MATHUSLA (FASER).

The long-lived particles hardly decay inside the detector volume when the boosted decay length is larger than $r_\mathrm{max}$: $r_\mathrm{min} < r_\mathrm{max} \lesssim d$. 
This determines another boundary of the sensitivity plots. 
In this case, $\epsilon_\mathrm{min}$ satisfies 
\eqs{
  \frac{r_\mathrm{max}-r_\mathrm{min}}{c \tau_{N_1} p^\ast_\mathrm{min} m_{N_1}^{-1}} \frac{N_{N_1}}{N_\mathrm{signal}} A_\mathrm{eff}(p^\ast_\mathrm{min}) \simeq 1 \,,
  \label{eq:epsilon_min_relation}
}
where $p^\ast_\mathrm{min}$ (subscript corresponds to $\epsilon_\mathrm{min}$, and does not necessarily coincide with the minimum energy) denotes momentum of $N_1$ leaving signals, which we will discuss later. 
Similarly to $\epsilon_\mathrm{max}$, we estimate $\epsilon_\mathrm{min}$ by considering a formula similar to \cref{eq:epsilon_min_relation} in the inelastic DM model instead of directly calculating $N_{N_1}$ and $A_\mathrm{eff}$. 
By using the similar formula in the inelastic DM, we can rewrite \cref{eq:epsilon_min_relation} as follows:
\eqs{
  \frac{r_\mathrm{max}-r_\mathrm{min}}{c \tau_{N_1} p^\ast_\mathrm{min} m_{N_1}^{-1}} \frac{N_{N_1}}{N^0_\chi} \frac{A_\mathrm{eff}(p^\ast_\mathrm{min})}{A^0_\mathrm{eff}(p_\mathrm{min})} = \frac{r_\mathrm{max}-r_\mathrm{min}}{d_0(p_\mathrm{min})} \,.
  \label{eq:fitting_epsilon_min}
}
Here, $d_0(p_\mathrm{min})$ is the boosted decay length of $\chi_2$ computed with the parameters in the inelastic DM that corresponds to $\epsilon$ from Ref.~\cite{Berlin:2018jbm} and with the same expression of momentum $p_\mathrm{min}$ as $p^\ast_\mathrm{min}$ (with proper replacement of model parameters).
Only the ratio $N_{N_1}/N^0_{\chi}$ appears again, and hence we do not care about the overall factor to the produced number, and we estimate it by $N_{A'}$ and its parameter dependence.
$A_\mathrm{eff}$ appears only as a ratio to $A^0_\mathrm{eff}$.
The validity of our approximation is demonstrated in \cref{app:fitting} by applying our formula to the inelastic DM model.

This approximation formula determines $\epsilon_\mathrm{min}$, and hence it seems to prefer to take $p^\ast_\mathrm{min}$ to be the minimum value with which the particles leave the visible signal above the energy threshold at the detector, $p_\mathrm{thr}$, since it leads to the least boosted particles.
However, the signal number of $N_1$ with a momentum is suppressed when the momentum of $N_1$ deviates from a typical momentum of $N_1$ leaving signals.
$N_1$ leaving signals typically has the momentum $p^\ast_\mathrm{geo} \simeq (m_{A'}^{\ast 2} - 4m_{N_1}^2)^{1/2}/2\theta_\mathrm{det}$ where $\theta_\mathrm{det}$ is the angle of $N_1$ with respect to the beam axis.
We basically take $p^\ast_\mathrm{min} = p^\ast_\mathrm{geo}$ to make the acceptance maximum as much as possible, but $p^\ast_\mathrm{min}$ does not coincide with $p^\ast_\mathrm{geo}$ when the particle with the typical momentum cannot leave a visible signal due to the energy threshold (\textit{i.e.}, $p^\ast_\mathrm{geo} < p_\mathrm{thr}$).
We will discuss the energy threshold for the visible decay products at each experiment later. 
Therefore, we take $p^\ast_\mathrm{min} = \mathrm{min}(p^\ast_\mathrm{max},\mathrm{max}(p_\mathrm{thr},p^\ast_\mathrm{geo}))$ unless $p_\mathrm{thr}$ exceeds $p^\ast_\mathrm{max}$ where nothing is detected.

The produced dark nucleons have to be energetic since decay products should be sufficiently energetic because of energy threshold at the detectors.
Heavier dark nucleons decay into a lighter dark nucleon and visible particles via small mixing angle. 
The final state dark nucleon takes over most of the energy that initial dark nucleon has, and then the SM particles can be less energetic.
Hence, the energy thresholds play a significant role to determine the sensitivity.
In this study, we take the energy threshold at MATHUSLA and FASER as follows.
\eqs{
  E_\mathrm{min}^\mathrm{thr} \simeq 
  \begin{cases}
    600\,\mathrm{MeV} & \text{(MATHUSLA)} \,, \\
    100\,\mathrm{GeV} & \text{(FASER)} \,. \\
  \end{cases}
}
In Ref.~\cite{Curtin:2018mvb}, the final state energy thresholds at MATHUSLA are discussed.
The thresholds range between $E_\mathrm{min}^\mathrm{thr} \simeq 200\,\mathrm{MeV}\text{--}1\,\mathrm{GeV}$ in order to detect displaced vertices efficiently at MATHUSLA, and then we take $E_\mathrm{min}^\mathrm{thr} \simeq 600\,\mathrm{MeV}$ per each track to follow the analysis by Ref.~\cite{Berlin:2018jbm}.
Meanwhile, the background events at FASER have been well discussed in Ref.~\cite{Ariga:2018zuc}.
Following the literature, we assume $E_\mathrm{min}^\mathrm{thr} \simeq 100\,\mathrm{GeV}$ for the total energy deposition in order to reduce the trigger rate and to remove low-energy backgrounds at FASER~\cite{Ariga:2018uku,Ariga:2018zuc}.
The corresponding momentum of $N_1$, which is denoted by $p_\mathrm{thr}$, is given by
\eqs{
  p_\mathrm{thr} \simeq
  \begin{cases}
    \displaystyle 
    \frac{2 E_\mathrm{min}^\mathrm{thr}}{\Delta_N} \simeq
    12\,\mathrm{GeV} \left( \frac{0.1}{\Delta_N} \right) & \text{(MATHUSLA)} \,, \\
    \displaystyle 
    \frac{E_\mathrm{min}^\mathrm{thr}}{\Delta_N} \simeq
    1\,\mathrm{TeV} \left( \frac{0.1}{\Delta_N} \right) & \text{(FASER)} \,. \\
  \end{cases}
}
Here, we multiply the factor 2 for MATHUSLA to make all the SM decay products sufficiently energetic since the energy threshold is imposed on each charged track.
Since the threshold value $p_\mathrm{thr}$ exceeds the maximal value, the FASER experiment decreases its sensitivity for the dark nucleon search if $\Delta_N \lesssim 0.03$.

In the case of MATHUSLA, for the fictitious dark photon $m^\ast_{A'} \lesssim 10\,\mathrm{GeV}$, the typical momentum that is close to the mass of dark photon is smaller than the threshold.
In this case, the signal number of $N_1$ decrease and depends $p_\mathrm{thr}/p^\ast_\mathrm{geo}$ after the final state phase space integration. 
We empirically find the efficiency $A_\mathrm{eff}$ related with the decrease of the signal number from \cite{Berlin:2018jbm} as follows.
\eqs{
  A^\mathrm{prod}(p^\ast) = \frac{2}{x^2 + x^{-2}} \,, \qquad 
  x = p^\ast/p^\ast_\mathrm{geo} \,.
}
$A^\mathrm{prod}(p^\ast)$ is normalized to be unity as the long-live lived particle has the typical momentum.
Though we take the MATHUSLA case as an example, we take into account the depletion of signals as $A^\mathrm{prod}(p^\ast)$ even for the FASER case when the momentum deviates from $p^\ast_\mathrm{geo}$. 
As far as we take $p^\ast_\mathrm{min} = p^\ast_\mathrm{geo}$, the produced particle comes in the detector. 
However, it may not come in the detector when $p^\ast_\mathrm{min}$ deviates from $p^\ast_\mathrm{geo}$.
We incorporate the angle acceptance as follows.
\eqs{
  A^\mathrm{geo}(p^\ast) = \frac{1}{1+ \theta^2/\theta_\mathrm{det}^2} \,, 
}
where $\theta = (m_{A'}^{\ast 2} - 4m_{N_1}^2)^{1/2}/2p^\ast$, and $\theta_\mathrm{det}$ is determined by each experiment: $\theta_\mathrm{det} = 0.5$ for the MATHUSLA and $\theta_\mathrm{det} = 2 \times 10^{-3}$ for FASER.
The total efficiency for the LHC lifetime frontier is defined by $A_\mathrm{eff} = A^\mathrm{geo} A^\mathrm{prod}$.

\begin{table}
	\centering
	\caption{
    A summary of features of visible decay searches: MATHUSLA, FASER, E137, and SeaQuest. 
    The top block corresponds to the dark nucleon searches, while the bottom block corresponds to the dark pion searches.
    The subscripts of momenta, $\mathrm{max}$ and $\mathrm{min}$, correspond to $\epsilon_\mathrm{max}$ and $\epsilon_\mathrm{min}$, respectively, and do not necessarily coincide with the maximum or minimum energy.
	}
	\label{tab:FittingParameters}
	\begin{tabular}{|c||c|c|c|c|c|}
		\hline
		& $r_\mathrm{min}$ & $r_\mathrm{max}$ & $\theta_\mathrm{det}$ & $p^\ast_\mathrm{max}$ & $p^\ast_\mathrm{min}$ \\ \hline
		MATHUSLA & $140\,\mathrm{m}$ & $340\,\mathrm{m}$ & 0.5 & $100\,\mathrm{GeV}$ & $p_\mathrm{thr}$ \\
		FASER & $470\,\mathrm{m}$ & $480\,\mathrm{m}$ & $2 \times 10^{-3}$ & $7\,\mathrm{TeV}$ & $p^\ast_\mathrm{geo}$ \\ \hline \hline
		& & & & $p^\ast_\mathrm{max}$ & $p^\ast_\mathrm{min}$\\ \hline
    E137 & $179\,\mathrm{m}$ & $383\,\mathrm{m}$ & $4 \times 10^{-3}$ & $20\,\mathrm{GeV}$ & $20\,\mathrm{GeV}$ \\
    SeaQuest & $5\,\mathrm{m}$ & $6\,\mathrm{m}$ & 0.05 & $120\,\mathrm{GeV}$ & $16\,\mathrm{GeV}$ \\ 
    MATHUSLA & $140\,\mathrm{m}$ & $340\,\mathrm{m}$ & 0.5 & $100\,\mathrm{GeV}$ & $p^\ast_\mathrm{geo}$ \\
		FASER & $470\,\mathrm{m}$ & $480\,\mathrm{m}$ & $2 \times 10^{-3}$ & $7\,\mathrm{TeV}$ & $p^\ast_\mathrm{geo}$ \\ \hline
	\end{tabular}
\end{table}

We summarize the parameters featuring the experimental setups of MATHUSLA and FASER in \cref{tab:FittingParameters}.
As for $p^\ast_\mathrm{min}$ in this table, we show just typical values: $p_\mathrm{thr}$ fro MATHUSLA and $p^\ast_\mathrm{geo}$ for FASER.
In this Table, we also show the parameters for the dark pion searches, which will be discussed in the next subsection.

\subsection{Dark Pion Searches \label{sec:pion_seaquest}}

We consider the dark-neutral pion decay with a specific mass spectrum $m_{A'} < m_{\pi'} < 2 m_{A'}$ in this subsection.
The dark-neutral pions can be produced via various mechanisms: 1) transition from the darkly-charged pions, 2) dark hadronization, and 3) via off-shell dark photon with the Wess-Zumino-Witten (WZW) term~\cite{Wess:1971yu,Witten:1983tw}.
The dark-neutral pions can be produced via the transition from darkly-charged pions that are produced from off-shell dark photon as in the dark nucleon production.
However, the transition rate from the darkly-charged pions is considerably tiny, and then the produced darkly-charged pions escape from the detectors without any signals.
In the SM, a hadronic channel $\pi^+ \pi^- \pi^0$ is opened with injected energy above the $\omega$-meson threshold~\cite{Achasov:2002ud,Achasov:2003ir,Ilten:2018crw}. 
In a similar manner, dark-neutral pions are produced via the hadronization when the injected energy is larger than the dark dynamical scale. 
We use the production via the dark hadronization for the LHC lifetime frontier, and use it even for fixed-target experiments when the injected energy is larger than the dark dynamical scale. 
When the injected energy is less than the dark dynamical scale, we utilize the production via off-shell dark photon.

We discuss each dark-pion production process in detail. 
As for the dark hadronization, we use an approximation formula for the dark pion production,
\eqs{
  N_{\pi'} \simeq 
  \left. \frac{n_{\pi'} \alpha'}{\pi} \frac53 N_{A'} \right|_{m_{A'}^\ast = m_{N_2}}
   \,.
   \label{eq:ProducedNumberPi_hadronization}
}
Here, $n_{\pi'}$ denotes the multiplicity of dark-neutral pions for the production through the dark photon mediation. 
The factor $5/3$ originates from the sum of the charge squared of dark quarks.
We take a similar value, $n_{\pi'} = 2.0$, to the multiplicity of pions by the $e^+ e^-$ annihilation for the center of mass energy just above $J/\psi$ threshold~\cite{DASP:1978ftr}.
Once the injected energy to the dark photon exceeds the dynamical scale, the dark hadrons are expected to be produced through the hadronization. 
In particular, the dark-neutral pions are expected to be produced above dark-isospin-singlet vector resonance (corresponding to $\omega$-meson in the SM), and therefore we take the fictitious dark photon mass to be $m_{A'}^\ast = m_{N_2}$.
We discuss how the change of the multiplicity affects the future sensitivity to the dark pion decay at the LHC lifetime frontier in \cref{sec:production}.

We introduce the relevant fixed-target experiments before discussing the production of dark pions at each visible decay search. 
As discussed in \cref{sec:darkphotonsearch}, there exist data from previous fixed-target experiments that have excluded visible decay of sub-GeV dark photons with $\epsilon \simeq 10^{-7}\text{--}10^{-5}$. 
The same data can be applicable for constraints on the three-body decay of the dark-sector particles when the final states include visible particles. 
The decay length of three-body decay is enhanced compared to that of two-body decay due to the final-state phase space, and hence a search for three-body decay favors the long-baseline fixed-target experiments.
Among the fixed-target experiments with electron beam, the E137 experiment is sensitive to longer lifetime than others since it has a very long natural shielding (hill of 179\,m) and huge decay volume before detector (open air region of 204\,m).
The E137 experiment utilizes a 20\,GeV electron beam, and the beam is dumped in a water-aluminum target. 
No signal events with the energy deposit above 3\,GeV were observed at E137, and therefore the expected signals from the dark-sector particles should be less than a few.
CHARM and U70/$\nu$Cal that are existing fixed-target experiments with proton beams have constrained the similar parameter space as E137.
The visible search at E137 has put a constraint on dark photon parameters~\cite{Bjorken:2009mm}, and hence we map the results of the existing constraint for the visible decay of sub-GeV dark photon into the visible decay of the dark pions.
In the literature, the on-shell dark photon is produced via the electron Bremsstrahlung, and the produced dark photon decays into the SM particles via kinetic mixing. 

The SeaQuest experiment is currently running at Fermilab with the 120\,GeV proton beam~\cite{Aidala:2017ofy}, 
and is originally designed to measure antiquark structure of nucleons via Drell-Yan dimuon production. 
A magnetized iron block of 5\,m is placed just downstream from a target to sweep away the soft SM radiation, and the detector is composed of a 3\,m magnet and four tracking stations, which allow us to reconstruct decay vertex and momentum precisely. 
The first tracking station is located 1\,m downstream from the beam-dump, and hence the decay volume of the SeaQuest is in the range of $r_\mathrm{min} = 5 \,\mathrm{m}$ and $r_\mathrm{max} = 6 \,\mathrm{m}$.
There is a plan to upgrade SeaQuest by installing an electromagnetic calorimeter that was utilized in PHENIX experiment at Brookhaven National Laboratory, which is named as ``DarkQuest''. 
This upgrade allows to detect electrons as decay products of dark-sector particles. 
Since the decay volume of SeaQuest is shorter than the other fixed-target experiments, the SeaQuest is sensitive to the dark photon parameter for shorter decay length.
The visible signal search at SeaQuest has been studied in the context of the strongly interacting massive particles (SIMP) model~\cite{Hochberg:2014dra,Hochberg:2014kqa}.
In particular, in Ref.~\cite{Berlin:2018tvf}, the authors consider the three-body decay of the dark vector mesons $V$ into the dark pions $\tilde \pi$, $V \to \tilde \pi + \ell^+ \ell^-$.
In the literature, dark photons are assumed to be heavier than dark mesons in the dark sector ($m_{A'}/m_{\tilde \pi}= 3\,, m_V /m_{\tilde \pi} = 1.8$).
The dark photon predominantly produced via Bremsstrahlung in a collision, and their prompt decay produces the dark vector mesons.
We map their results into the dark-neutral pion decay.

\begin{figure}
  \centering
  \includegraphics[width=0.7\textwidth]{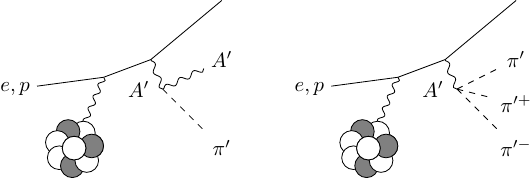}
  \caption{
    Bremsstrahlung production of dark-neutral pions via off-shell dark photons. 
  }
  \label{fig:pionprod}
\end{figure}

Concerning the production through the off-shell dark photon, we use \cref{eq:hadron_producednumber} and compute the off-shell dark photon decay rate $\Gamma_{A'}$ through the WZW term~\cite{Wess:1971yu,Witten:1983tw}%
\footnote{
  Our definition of pion decay constant differs from the literature~\cite{Witten:1983tw} by a factor 2.
}.
\eqs{
  \mathcal{L}_\mathrm{eff} 
  \supset - \frac{N_C \alpha'}{24 \pi f_{\pi'}} \pi'^3 F'^{\mu\nu} \tilde F'_{\mu \nu} 
  - \frac{i N_C e'}{12 \pi^2 f_{\pi'}^3} \epsilon^{\mu\nu\rho\sigma} A'_\mu \partial_\nu \pi'^+ \partial_\rho \pi'^- \partial_\sigma \pi'^3 \,.
  \label{eq:WZWinteraction}
}
The production processes of dark-neutral pions through the interactions are depicted in \cref{fig:pionprod}.
Unlike elementary particles, the production of the dark-neutral pions increases following the injected energy due to the derivative couplings.
The produced numbers of the dark-neutral pions would be dominated by the UV contribution.
The decay rates of the off-shell dark photon (with the fictitious mass $m_{A'}^\ast$) are given by
\eqs{
  \Gamma_{A'}(A'^\ast \to A' \pi'^3) & = 
  \frac{\alpha'^2 m_{A'}^{\ast 3}}{512 \pi^3 f_{\pi'}^2} \frac14 \left[1-\frac{(m_{A'}+m_{{\pi'}})^2}{m_{A'}^{\ast 2}}\right]^{3/2}\left[1-\frac{(m_{A'}-m_{{\pi'}})^2}{m_{A'}^{\ast 2}}\right]^{3/2} \,, \\
  \Gamma_{A'}(A'^\ast \to \pi'^+ \pi'^- \pi'^3) & \simeq 
  \frac{\alpha'}{32 \pi^6} \frac{m_{A'}^{\ast 7}}{32 \cdot 480 f_{\pi'}^6} \,. 
  \label{eq:offshellAdecay}
}
Here, we assume $m^\ast_{A'} \gg m_{\pi'}$ for the decay $A'^\ast \to \pi'^+ \pi'^- \pi'^3$. 
The derivative interaction of dark photons and dark pions lead to the enhancement of the production cross section when a higher energy is injected to the dark sector.
The three-pion production is negligible since the production is highly suppressed by high power of $m^\ast_{A'}/f_{\pi'}$ and since we use the dark hadronization when the $m^\ast_{A'}$ is larger than $f_{\pi'}$.

We discuss the production of the dark-neutral pions via off-shell dark photons at the fixed-target experiments, E137 and SeaQuest.
As for E137, the scaling of $N_{A'}$ via electron Bremsstrahlung is $\epsilon^2 m_{A'}^{-2}$~\cite{Bjorken:2009mm}.
Since each logarithmic bin of $m_{A'}^\ast$ equally contributes to the dark pion production, we cannot identify a certain $m_{A'}^\ast$ at which the dark pions are predominantly produced.
Hence, we do not factorize the signal number \cref{eq:Signal_Number} as the product of the produced number and the efficiency at a certain $m_{A'}^\ast$.
The high $m_{A'}^\ast$ contribution to the signal number is suppressed due to the angle acceptance and the loss of the nuclear coherence.
Hence, the signal number of dark pions is dominated by the IR contribution.

As for SeaQuest, the dark photons with the mass lighter than 2 GeV are predominantly produced via meson decays and proton Bremsstrahlung.
The dark photons via meson decays are less boosted compared to those produced via Bremsstrahlung in a proton beam and have a large angle from the beam axis.
The dark photon production via the Bremsstrahlung is dominated near the $\rho$ meson resonance at $m_{A'} \sim m_\rho$, and is nearly constant for $m_{A'} \lesssim 500 \,\mathrm{MeV}$.
It is reasonable to assume that the dark pion production is determined by a virtual dark photon production with a fictitious mass of $m_{A'}^\ast = m_\rho$.
Since the $\rho$ meson has a broad decay width $\Gamma_\rho$, we integrate the produced number of $A'$ within the range of $m_\rho \pm \Gamma_\rho$.
Though we compute the signal number of dark pions by use of \cref{eq:Signal_Number} numerically, the $\pi'$ yield is approximated by
\eqs{
  N_{\pi'} \simeq 
  \left. \frac{\alpha'^2}{2048 \pi^4} \left( \frac{m_\rho}{f_{\pi'}} \right)^2 \frac{\Gamma_\rho}{m_\rho} N_{A'} \right|_{m_{A'}^\ast = m_\rho}
  + \left. \frac{\alpha'}{32 \cdot 32 \cdot 480 \pi^7} \left( \frac{m_\rho}{f_{\pi'}} \right)^6 \frac{\Gamma_\rho}{m_\rho} N_{A'} \right|_{m_{A'}^\ast = m_\rho} \,.
  \label{eq:ProducedNumberPi_FixedpBrems}
}
Here, the first term comes from the left process, while the second term comes from the right process in \cref{fig:pionprod}.
The dark pions can be produced via hadronization since the beam energy of SeaQuest is $120\,\mathrm{GeV}$, which is much higher than the dark dynamical scale.
However, the Bremsstrahlung production gets suppressed above $1\,\mathrm{GeV}$ due to the proton form factor, and then the Drell-Yan production of dark photons begin to be relevant above $2 \,\mathrm{GeV}$.
Compared with the Bremsstrahlung at low $m_{A'}$, the $A'$ yield via Drell-Yan process is much smaller at high $A'$ mass even though the Drell-Yan dominates the production.
For the production of the dark pions at the SeaQuest, therefore, we assume that the dark pions are predominantly produced via proton Bremsstrahlung near the $\rho$-resonance. 
As for the scaling of $N_{A'}$ via proton Bremsstrahlung, we use $N_{A'}$ at SeaQuest computed in Ref.~\cite{Berlin:2018pwi}.


Now, let us discuss the sensitivity curves of the visible decay searches for the dark-neutral pions. 
We focus on the process $\pi' \to A' + \ell^+ \ell^-$, but the final state $A'$ would promptly decay into leptons when the kinetic mixing $\epsilon \gtrsim 10^{-4}$.
In this study, we ignore the decay of final state $A'$ to a lepton pair.
As in the dark nucleon searches [see \cref{eq:fitting_epsilon_max,eq:fitting_epsilon_min}], we use approximation formulae for the sensitivity boundaries instead of simulating the production number and the signal efficiency, again.
\begin{align}
  \ln \frac{A_\mathrm{eff}(p^\ast_\mathrm{max})}{A^0_\mathrm{eff}(p_\mathrm{max})} + \ln \frac{N_{\pi'}}{N^0_\mathrm{LLP}} + \frac{r_\mathrm{min}}{d_0(p_\mathrm{max})} & = \frac{r_\mathrm{min}}{c \tau_{\pi'} p^\ast_\mathrm{max} m_{\pi'}^{-1}} \,,
  \label{eq:fit_epsilon_max_pion} \\
  \frac{r_\mathrm{max}-r_\mathrm{min}}{c \tau_{\pi'} p^\ast_\mathrm{min} m_{\pi'}^{-1}} \frac{N_{\pi'}}{N^0_\mathrm{LLP}} \frac{A_\mathrm{eff}(p^\ast_\mathrm{min})}{A^0_\mathrm{eff}(p_\mathrm{min})} & = \frac{r_\mathrm{max}-r_\mathrm{min}}{d_0(p_\mathrm{min})} \,.
  \label{eq:fit_epsilon_min_pion}
\end{align}
The first equation gives a lower limit on the boosted decay length of $\pi'$, while the second gives an upper limit.
$N^0_\mathrm{LLP}$ and $A^0_\mathrm{eff}$ denote the produced number of the long-lived particles and the efficiency in the references, respectively.
The produced number $N_{\pi'}$ and the efficiency $A_\mathrm{eff}$ appears as the ratios to those in the reference model, and hence we only have to care about the parameter dependence of the produced number and the efficiency.

As for the LHC lifetime frontier, we use the same reference to estimate the boosted decay lengths $d_0(p_\mathrm{min})$ and $d_0(p_\mathrm{max})$ as in the previous subsection, and use the produced number of long-lived particles $N_\chi^0$ in the inelastic DM model for $N^0_\mathrm{LLP}$.
We take the momentum $p^\ast_\mathrm{max}$ to be the same as that of the dark nucleon, which are shown in \cref{tab:FittingParameters}.
Meanwhile, the momentum $p^\ast_\mathrm{min}$ is different from the dark nucleon case. 
Since the mass difference of dark-neutral pion and dark photon is larger than that of dark nucleons, the decay products are more energetic.
The dark pions leaving signals typically have the momentum of $p^\ast_\mathrm{geo} = m_{A'}^\ast/\theta_\mathrm{det}$ with the fictitious mass $m_{A'}^\ast$ which is comparable to the dark dynamical scale and the detector angle $\theta_\mathrm{det}$ with respect to the beam axis.
The typical momentum is larger than the threshold momentum as long as the dark dynamical scale is larger than 1\,GeV, and then we take $p^\ast_\mathrm{min} = p^\ast_\mathrm{geo}$ at MATHUSLA and FASER. 

As for E137, we take $r_\mathrm{min} = 179 \,\mathrm{m}$ and $r_\mathrm{max} = 383 \,\mathrm{m}$ as shown in \cref{tab:FittingParameters}. 
We map the result of the visible decay search for dark photon at E137~\cite{Bjorken:2009mm} into the dark-neutral pion search, and $N^0_\mathrm{LLP}$ is the number of dark photon produced at the E137.
We read out two reference parameter sets $(\epsilon, m_{A'})$ which, respectively, correspond to reference points on the upper and lower boundaries of sensitivity area from Ref.~\cite{Bjorken:2009mm}.
$p^\ast_\mathrm{max} = 20\,\mathrm{GeV}$ is given by the maximum momentum of produced dark pions that corresponds to the electron beam energy.
Meanwhile, $p^\ast_\mathrm{min}$ corresponds to the momentum of dark pions that typically comes to the decay volume at E137.
The differential production cross section of the dark photon has the collinear singularity, and it is regularized by the dark photon mass~\cite{Bjorken:2009mm}.
The momentum of the off-shell dark photon peaks near the beam energy,%
\footnote{
  Indeed, the energy distribution of the production cross section implies that the cost for the production of $A'^{(\ast)}$ with $p_{A'}$ is $A^\mathrm{prod} = (m_{A'}/p_\mathrm{beam})^2(1-p_{A'}/p_\mathrm{beam})^{-1}$ where $p_\mathrm{beam} = 20 \,\mathrm{GeV}$.
  Here, $A^\mathrm{prod}$ is normalized by the IR regulator $p_{A'}|_\mathrm{max} = p_\mathrm{beam} - m_{A'}^2/p_\mathrm{beam}$.
} 
and hence we take $p^\ast_\mathrm{min} = 20\,\mathrm{GeV}$ and the production efficiency is not suppressed.
We include the angle acceptance at E137 as $A_\mathrm{eff}(p^\ast) = (1+\theta^2/\theta_\mathrm{det}^2)^{-1}$ with $\theta = m_{A'}^\ast/p^\ast$ and $\theta_\mathrm{det} \simeq 0.004$ in the integral in \cref{eq:Signal_Number}.
Due to the loss of nuclear coherence at the high mass $m_{A'}^\ast$, the contribution to the signal number is suppressed for the high $m_{A'}^\ast$.
We do not take into account the decoherence in the integral in \cref{eq:Signal_Number}, and hence we take the high-energy cut of the integral to be at $m_{A'}^\ast \simeq 1 \, \mathrm{GeV}$.
We take the same momentum for the on-shell dark photon, $p_\mathrm{max} = p_\mathrm{min} = 20\,\mathrm{GeV}$, to evaluate $d_0(p_\mathrm{min})$ and $d_0(p_\mathrm{max})$ as for the mapping of the visible decay search for dark photon.

Concerning SeaQuest, we take $r_\mathrm{min} = 5 \,\mathrm{m}$ and $r_\mathrm{max} = 6 \,\mathrm{m}$ as shown in \cref{tab:FittingParameters}.
$p^\ast_\mathrm{max} = 120\,\mathrm{GeV}$ is the momentum of produced dark pions that corresponds to the proton beam energy.
For the SeaQuest sensitivity, we incorporate dark photon production only via proton Bremsstrahlung, and the off-shell dark photons are predominantly produced at the resonance, $m_{A'}^\ast = m_\rho$. 
Meanwhile, the angular scale of the SeaQuest spectrometer is $\theta_\mathrm{det} = 0.05$~\cite{Berlin:2018pwi}.
Hence, the dark pions with the momentum $p^\ast_\mathrm{min} = p^\ast_\mathrm{geo} = m_\rho/\theta_\mathrm{det} \simeq 16 \, \mathrm{GeV}$ predominantly come to the detector at the SeaQuest.
We incorporate the geometric acceptance as $A^\mathrm{geo}(p^\ast) = (1+\theta^2/\theta^2_\mathrm{det})^{-1}$ with $\theta = m_\rho/p^\ast$. 
As for the production through the proton Bremsstrahlung, the differential production cross section peaks at the soft dark photon emission, and hence we take $A^\mathrm{prod}(p^\ast) = p_\mathrm{beam}/p_{A'}^\ast$ where $p_{A'}^\ast$ is the momentum sum of the final states of Bremsstrahlung production and is constant when the momentum $p^\ast$ is fixed.
We map the result of the search for the three-body decays of dark vector mesons at SeaQuest~\cite{Berlin:2018tvf} into the dark-neutral pion decay.
We take $p_\mathrm{max} = 120\,\mathrm{GeV}$ as for the momentum of dark vector mesons.
Meanwhile, due to the IR regulator of the soft emission via the proton Bremsstrahlung, we take the typical momentum of dark vector meson leaving signal to be $p_\mathrm{min} = \mathrm{max} (m_{A'}/\theta_\mathrm{det}, 10\,\mathrm{GeV})$ where $10\,\mathrm{GeV}$ is assumed as the minimum momentum in the literature~\cite{Berlin:2018pwi}.
We read out two reference parameter sets $(\epsilon, m_{A'})$ which, respectively, correspond to reference points on the upper and lower boundaries of sensitivity area from Ref.~\cite{Berlin:2018tvf}, and then we obtain $d_0(p_\mathrm{min})$ and $d_0(p_\mathrm{max})$ for the dark vector meson searches at SeaQuest.

\section{Results \label{sec:results}}

As shown in the previous section, there are three kinds of long-lived particles in the composite ADM models: dark nucleons, dark pions, and dark photon.
We take the ADM mass to be $m_{N_2} = 8.5/n_{g'}\,\mathrm{GeV}$. 
In particular, we take $n_{g'} = 1 \,, 2 \,, 4 \,, $ and 8 ($m_{N_2} = 8.5\,, 4.3 \,, 2.1 \,, $ and $1.1\,\mathrm{GeV}$).
$n_{g'} = 8 $ corresponds to the maximum number of generations with which the dark QCD is asymptotically free.
The dark QED coupling is assumed to be $\alpha' = 0.05 \,, 0.03 \,, 0.01\,,$ and $ 7 \times 10^{-3}$, respectively, for $n_{g'} = 1 \,, 2 \,, 4 \,, $ and 8 in order to avoid the Landau pole up to the Planck scale.
Since we assume $m_{A'} < m_{\pi'} < 2 m_{A'}$ in this article, dark photons are the lightest particle in the dark sector.
We place the constraints and the future sensitivities from the dark photon searches in the same plots.

\begin{figure}
  \centering
  \includegraphics[width=0.48\textwidth]{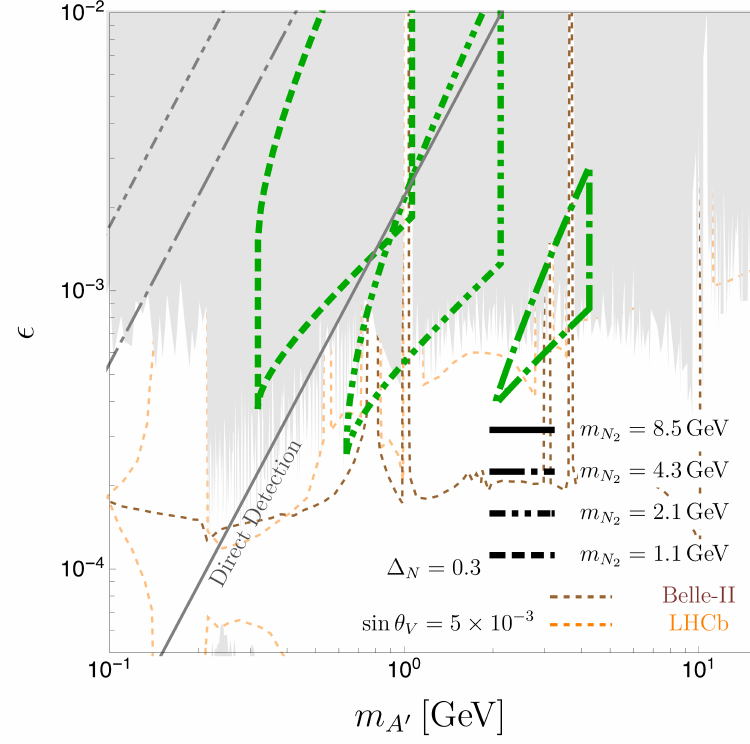}
  \includegraphics[width=0.48\textwidth]{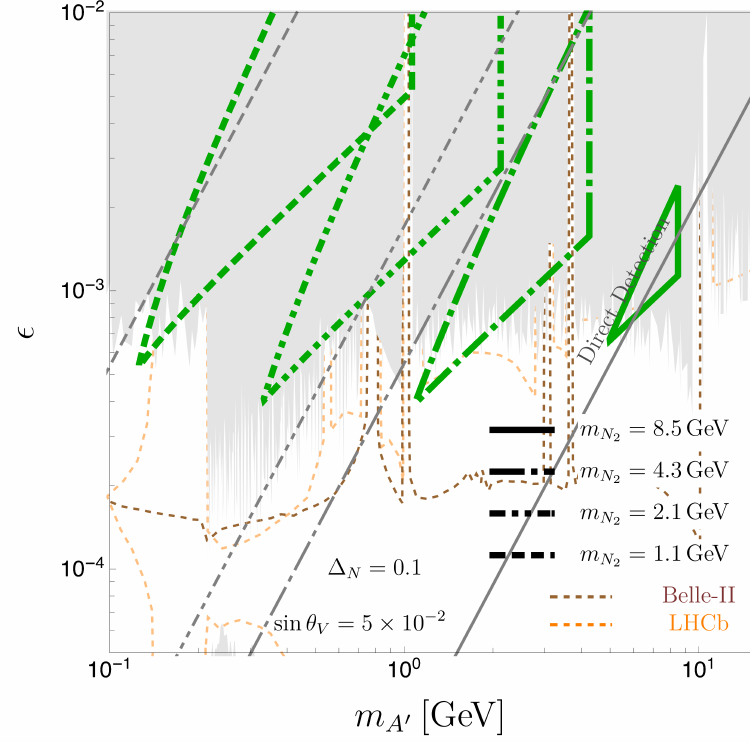}
  \includegraphics[width=0.48\textwidth]{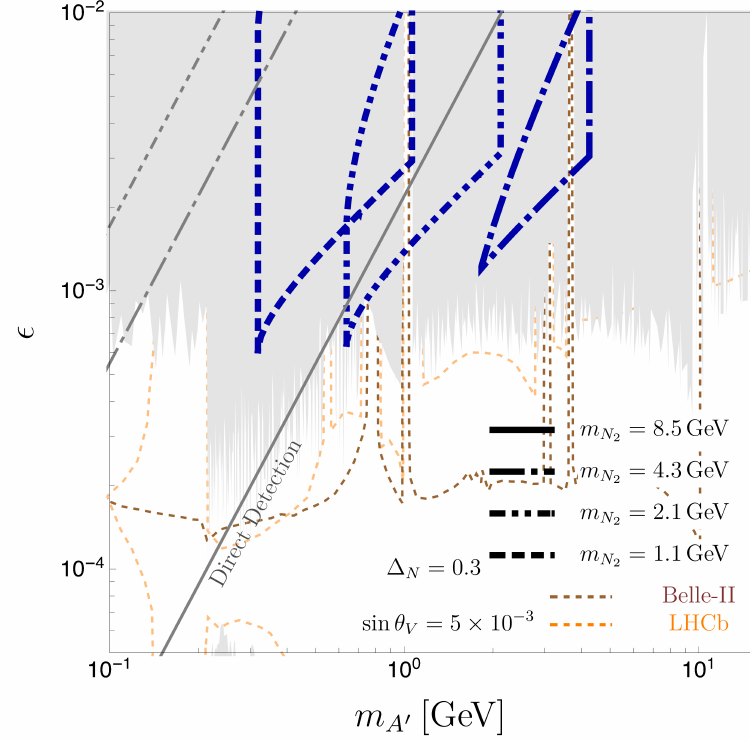}
  \includegraphics[width=0.48\textwidth]{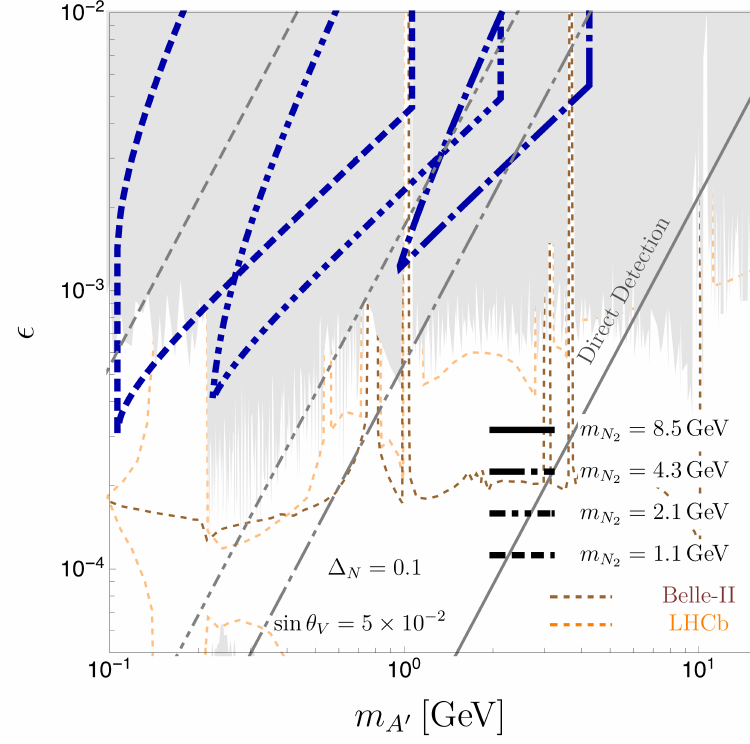}
  \caption{
    Dark nucleon searches at the LHC lifetime frontier on $\epsilon$-$m_{A'}$ plane: $m_{N_2} = 8.5\,, 4.3\,, 2.1\,, $ and $1.1 \,\mathrm{GeV}$ from right to left that correspond to $n_{g'} = 1 \,, 2 \,, 4\,,$ and $8$, the mixing angle $\sin\theta_V = 5 \times 10^{-3}$ (left) and $\sin\theta_V = 5 \times 10^{-2}$ (right), $\Delta_N = 0.1$ (left) and $\Delta_N = 0.3$ (right).
    The $U(1)_D$ coupling to be $\alpha' = 0.05 \,, 0.03 \,, 0.01\,,$ and $ 7 \times 10^{-3}$ from right to left.
    The top (bottom) panels show the future sensitivity at MATHUSLA (FASER): the area inside the lines can be explored by the experiments.
    The gray shaded region is excluded by the existing constraints on visible dark photon decay: (top region) the prompt decay searches by BaBar~\cite{Aubert:2009cp,Lees:2014xha}, KLOE~\cite{Archilli:2011zc,Babusci:2012cr,Anastasi:2015qla,Anastasi:2016ktq}, and LHCb~\cite{Aaij:2017rft,LHCb:2019vmc}; (left-bottom region) the fixed-target experiments  $\nu$Cal~\cite{Blumlein:2011mv,Blumlein:2013cua} and CHARM~\cite{Bergsma:1985is,Gninenko:2012eq}.
    The future sensitivity of Belle-II (LHCb) to visible dark photon decay is shown as brown (orange) dashed lines on middle parts of the panels~\cite{Ilten:2016tkc,Kou:2018nap}.
    The gray diagonal lines shows the direct detection bound on the composite ADM, and the parameter space above the line is already excluded.
    The sensitivity of MATHUSLA and FASER sharply cut at the left side of sensitivity because the decay into the on-shell dark photon opens.
  }
  \label{fig:nucleon_epsilon-mA}
\end{figure}

First, we discuss searches for the dark nucleon decay, $N_1 \to N_2 f \bar f$, at the LHC lifetime frontier. 
\cref{fig:nucleon_epsilon-mA} shows the future sensitivities along with the dark photon searches in the dark photon parameter $\epsilon$-$m_{A'}$ plane. 
In this plot, we take the nucleon mass difference to be $\Delta_N = 0.3$ (left) and $\Delta_N = 0.1$ (right).
We take different mixing angle $\theta_V$ in the left and right panels: $\sin\theta_V = 5 \times 10^{-3}$ (left) and $\sin\theta_V = 5 \times 10^{-2}$ (right).
As for the dark photon constraint, the shaded region is excluded by the existing constraints: the top of figure is already excluded by BaBar, KLOE, and LHCb. 
The future prospects of the dark photon searches are shown as thin-dashed lines in the figure: Belle-II (brown) and LHCb (orange) on the top of panels. 

The thick lines show the sensitivity by the LHC lifetime frontier with the ADM mass to be fixed, MATHUSLA (green) and FASER (blue).
We plot the cases with $n_{g'} = 1\,, 2\,, 4 \,,$ and $8$ in each panel, which correspond to the dark nucleon mass to be $m_{N_2} = 8.5\,, 4.3\,, 2.1\,, $ and $1.1 \,\mathrm{GeV}$, respectively.
The sensitivity curves for different $n_{g'}$ are differentiated from each other by their line types.
We use the approximation formulae \cref{eq:fitting_epsilon_max,eq:fitting_epsilon_min} to draw these curves. 
Since the dark nucleon mass is fixed, the lifetime of $N_1$ gets longer as the dark photon mass gets larger; $\epsilon$ that is accessible at the LHC lifetime frontier gets larger when the dark photon mass increases.
The decay mode $N_1 \to N_2 + A'$ opens when the dark photon mass is lighter than $\Delta_N m_{N_2}$, and therefore the plots are sharply cut at the left side of the sensitivity region.
We assume that the dark photon is the lightest particle in the dark sector, and therefore we cut the sensitivity region at $m_{N_2} = m_{A'}$.

The mixing angle $\theta_V$ measures the fraction of darkly charged baryons in the lightest dark nucleon.
As discussed in \cref{sec:directdetection}, direct detection experiments have put constraint on mixing parameters $\epsilon$ and $\theta_V$ since the darkly charged baryons can interact with the SM particles via dark photon.
The direct detection bounds are placed the left-top region of each panel, and the region above the lines are excluded. 
The four line types correspond to that of sensitivity plots: the mass of ADM is $8.5 \,, 4.3 \,, 2.1\,,$ and $1.1\,\mathrm{GeV}$ from right to left.
Since the direct detection constraints put an upper bound on $\epsilon \sin^2 \theta_V$, the direct detection bound on $\epsilon$ gets weaker for the smaller $\sin\theta_V$ [see \cref{eq:DDBound}].
We use the constraints by Panda-X II~\cite{Ren:2018gyx} (for $m_{N_2} = 8.5 \,\mathrm{GeV}$), DarkSide-50~\cite{DarkSide:2018bpj} (for $m_{N_2} = 4.3$ and $ 2.1 \,\mathrm{GeV}$), and CRESST-III~\cite{CRESST:2019jnq} (for $m_{N_2} = 1.1 \,\mathrm{GeV}$).

The produced particles leaving signals are less boosted for the MATHUSLA sensitivities than for the FASER sensitivities since MATHUSLA will be located in the off-axis direction from the LHC beam line. 
Thus, MATHUSLA has the sensitivity to the longer lifetime, namely smaller $\epsilon$, than FASER even though the both detectors will be located at the similar distance from the LHC collision point.
Since the DM is produced via the Drell-Yan mechanism, the produced number enhances for the lighter DM.
Besides, the DM is produced more preferably in the forward direction to the beam line as the DM gets lighter, and hence the FASER sensitivity area gets much wider than the MATHUSLA one. 
The dark nucleon searches at the LHC frontier have the sensitivity to $\epsilon \gtrsim 10^{-4}$. 
In other words, some portion of the sensitivity to the dark nucleon visible decay is comparable to the future sensitivity at the LHC-b and Belle-II that searches for prompt decay of dark photons.
In the plots, the upper and lower sensitivity curves sharply cross each other at the left-bottom point. 
This is because we use the approximation formulae the sensitivity curves, \cref{eq:fitting_epsilon_max,eq:fitting_epsilon_min}, and then it is expected that the shape of sensitivity curves of MATHUSLA and FASER will be round on the edge.
Most of the parameter space where the approximation formulae are robust has been excluded by the current bound on the visible dark photon decay. 
The dedicated analysis may lower the lower sensitivity curves since our approximation formula provides the conservative lower bound for multi-GeV dark nucleon (see \cref{app:fitting}).

We show the parameter dependence in the figures; in particular, we change the mixing angle $\theta_V$, the mass difference $\Delta_N$, and the mass of DM $m_{N_2}$.
In the following, we discuss how the sensitivity curves change when we make them larger.
The lifetime gets shorter as the values of these parameters get larger.
Since the upper boundary of the sensitivity area is mainly determined by the lifetime, the smaller $\epsilon$ will be explored in order to fix the lifetime. 
On the other hand, the lower boundary of the sensitivity area is basically determined by the product of lifetime and production cross section. 
The production cross section does not significantly depend on $\theta_V$ and $\Delta_N$, but does depend on $\epsilon^2$.
The lower boundary also gets lower in $\epsilon$ as the values of these parameters get larger, but not so much as the upper boundary. 
As a result, the sensitivity range of $\epsilon$ shrinks for large values of parameters. 
For instance, \cref{fig:nucleon_epsilon-mA} shows the sensitivity area gets shrinking for the heavier DM mass (smaller $n_{g'}$).
In particular, sensitivities disappear for $n_{g'} = 1$ except for the right top panel of \cref{fig:nucleon_epsilon-mA}.
Once $\Delta_N m_{N_2}$ exceeds the dark photon mass, the decay mode with the on-shell dark photon opens. 
The sensitivity range of $m_{A'}$ also shrinks for large $\Delta_N$.
We also show different choices of parameters in \cref{fig:nucleon_epsilon-mA_Parameters} in \cref{sec:production}.

We take the multiplicity of dark nucleon production to be a similar to that of nucleons in the SM near $J/\psi$ threshold. 
In the SM, pions are much lighter than nucleons, and hence it would be expected that pions are likely to be produced more than nucleons. 
In our model, on the other hand, the dark pion mass can get closer to the dark nucleon mass compared to the SM case. 
The multiplicity of dark nucleons can be similar to the dark pions, and then the sensitivity range would be enhanced.
We discuss the change of multiplicity in \cref{sec:production}.

We note that the fixed-target experiments, such as E137 and SeaQuest, would also be available to searching for the dark nucleon below a few GeV (\textit{c.f.}, sub-GeV inelastic DM searches~\cite{Berlin:2018pwi,Berlin:2018jbm}). 
We do not investigate this possibility for the lighter dark nucleons further in this study. 

\begin{figure}
  \centering
  \includegraphics[width=0.48\textwidth]{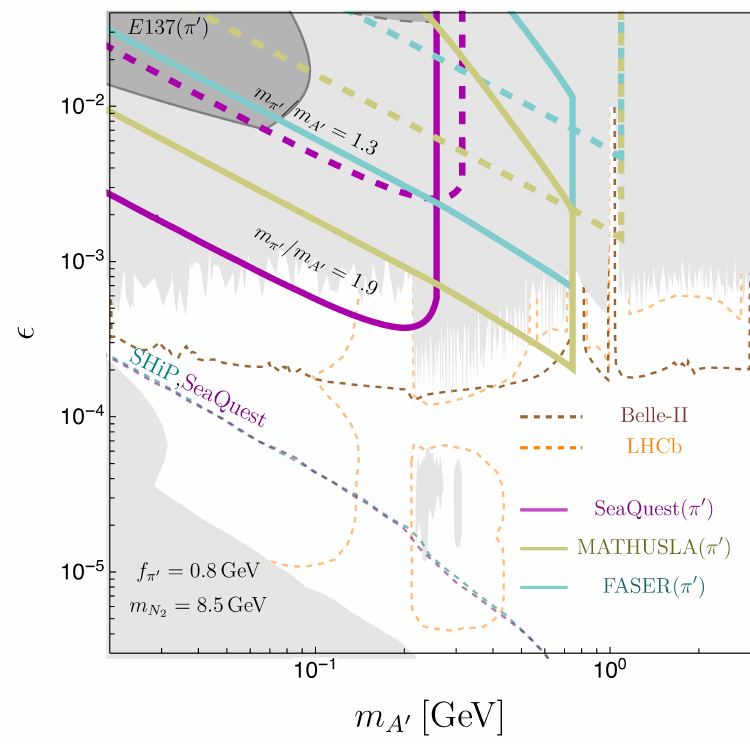}
  \includegraphics[width=0.48\textwidth]{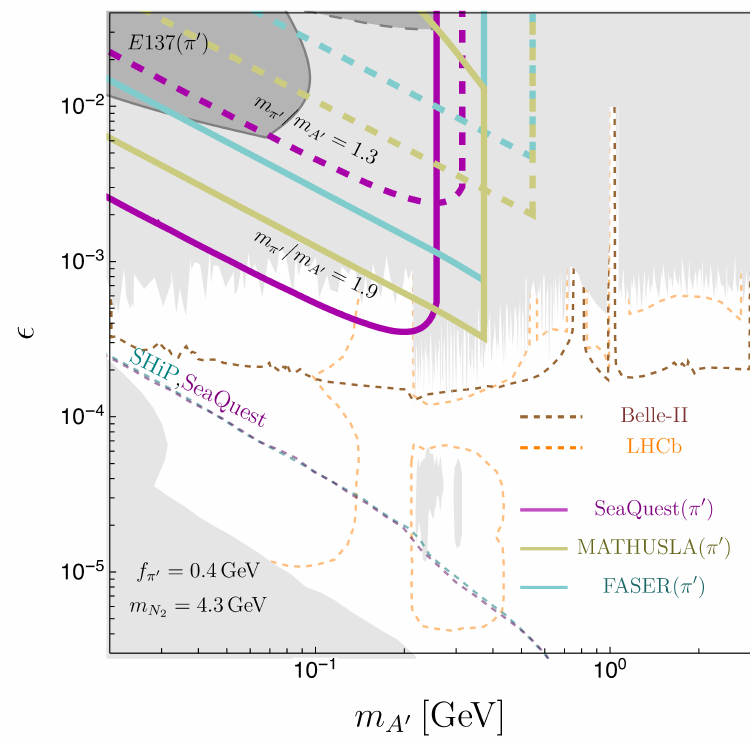}
  \includegraphics[width=0.48\textwidth]{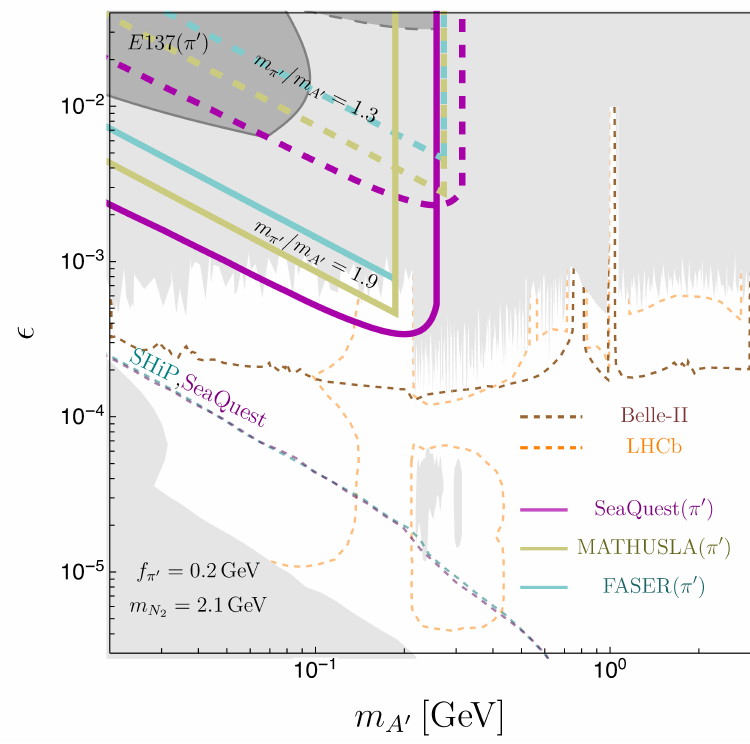}
  \includegraphics[width=0.48\textwidth]{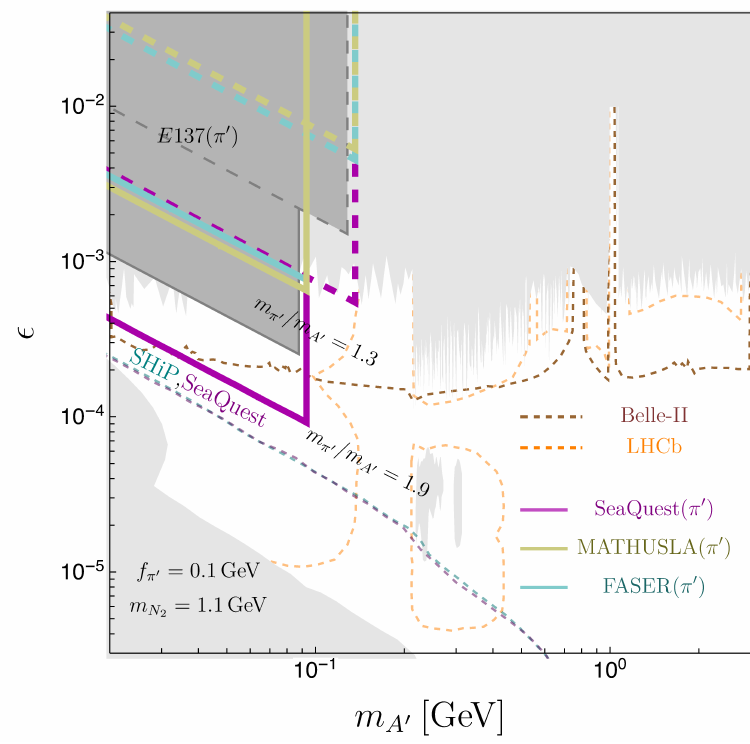}
  \caption{
    Dark pion searches at fixed-target and collider experiments on $\epsilon$-$m_{A'}$ plane with different values of $m_{\pi'}/m_{A'} =$ 1.9 (solid) and 1.3 (dashed).
    We take the different choice of pion decay constants $f_{\pi'}$: $f_{\pi'} = 0.8 \,, 0.4 \,, 0.2\,,$ and $0.1 \, \mathrm{GeV}$.
    The future sensitivities to the three-body dark pion decay are shown by magenta lines (SeaQuest), yellow lines (MATHUSLA), and light blue lines (FASER).
    The dark-shaded region in the left-top of the panels is excluded by E137 experiment~\cite{Bjorken:1988as,Batell:2014mga}: with dashed boundary ($m_{\pi'}/m_{A'} = 1.3$) and with solid boundary ($m_{\pi'}/m_{A'} = 1.9$).
    The existence constraints are depicted as the shaded region on top (BaBar, KLOE, and LHCb) and left-bottom ($\nu$Cal and CHARM) in each panel, and the future sensitivities to the visible dark photon decay are shown as thin-dashed lines, Belle-II (brown), LHCb (orange), SHiP (cyan), and SeaQuest (magenta).
  }
  \label{fig:pion_epsilon-mA}
\end{figure}

\cref{fig:pion_epsilon-mA} shows the future sensitivity and the existing constraint on the dark pion three-body decay, $\pi'^3 \to A' + f \bar f$. 
There are three dimensionful parameters associated with the dark pion decay: dark photon mass $m_{A'}$, dark pion mass $m_{\pi'}$, and dark pion decay constant $f_{\pi'}$.
We assume the dark pion mass in $m_{A'} < m_{\pi'} < 2 m_{A'}$ in order that the three-body decay dominates the dark pion lifetime, in particular we take $m_{\pi'}/m_{A'} = 1.9$ (solid lines) and $1.3$ (dashed lines).
Since the dark dynamical scale is fixed by the ADM mass, the decay constant can be scaled as a function of dark nucleon mass,
\eqs{
  f_{\pi'} \simeq f_{\pi} \frac{m_{N_2}}{m_N} 
  \simeq 0.8\,\mathrm{GeV} \left( \frac{m_{N_2}}{8.5\,\mathrm{GeV}} \right)\,,
  \label{eq:fpimN}
}
where we use the SM values, $f_{\pi} = 92 \,\mathrm{MeV}$ and $m_N = 938 \, \mathrm{MeV}$. 
In these plots, we take $f_{\pi'} = 0.8 \,\mathrm{GeV}$ (left-top), $0.4 \,\mathrm{GeV}$ (right-top), $0.2 \,\mathrm{GeV}$ (left bottom), and $0.1 \,\mathrm{GeV}$ (right bottom), which correspond to the dark nucleon mass $8.5 , 4.3 , 2.1,
$ and $1.1\mathrm{GeV}$ used in \cref{fig:nucleon_epsilon-mA}.
We also use the same value for the $U(1)_D$ coupling: $\alpha' = 0.05$ (left-top), 0.03 (right-top), 0.01 (left-bottom), and $ 7 \times 10^{-3}$ (right-bottom).
The existing constraints and the future sensitivities of the dark photon searches are also shown in the figures.

The shaded region on the top-left of each panel is excluded by the E137 bound: we take $m_{\pi'}/m_{A'} = 1.3$ (with dashed-line boundary) and $m_{\pi'}/m_{A'} = 1.9$ (with solid-line boundary).
E137 uses the electron beam of $20\,\mathrm{GeV}$, and hence it is hard to produce heavy particles of multi-GeV through the electron Bremsstrahlung at E137.
We take the fictitious dark photon mass to be less than $1\,\mathrm{GeV}$ where the form factor of target nucleus will be $\mathcal{O}(1)$. 
When the energy injection to the dark sector is above the dark dynamical scale, it is possible to produce dark pions through dark hadronization.
We use the hadronization for the dark pion production \cref{eq:ProducedNumberPi_hadronization} with $f_{\pi'} = 0.1 \,\mathrm{GeV}$, while use the production via the off-shell dark photon \cref{eq:WZWinteraction} for other $f_{\pi'}$. 
The dark hadronization must produce six dark pions at one hadronic event due to the isospin symmetry and the multiplicity of the dark pion production $n_{\pi'} = 2$.
The exclusion area for $f_{\pi'} = 0.1 \,\mathrm{GeV}$ is sharply cut at $0.2 \, m_{A'} /m_{\pi'} \, \mathrm{GeV}$ where the total mass of the final states via the dark hadronization, $6 m_{\pi'}$, equals to the high mass cut 1\,GeV as discussed in \cref{sec:pion_seaquest}.

The magenta lines show the future sensitivity of SeaQuest to the three-body decay, while the cyan and the yellow lines show the future sensitivity at the LHC lifetime frontier, FASER and MATHUSLA, respectively.
At SeaQuest, we focus only on off-shell dark photon production via Bremsstrahlung in a proton--nucleus collision though it would be also important to take into account the dark photon production via Drell-Yan processes above a few GeV. 
Since the off-shell dark photon production is sharply dropped above the fictitious mass $m_{A'}^\ast \simeq m_\rho$, which we use to estimate the signal number of dark pions in \cref{eq:Signal_Number}, the SeaQuest abruptly loose their sensitivity to the dark pion decay near a few hundred MeV.
Similarly to the E137 constraint, we use the hadronization for the dark-neutral pion production \cref{eq:ProducedNumberPi_hadronization} with $f_{\pi'} = 0.1 \,\mathrm{GeV}$, while use the production via the off-shell dark photon \cref{eq:WZWinteraction} for others.

As shown in \cref{fig:pion_epsilon-mA}, the sensitivity curves of SeaQuest and the existing constraints by E137 drastically change since the production channel of dark-neutral pions via dark hadronization opens as the injected energy is higher than the dark dynamical scale.
Even in the SM, the production cross section with the neutral pion final states drastically change near the dynamical scale.
As for the charged pions in the SM, the production cross section gradually increases due to the $\rho$-meson broad width as the injected energy to the hadronic sector increases. 
Meanwhile, the production of neutral pions opens at the $\omega$-meson threshold~\cite{Achasov:2002ud,Achasov:2003ir,Ilten:2018crw}. 
Due to the narrow width of the $\omega$-meson, the production cross section sharply increases at the dynamical scale. 
We do not go further into the hadronization in this study since we are agnostic about the hadronization.

At LHC lifetime frontier, the dark pions are produced via three dark pion production with an injected energy to the dark sector above the dark dynamical scale.
It is required to inject more energy in order to produce dark pions when $6 m_{\pi'}$ exceeds the dark dynamical scale.
In this study, we assume $6 m_{\pi'} \leq m_{N_2}$ instead of dedicated analysis of hadronization in the dark sector, and hence the sensitivity region is sharply cut at the threshold
As we discuss in the dark nucleon searches, MATHUSLA has the sensitivity to the longer lifetime, namely smaller $\epsilon$, than FASER.
Similarly to the dark nucleon searches, the dark pion searches at SeaQuest and LHC lifetime frontier are comparable to the future sensitivity of prompt decay searches for dark photons.

\begin{figure}
  \centering
  \includegraphics[width=0.48\textwidth]{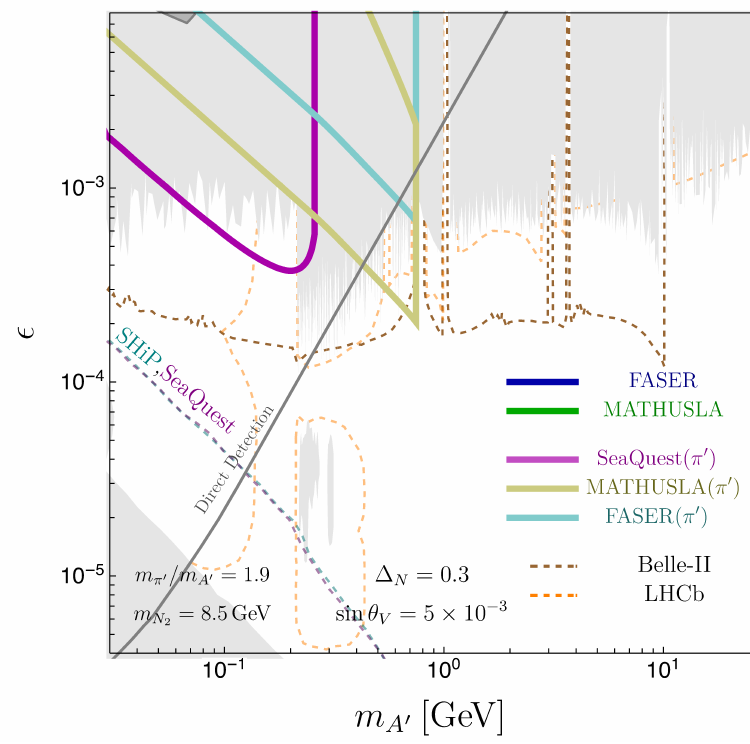}
  \includegraphics[width=0.48\textwidth]{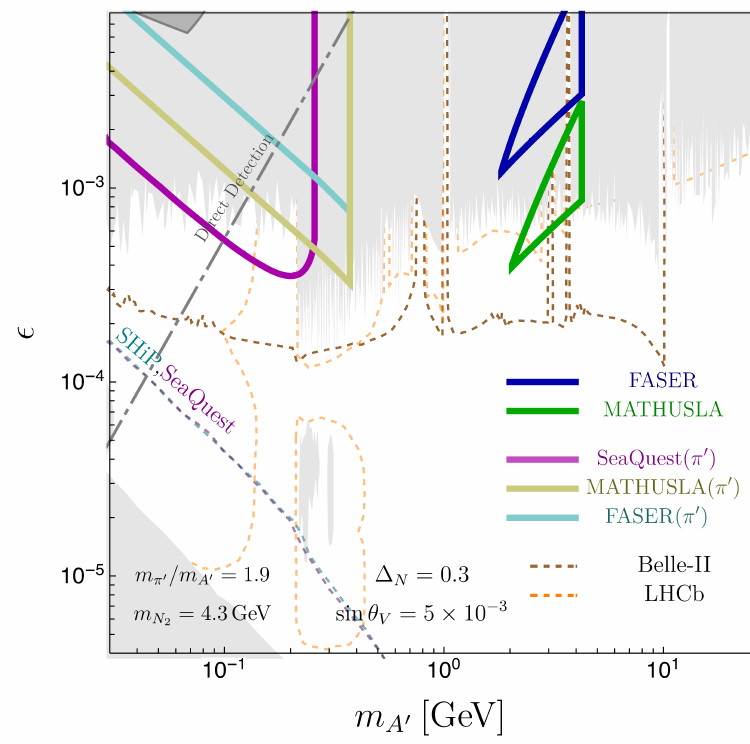}
  \includegraphics[width=0.48\textwidth]{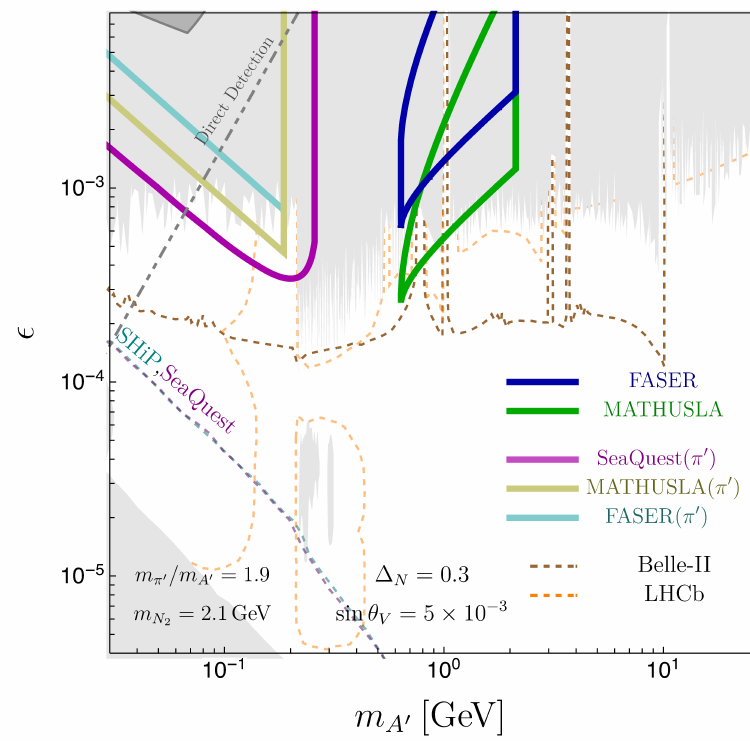}
  \includegraphics[width=0.48\textwidth]{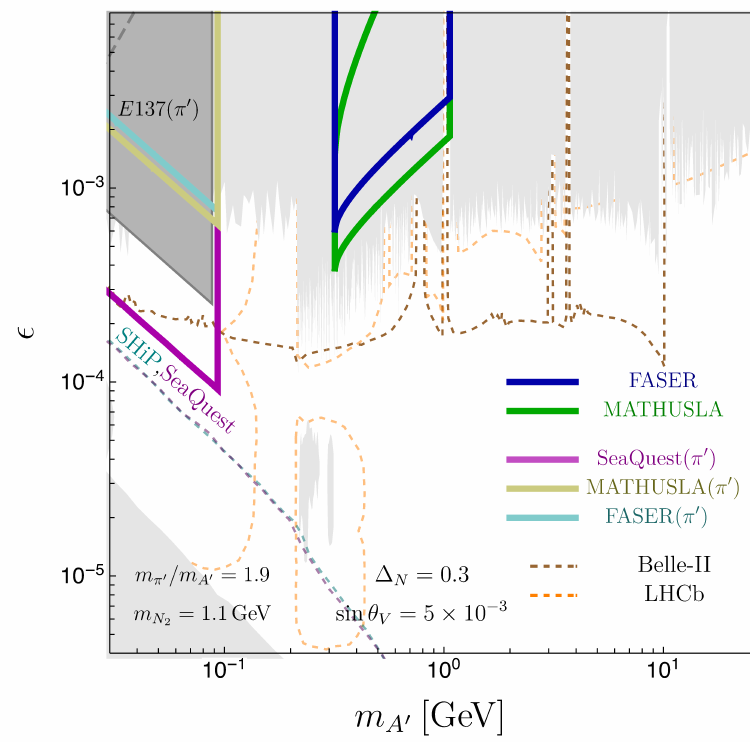}
  \caption{
    Summary plots of the visible decay searches of the composite ADM model at the visible decay searches.
    Blue and green solid lines show the future sensitivity to the dark nucleon transition, $N_1 \to N_2 + f \bar f$, at the LHC lifetime frontier: MATHUSLA (green) and FASER (blue).
    Cyan and yellow lines show the future sensitivity to the dark pion decay, $\pi'^3 \to A' + f \bar f$, at the LHC lifetime frontier, MATHUSLA (yellow) and FASER (cyan), while magenta line shows the future sensitivity to the decay at SeaQuest. 
    The dark shaded region is excluded by E137 constraint on the dark pion decay.
    We take the dark nucleon mass $m_{N_2} = 8.5 \,, 4.3 \,, 2.1 \,, 1.1\mathrm{GeV}$ and pion decay constant $f_{\pi'} = 0.8 \,, 0.4 \,, 0.2 \,, 0.1 \,\mathrm{GeV}$.
    We take the dark nucleon parameters to be $\Delta_N = 0.3$ and $\sin\theta_V = 5 \times 10^{-3}$, while the dark pion mass is $m_{\pi'}/m_{A'} = 1.9$ in all panels.
    The existence constraints and the future sensitivities to the visible dark photon decay (the top and left-bottom shaded areas and the future sensitivity curves from Belle-II, LHCb, SHiP, and SeaQuest) are the same as \cref{fig:nucleon_epsilon-mA}.
  }
  \label{fig:summaryplot}
\end{figure}
\cref{fig:summaryplot} compares the visible decay searches in the composite ADM models with $m_{N_2} = 8.5\,\mathrm{GeV}$ (left top), $m_{N_2} = 4.3 \,\mathrm{GeV}$ (right top), $m_{N_2} = 2.1\,\mathrm{GeV}$ (left bottom), and $m_{N_2} = 1.1\,\mathrm{GeV}$ (right bottom).
The dark nucleon parameters are assumed to be $\Delta_N = 0.3$ and $\sin\theta_V = 5 \times 10^{-3}$, while the dark pion mass is fixed to be $m_{\pi'}/m_{A'} = 1.9$ and  the dark pion decay constant is determined by \cref{eq:fpimN} in each panel.
In this plot, we take $\alpha' = 0.05$ (left top), $\alpha' = 0.03$ (right top), $\alpha' = 0.01$ (left bottom), and $\alpha' = 7 \times 10^{-3}$ (right bottom). 
Throughout this study, we take the dark QED coupling not to be diverged up to the Planck scale.
We can take a larger coupling when some of generations are decoupled from the low-energy theory and the dark QED is unified into Yang-Mills theory with a larger gauge group which is asymptotically free at the high-energy scale.%
\footnote{
  The dark nucleon mass is determined by the degree of freedom in the dark sector at the decoupling temperature of the intermediate-scale portal interaction \cref{eq:Intermediate_Portal}. 
  Therefore, the dark quarks should be decoupled (just) below the decoupling temperature scale.
  Meanwhile, in order not to change the degrees of freedom in the dark sector, the dark QED is unified into Yang-Mills theory above the decoupling temperature scale.
} 
In that case, both of the decay rate and the production cross section of the dark-sector particles increase, and hence we can explore more smaller kinetic mixing $\epsilon$.

Furthermore, the dark hadrons produced at the visible decay searches are assumed to be made only of one of the $n_{g'}$ generations in this study.
In other words, we assume that all dark quarks except for the lightest generation have masses larger than $\mathcal{O}(10)$ GeV or more. 
The production rate of dark hadrons enhances due to transition via the dark strong force from other dark hadrons when all $n_{g'}$ generations are produced.
In this case, the production rate is na\"ively expected to be multiplied by $n_{g'}$, and hence it makes the lower boundary of the sensitivity curves a factor of $n_{g'}^{1/4}$ smaller than what we have shown.

\section{Conclusion \label{sec:conclusion}}

Composite ADM models with dark photon have multiple particles of GeV or sub-GeV scale: dark photons, dark pions, and dark nucleons. 
Each of particles plays significant roles in the composite ADM framework: the lightest dark nucleon is the very ADM, the strong annihilation into dark pions depletes the symmetric component of dark nucleons, and the dark photons release a huge entropy in the dark sector to the visible sector.
The dark sector connects to the visible sector via the dark photon portal only with small kinetic mixing at a low energy, and thus the dark particles tend to be long-lived and leave visible signals. 
The lightest dark nucleon must have the mass of GeV to have the correct relic abundance in the composite ADM framework, and therefore it is certainly a good target of the LHC lifetime frontier.

In this study, we have focused on the case that the lightest dark nucleon mainly consists of dark neutron but is slightly mixed with dark proton due to $U(1)_D$ breaking. 
The constraint from the direct detection on the kinetic mixing $\epsilon$ gets milder thanks to the nucleon mixing.
In this case, the heavier nucleon that is mostly composed of dark proton can decay into the lightest dark nucleon through three-body decay.
\cref{fig:nucleon_epsilon-mA} summarizes the dark nucleon searches at the LHC lifetime frontier, FASER and MATHUSLA.
They will explore dark photon mass above sub-GeV and kinetic mixing $\epsilon \gtrsim 10^{-4}\text{--}10^{-3}$ for the DM of multi-GeV.

Meanwhile, dark pions are lighter than dark nucleons, which means that the dark pions have the mass of sub-GeV in the composite ADM framework.
In this study, we have considered dark pions with the mass in $m_{A'} < m_{\pi'} < 2 m_{A'}$ that is optimized for the visible signals from dark pions via three-body decay $\pi'^3 \to A' + f \bar f$.
E137 puts the most strong constraint on the decay, and SeaQuest have a great sensitivity to the decay in near future exploring dark photon mass below GeV and kinetic mixing $\epsilon \gtrsim 10^{-5}\text{--}10^{-4}$
Besides, we have discussed the dark pion searches at the LHC lifetime frontier. 
The dark pions have been assumed to be produced via hadronization in the dark sector at the LHC.
Similarly to the sensitivity of SeaQuest, we have found that the LHC lifetime frontier has the potential to explore the kinetic mixing of $\epsilon \gtrsim 10^{-4}$.

Dark strong dynamics naturally provides rich structure in the dark sector.
Hence, we have various decay signals from dark hadrons unlike, for example, the visible decay of dark photon.
Interestingly, the visible signals from dark hadrons and dark photon will be tested at various experiments in near future: dark nucleons at LHC, dark pions at LHC and SeaQuest (fixed-target experiments), and dark photons at Belle-II and LHCb.

In this study, we use the approximation formulae to estimate the sensitivity boundaries.
In order to precisely determine the future sensitivity, we need to understand the produced numbers of dark hadrons and the efficiencies. 
In particular, dark hadrons are assumed to be produced via the off-shell dark photon or via the dark hadronization in this study. 
The dark-sector particle production through the off-shell $A'$ has been less studied compared to the production through the decay of on-shell $A'$. 
We evaluate the produced numbers of dark hadrons via the off-shell dark photons by use of the on-shell dark photon production at the fictitious mass $m_{A'}^\ast$ and by considering the dominant contribution to the numbers.
It is important to study the off-shell production in more detail, in particular by simulating the produced numbers of dark hadrons at LHC and at the fixed-target experiments, for the future sensitivity. 
Besides the dedicated analysis for the off-shell production, it is also required to understand hadron physics with different parameters from the SM ones. 
We have na\"ively used empirical values of hadron production in the SM in order to estimate produced numbers of dark hadrons.
It is worth studying more dedicated analysis with inclusion of dark hadron physics (\textit{e.g.}, R-ratio of dark hadrons, multiplicity of dark hadrons, etc.), but we leave them for future study.

In this study, we have focused on a specific mass spectrum of the dark pion, $m_{A'} < m_{\pi'} < 2 m_{A'}$, which is relevant for visible decay searches.
For the dark pions with $2 m_{A'} < m_{\pi'}$, produced dark pions promptly decay into dark photons through the anomaly-induced interaction, $\pi'^3 \to A' A'$, and then we lose the future sensitivity and the existing constraints from fixed-target experiments.
When the dark pions are the lightest particle in the dark sector, $m_{\pi'} < m_{A'}$, dark pions can decay into the visible particle with lifetime longer than $1\,\mathrm{s}$.
Instead, dark photon would dominantly decay into dark pions providing invisible signals of dark photons at the fixed-target experiments such as LDMX~\cite{Izaguirre:2014bca,Akesson:2018vlm}. 
Furthermore, since the dark sector consists of the confining gauge dynamics, dark vector mesons would exist with the mass of dark dynamical scale. 
If the dark vector mesons have a broad resonance similar to the SM $\rho$ mesons, the production of darkly charged pions would change due to the broad resonance and kinetic mixing between dark photons and dark $\rho$ mesons. 
We leave the long-lived particle searches with different mass spectra for future work.

\subsection*{Acknowledgement}
The work of A. K. and T. K. is partly supported by IBS under the project code, IBS-R018-D1.
A. K. acknowledges partial support from Grant-in-Aid for Scientific Research from the Ministry of Education, Culture, Sports, Science, and Technology (MEXT), Japan, 18K13535 and 19H04609; from World Premier International Research Center Initiative (WPI), MEXT, Japan; from Norwegian Financial Mechanism for years 2014-2021, grant nr 2019/34/H/ST2/00707; and from National Science Centre, Poland, grant DEC-2018/31/B/ST2/02283.
T. K. is grateful to Nagoya University for their hospitality during the COVID-19 pandemic.

\newpage
\appendix

\section{Linear Sigma Model \label{app:LSM}}

We analyze chiral symmetry breaking and scalar dark mesons by use of the linear sigma model (LSM) that is a low-energy effective theory of two-flavor dark QCD.
The Lagrangian density of the LSM is constructed to be invariant under the flavor transformation, $SU(2)_L \times SU(2)_R \times U(1)_V$, which reflects the symmetries of the dark QCD.
$U(1)_A$ is explicitly broken by an axial anomaly term.
Let $\Phi$ be an LSM field that is an $2 \times 2$ matrix scalar field.
$\Phi$ is identified with the dark-quark bilinear as $\Phi_{ij} \sim \bar q_j q_i$ with the flavor subscripts $i,j$.
According to this identification, $\Phi$ transforms under the flavor transformation, $SU(2)_L \times SU(2)_R \times U(1)_A$, as follows.
\eqs{
  \Phi \to e^{i \alpha} L \Phi R^\dag \,,
}
where $L$ and $R$ are the unitary matrices of $SU(2)_L \times SU(2)_R$.
$e^{i \alpha}$ is an elements of $U(1)_A$. 
$\Phi$ does not transform under $U(1)_V$.
Transformation properties of the dark-quark bilinear determine $C$ and $P$ transformation of $\Phi$ as follows.
\eqs{
  C\,:\, & \hat \Phi \to \mathcal{C} \hat \Phi \mathcal{C}^{-1} = \hat \Phi^T \,, \\
  P\,:\, & \hat \Phi \to \mathcal{P} \hat \Phi \mathcal{P}^{-1} = \hat \Phi^\dag \,.
}
Here, we define the discrete transformation for the field in the charge basis, which is denoted by the hatted fields.

We decompose the matrix field $\Phi$ into scalar fields $\sigma^a$ and pseudo-scalar fields $\pi^a$ as 
\eqs{
  \Phi = (\sigma^a + i \pi^a) T^a \,,
}
where $T^a ~ (a = 0, 1, 2, 3)$ are a unit matrix and $SU(2)$ generators that satisfy 
\eqs{
  [T^a\,, T^b] = i f_{abc} T^c \,, ~~~~~
  \{ T^a \,, T^b \} = d_{abc} T^c \,, ~~~~~
  \mathrm{tr}(T^a T^b) = \frac12 \delta^{ab} \,.
}
$f_{abc}$ is totally antisymmetric in all indices, while $d_{abc}$ is totally symmetric.
They are just structure constants of $SU(2)$ when $a, b, c \neq 0$, while we have $f_{ab0} = 0 \,, d_{ab0} = \delta_{ab}$.

The Lagrangian density is 
\eqs{
  \mathcal{L} = \mathrm{tr}(\partial_\mu \Phi^\dag \partial^\mu \Phi) 
  - V(\Phi) + \mathrm{tr}(H \Phi + H^\dag \Phi^\dag) \,.
}
In this sub section, we do not include the gauge interactions of the LSM field.
Here, the second term is the potential for $\Phi$ that is given by
\eqs{
  V(\Phi) & = - \mu^2 \mathrm{tr}(\Phi^\dag \Phi) 
  + \lambda_1 [\mathrm{tr}(\Phi^\dag \Phi) ]^2 + \lambda_2 \mathrm{tr}(\Phi^\dag \Phi)^2 
  - \left[ c \det \Phi + c^\ast \det \Phi^\dag \right] \,.
}
We assume parameters $\lambda_1\,, \lambda_2$\,, and $\mu^2$ to be positive.
The last term breaks $U(1)_A$, and the coefficient $c$ is a complex number with mass dimension two in general.
Throughout this paper, we fix the coefficient $c$ to be positive by $U(1)_A$ rotation of $\Phi$.

The third term of the Lagrangian reflects the explicit breaking of the chiral symmetry by the dark quark masses.
We refer to $H \equiv j^a T^a$ as a source matrix that is proportional to the mass matrix of the current dark quarks.
It would be sufficient to consider the linear term in $H$ as far as the dark quark masses are sufficiently small and its perturbation works well.

The Lagrangian is rewritten in terms of the components scalars, $\sigma^a$ and $ \pi^a$, as follows.
\eqs{
  \mathcal{L} = & \frac12 \left[ \partial_\mu \sigma^a \partial^\mu \sigma^a + \partial_\mu \pi^a \partial^\mu \pi^a \right] - U(\sigma,\pi) \,.
}
The scalar potential $U(\sigma,\pi)$ is 
\eqs{
  U(\sigma,\pi) \equiv & \frac12 \sigma^a \left[ (-\mu^2+c) \delta_{ab} - 2 c \delta_{a,0} \delta_{b,0} \right] \sigma^b
  + \frac12 \pi^a \left[ -(\mu^2+c) \delta_{ab} + 2 c \delta_{a,0} \delta_{b,0} \right] \pi^b \\
  & + 2 \mathcal{H}_{abcd} \sigma^a \sigma^b \pi^c \pi^d 
  + \frac13 \mathcal{F}_{abcd} \left[ \sigma^a \sigma^b \sigma^c \sigma^d + \pi^a \pi^b \pi^c \pi^d \right] \\
  & - \frac12 \left[ (\sigma^a + i \pi^a) j^a + (\sigma^a - i \pi^a) j^{a\ast} \right] \,, \\
}
with the coefficients
\eqs{
  \mathcal{H}_{abcd} & = \frac{\lambda_1}{4} \delta_{ab}\delta_{cd} + \frac{\lambda_2}{8} (d_{abn}d_{cdn} + f_{acn}f_{bdn} +f_{adn}f_{bcn} ) \,, \\
  \mathcal{F}_{abcd} & = \frac{\lambda_1}{4} (\delta_{ab}\delta_{cd} + \delta_{ac}\delta_{bd} + \delta_{ad}\delta_{bc}) + \frac{\lambda_2}{8} (d_{abn}d_{cdn} + d_{acn}d_{bdn} +d_{adn}d_{bcn} ) \,.
}
$\sigma^0$ and $\pi^a~(a =1,2,3)$ may obtain the dominant vacuum expectation value (VEV) since they apparently have negative mass squared. 
We simply set $\pi^a$ that obtains its VEV to be $\pi^3$ by use of the vectorial flavor rotation $SU(2)_V$ of $\Phi$. 
We parametrize masses of scalars after the symmetry breaking as follows.
\eqs{
  U(\sigma+\bar\sigma,\pi+\bar\pi) = 
  \frac12 (m_\pi^2)_{ab} \pi^a \pi^b + \frac12 (m_\sigma^2)_{ab} \sigma^a \sigma^b + (m_{\pi\sigma}^2)_{ab} \pi^a \sigma^b + \cdots \,.
}
Here, letters with bars denote their VEVs, and the higher terms are abbreviated.

Let us compute the mass spectra under some assumptions. 
We can generally diagonalize the source term by use of $SU(2)_L \times SU(2)_R$ transformation of $\Phi$.
\eqs{
  H = e^{i \alpha_H} (j^0 T^0 + j^3 T^3) \,.
}
We can utilize $U(1)_A$ transformation to remove the overall phase $\alpha_H$, but the phase of the anomaly term $c$ appears, again.
To avoid the phase of $c$, we keep $\alpha_H$ unless we assume the source term to be real.
The kinetic term and the potential $V(\Phi)$ is invariant under $\Phi \to - \Phi$ as the number of flavors $N_f$ is even.
Therefore, we can set $j^0 > 0$ without loss of generality for $N_f = 2$.

\subsection{Decay Constant}
Let us discuss the decay constant in the LSM. 
In particular, we clarify the normalization of the decay constant by use of the commutation relations of the charges and partially-conserved axial current (PCAC) relation.
The infinitesimal axial-vector ($L = R^\dag = e^{i \alpha^a T^a}$) and vector ($L = R = e^{i \beta^a T^a}$) transformations define their Noether currents, $J^\mu_{Aa}$ and $J^\mu_{Va}$.
 \begin{align}
  J^\mu_{Aa} 
  & = - \frac12 d_{abc} (i \phi^{b \ast} \partial^\mu \phi^c - i \phi^{b} \partial^\mu \phi^{c\ast}) \,, &
  J^\mu_{Va} 
  & = \frac12 f_{abc} (\phi^{b \ast} \partial^\mu \phi^c + \phi^{b} \partial^\mu \phi^{c\ast}) \,.
\end{align}
Here, we define $\Phi \equiv \phi^a T^a$. 
The PCAC relation gives 
\eqs{
  \langle 0 | J^{\mu}_{Aa} (x) | \pi^b (p)\rangle = i p^\mu f^{ab} e^{- i p \cdot x} \,, 
  \label{eq:PCAC_LSM}
}
where $f^{ab}$ is the decay constant. 
In terms of the component fields, $\sigma^a$ and $\pi^a$, the axial current is $J^\mu_{Aa} = d_{abc} (\sigma^b \partial^\mu \pi^c - \pi^b \partial^\mu \sigma^c)$.
When the sigma fields get their VEVs, $\langle\sigma^a\rangle = \bar \sigma^a$, the PCAC relation gives the decay constant as follows. 
\eqs{
  f^{ab} = \sum_c d_{abc} \bar \sigma^c \,.
}
In particular, when the VEV is universal, $\bar \sigma^a = \bar \sigma $, the decay constant is diagonalized: $f^{ab} = \bar \sigma^0 \delta^{ab}$.

We can also define the currents in terms of quarks denoted by $Q$: 
\begin{align}
  j^{\mu a}_V(x) & = \overline Q \gamma^\mu T^a Q  \,, &
  j^{\mu a}_A(x) & = \overline Q \gamma^\mu \gamma_5 T^a Q \,.
\end{align}
This quark axial current annihilates the one-pion state, and it also defines the pion decay constant.
\eqs{
  \langle0| j^{\mu a}_A(x)|\pi^b(p)\rangle = i p^\mu f_\pi \delta^{ab} e^{- i p \cdot x} \,,
}
In the SM, the observation of the charged pion decay, $\pi^+ \to \mu^+ \nu$, determines the decay constant. 
In this normalization, we have $f_\pi = 92\,\mathrm{MeV}$ from the SM pion decay.
This normalization provides a commutation relation of the charges as follows.
\begin{align}
  [q_A^a,q_A^b] & = i f_{abc} q^c_V \,, & 
  q^{a}_V & \equiv \int d^3 x j^{0a}_V(x) \,, & 
  q^{a}_A & \equiv \int d^3 x j^{0a}_A(x) \,.
\end{align}

In the LSM, the axial charge and the vector charge are defined by
\eqs{
  Q_A^a \equiv \int d^3 x \, J^0_{Aa}(x) \,, ~~~
  Q_V^a \equiv \int d^3 x \, J^0_{Va}(x) \,.
}
These charges obey the following commutation relation.
\begin{align}
  [Q_A^a,Q_A^b] & = i f_{abc} Q^c_V \,,  
\end{align}
The charges in the LSM satisfy the same algebra as the charges defined in terms of quark currents.
Thus, we conclude that $f^{ab}$ defined in \cref{eq:PCAC_LSM} follows the same normalization as the decay constant defined by quark currents.

\subsection{Universal Source \label{app:UniV}}

In this subsection, we compute the spectrum with the source term $H = e^{i \alpha_H} j^0 T^0 ~ (j^0 > 0)$.
First, we consider the case with the real source ($\alpha_H = 0$).
We set $\langle \sigma^a \rangle = \bar\sigma^0 \delta_{a,0}$ and $\langle \pi^a \rangle = 0$.

The tree-level effective potential is given by
\eqs{
  U(\bar \sigma^0) = - \frac12 (\mu^2 + c)(\bar \sigma^0)^2 + \frac14 \lambda (\bar \sigma^0)^4 - j^0 \bar \sigma^0 \,, ~~~~~
  \lambda \equiv \lambda_1 + \frac{\lambda_2}{2} \,.
}
Since the potential is tilted in the presence of $j^0 > 0$, the stationary point condition for this potential gives the VEV up to $\mathcal{O}(j^0)$ as 
\eqs{
  \bar \sigma^0 = \sqrt{\frac{c+\mu^2}{\lambda}} + \frac{j^0}{2 (c + \mu^2)} \,.
  \label{eq:VEV_UniversalMass}
}
We define the decay constant $f \equiv \bar \sigma^0$.
We get the mass spectrum up to $\mathcal{O}(j^0)$ as follows.
\begin{align}
  (m_\sigma^2)_{00} & = 2 \lambda f^2 + \frac{j^0}{f}\,, &
  (m_\sigma^2)_{ii} & = 2 c + \lambda_2 f^2  + \frac{j^0}{f} \,, 
  \label{eq:ScalarMass_UniversalReal}
  \\ 
  (m_\pi^2)_{00} & = 2 c + \frac{j^0}{f} \,, &
  (m_\pi^2)_{ii} & = \frac{j^0}{f} \,.
  \label{eq:PseudoScalarMass_UniversalReal}
\end{align}
Here, the subscript $i$ runs $1\,, 2\,,$ and $3$.
There is no off diagonal components of mass matrices, and thus the mass spectrum is already diagonalized.
Since the source $j^0$ is proportional to the current quark mass, $\pi^i~(i=1,2,3)$ are identified to be the pseudo Nambu-Goldstone bosons.
We can identify $\sigma^0$ to be the so-called $\sigma$ meson, and $\pi^0$ to be the so-called $\eta'$ meson.
The $U(1)_A$ anomaly provides the sizable contribution to the mass of $\pi^0$, and therefore $\pi^0$ is heavier than the other pseudo scalars.

When the phase $\alpha_H$ is turned on, the potential is also tilted to the $\pi^0$ direction.
Therefore, $\pi^0$ acquires its VEV that proportional to the source, and we set $\langle \sigma^a \rangle = \bar\sigma^0 \delta_{a,0}$ and $\langle \pi^a \rangle = \bar\pi^0 \delta_{a,0}$. 
The effective potential is given by
\eqs{
  U(\bar \sigma^0,\bar \pi^0) 
  & = - \frac12 (\mu^2+c) (\bar \sigma^0)^2 + \frac12 (-\mu^2+c) (\bar \pi^0)^2 
  + \frac{\lambda}{4} [(\bar \sigma^0)^2 + (\bar \pi^0)^2]^2 \\
  & \quad - (j^0 \cos \alpha_H \bar \sigma^0 - j^0 \sin\alpha_H \bar \pi^0) \,.
}
As we mentioned above, we can restrict the parameter space to $j^0 \cos \alpha_H > 0$ by flipping the sign of $\Phi$ without loss of generality. 
The stationary point condition for the potential gives their VEVs as follows.
\begin{align}
  \bar \sigma^0 & = \sqrt{\frac{c+\mu^2}{\lambda}} + \frac{j^0}{2 (c + \mu^2)} \cos \alpha_H \,, & 
  \bar \pi^0 & = - \frac{j^0}{2 c} \sin \alpha_H \,. 
\end{align}
Since both $\sigma^0$ and $\pi^0$ get their VEVs, there is mass mixing among $\sigma^a$ and $\pi^a$ at the order of $j^0$.
This leads corrections of $\mathcal{O}[(j^0)^2]$ to the mass eigenvalues.
At the leading order of $j^0$, we have the similar mass spectrum of scalars to the one without the phase $\alpha_H$.
\begin{align}
  (m_\sigma^2)_{00} & = 2 \lambda f^2 + \frac{j^0}{f} \cos \alpha_H\,, &
  (m_\sigma^2)_{ii} & = 2 c + \lambda_2 f^2  + \frac{j^0}{f} \cos \alpha_H\,, 
  \label{eq:ScalarMass_UniversalComplex} \\ 
  (m_\pi^2)_{00} & = 2 c + \frac{j^0}{f} \cos \alpha_H\,, &
  (m_\pi^2)_{ii} & = \frac{j^0}{f} \cos \alpha_H\,.
  \label{eq:PseudoScalarMass_UniversalComplex} 
\end{align}
We define the decay constant $f \equiv \bar \sigma^0$, again.

\subsection{Isospin-violating Source \label{app:ISoV}}

In this subsection, we compute the spectrum with the source term $H = e^{i \alpha_H} j^3 T^3 ~ (j^3 > 0)$.
As with the previous subsection, we first consider the case with the real source ($\alpha_H = 0$).
Since the potential is tilted in the $\sigma^3$ direction, $\sigma^3$ must have its VEV that is proportional to the source. 
However, the dominant VEV is determined irrespective to the source term since the mass squared of $\sigma^3$ is positive.
As discussed in the introduction of this appendix, there are two possible dominant VEVs: one is $\bar \sigma^0$ and another is $\bar \pi^3$.

First, we discuss the former case; $\langle \sigma^a \rangle = \bar\sigma^0 \delta_{a,0} + \bar\sigma^3 \delta_{a,3}$ and $\langle \pi^a \rangle = 0$.
The effective potential for this vacuum choice is 
\eqs{
  U(\bar \sigma^0, \bar \sigma^3)
  & = - \frac12 (\mu^2+c) (\bar \sigma^0)^2 + \frac12 (-\mu^2+c) (\bar \sigma^3)^2 
  + \frac14 \left(\lambda + \frac{\lambda_2}{2} \right) [(\bar \sigma^0)^4 + (\bar \sigma^3)^4] \\
  & \quad + \frac12 \left(\lambda + \frac{3\lambda_2}{2} \right) (\bar \sigma^0)^2 (\bar\sigma^3)^2 
  - j^3 \bar\sigma^3 \,.
}
From the stationary point condition of this potential, we determine the VEVs at the order of $\mathcal{O}[(j^3)^2]$.
\begin{align}
  \bar \sigma^0 & = \sqrt{\frac{c+\mu^2}{\lambda}} - \sqrt{\frac{\lambda}{c+\mu^2}} \left( 1 + \frac{\lambda_2}{\lambda} \right) \frac{(\bar \sigma^3)^2}{2} \,, &
  \bar \sigma^3 = \frac{\lambda}{2 c \lambda + \lambda_2(c+\mu^2)} j^3 \,.
\end{align}
We note that $\bar \sigma^3$ is positive for $j^3 > 0$ and is proportional to $j^3$.
There is no linear correction of $j^3$ to the scalar masses since the source $j^3$ appears in $\bar \sigma^0$ from the second orders of $j^3$.
The pseudo Nambu-Goldstone bosons are massless up to $\mathcal{O}(j^3)$.
The scalar mass spectrum up to $\mathcal{O}(j^3)$ is
\begin{align}
  (m_\sigma^2)_{00} & = 2 \lambda f^2\,, &
  (m_\sigma^2)_{ii} & = 2 c + \lambda_2 f^2 \,, \\ 
  (m_\pi^2)_{00} & = 2 c \,, &
  (m_\pi^2)_{ii} & = 0 \,.
\end{align}
Since there is the mass mixing of $\mathcal{O}(j^3)$ between $\pi^0$ and $\pi^3$, we have to take care of corrections of $\mathcal{O}[(j^3)^2]$.
In particular, since the dark pions are massless even at the order of $\mathcal{O}(j^3)$, it is not obvious if the potential for pseudoscalars is stabilized.
The lightest pseudoscalar mass is indeed tachyonic:
\eqs{
  m_\pi^2 = - \frac{\lambda_2 j^3}{2c} \bar \sigma^3 \,, 
}
and hence, the lightest pseudoscalar developes its VEV.
We conclude that the choice of VEVs, $\langle \sigma^a \rangle = \bar\sigma^0 \delta_{a,0} + \bar\sigma^3 \delta_{a,3}$ and $\langle \pi^a \rangle = 0$, does not provide a proper vacuum.

Instead, we discuss the latter case; $\langle \sigma^a \rangle = \bar\sigma^3 \delta_{a,3}$ and $\langle \pi^a \rangle = \bar\pi^3 \delta_{a,3}$.
The effective potential for this vacuum choice is given by
\eqs{
  U(\bar \sigma^3, \bar \pi^3)
= \frac12 (-\mu^2+c) (\bar \sigma^3)^2 - \frac12 (\mu^2+c) (\bar\pi^3)^2 + \frac14 \lambda [(\bar \sigma^3)^2 + (\bar \pi^3)^2]^2
- j^3 \bar\sigma^3 \,.
}
From the stationary point condition of this potential, we determine the VEVs at the order of $\mathcal{O}[(j^3)^2]$.
\begin{align}
  \bar \sigma^3 & = \frac{j^3}{2c} \,, &
  \bar \pi^3 = \sqrt{\frac{c+\mu^2}{\lambda}} - \sqrt{\frac{\lambda}{c+\mu^2}} \frac{(\bar\sigma^3)^2}{2}\,.
  \label{eq:VEV_Isospin}
\end{align}
There is no linear correction of $j^3$ to the scalar masses since the source $j^3$ appears in $\bar \pi^3$ from the second orders of $j^3$.
The pseudo Nambu-Goldstone bosons are massless up to $\mathcal{O}(j^3)$.
The scalar mass spectrum up to $\mathcal{O}(j^3)$ is given by
\begin{align}
  (m_\pi^2)_{33} & = 2\lambda f^2 \,, &
  (m_\pi^2)_{00} & = (m_\sigma^2)_{11} =  (m_\sigma^2)_{22} = 2 c + \lambda_2 f^2 \,, 
  \label{eq:ScalarMass_IsospinMass} \\ 
  (m_\sigma^2)_{33} & = 2c \,, &
  (m_\sigma^2)_{00} & = (m_\pi^2)_{11} = (m_\pi^2)_{22} = 0 \,.
  \label{eq:PseudoScalarMass_IsospinMass}
\end{align}
Here, we take $f = \bar \pi^3$ as with the previous case: $f^2 \simeq \lambda^{-1} (c+\mu^2)$ of the order of $\mathcal{O}[(j^3)^0]$.
This spectrum coincides with that with the previous vacuum choice up to $\mathcal{O}(j^3)$.
As with the previous case, we have to care about correction up to $\mathcal{O}[(j^3)^2]$ since we have massless scalars.
There is mass mixing between $\sigma$ and $\pi$ since both $\sigma$ and $\pi$ get their VEVs in this vacuum choice. 
Every massless scalars obtains the mass of $\mathcal{O}[(j^3)^2]$:
\begin{align}
  m^2 & = \frac{j^3 \bar \sigma^3 \lambda_2}{M^2} \,, &
  M^2 & \equiv 2 c + \lambda_2 f^2 \,.
\end{align}
Here, $j^3$ and $\bar \sigma^3$ have the same sign, and therefore all massless scalars get positive mass of $\mathcal{O}[(j^3)^2]$.
We conclude that the vacuum choice, $\langle \sigma^a \rangle = \bar\sigma^3 \delta_{a,3}$ and $\langle \pi^a \rangle = \bar\pi^3 \delta_{a,3}$, provides a proper vacuum up to $\mathcal{O}[(j^3)^2]$ for the isospin-breaking source $H = j^3 T^3 ~ (j^3 > 0)$.

We note that mass spectrum \cref{eq:ScalarMass_IsospinMass,eq:PseudoScalarMass_IsospinMass} is similar to the spectrum \cref{eq:ScalarMass_UniversalComplex,eq:PseudoScalarMass_UniversalComplex} with $\alpha_H = \pi/2$.
In particular, the spectrums coincide with each other by replacing $\pi^3 \leftrightarrow \sigma^0$ and $\pi^0 \leftrightarrow \sigma^3$.
The universal mass with $\alpha_H = \pi/2$ implies the source term is pure imaginary. 
Since we can rotate the LSM field by a specific flavor rotation: 
\begin{align}
  \widetilde \Phi & = 
  U_\Phi \Phi \,, &
  U_\Phi & =
  \begin{pmatrix}
    i & 0 \\
    0 & - i 
  \end{pmatrix} \,,
\end{align}
mass spectrums have correspondence between real and pure imaginary sources.
This rotation implies that the mass spectrum of pure imaginary source $j^3$ coincides with \cref{eq:ScalarMass_UniversalReal,eq:PseudoScalarMass_UniversalReal}.
Let us confirm this correspondence explicitly by compute the spectrum with the complex isospin breaking source $H = e^{i \alpha_H} j^3 T^3 ~ (j^3 > 0)$.
We again take the same vacuum $\langle \sigma^a \rangle = \bar\sigma^3 \delta_{a,3}$ and $\langle \pi^a \rangle = \bar\pi^3 \delta_{a,3}$ as before. 
\eqs{
  U(\bar \sigma^3, \bar \pi^3)
  & = \frac12 (-\mu^2+c) (\bar \sigma^3)^2 - \frac12 (\mu^2+c) (\bar\pi^3)^2 + \frac14 \lambda [(\bar \sigma^3)^2 + (\bar \pi^3)^2]^2 \\ 
  & - j^3 \cos \alpha_H \bar\sigma^3 - j^3 \sin \alpha_H \bar \pi^3 \,,
}
This determines the VEVs from the stationary point condition of the potential as follows:
\begin{align}
  \bar \sigma^3 & = \frac{j^3}{2c} \cos \alpha_H \,, &
  \bar \pi^3 = \sqrt{\frac{c+\mu^2}{\lambda}} + \frac{j^3}{2(c+\mu^2)} \sin \alpha_H \,.
\end{align}
These VEVs give the mass spectrum of scalars up to $\mathcal{O}(j^3)$ as follows:
\begin{align}
  (m_\pi^2)_{33} & = 2 \lambda f^2 + \frac{3 j^3}{f} \sin\alpha_H \,, &
  (m_\pi^2)_{00} & = (m_\sigma^2)_{11} =  (m_\sigma^2)_{22} = 2 c + \lambda_2 f^2 + \frac{j^3}{f} \left( 1 + \frac{\lambda_2}{\lambda} \right) \sin\alpha_H\,, 
  \label{eq:ScalarMass_IsospinComplex} \\ 
  (m_\sigma^2)_{33} & = 2c + \frac{j^3}{f} \sin\alpha_H \,, &
  (m_\sigma^2)_{00} & = (m_\pi^2)_{11} = (m_\pi^2)_{22} = \frac{j^3}{f} \sin\alpha_H \,.
  \label{eq:PseudoScalarMass_IsospinComplex}
\end{align}
When we take $\alpha_H = \pi/2$ (pure imaginary $j^3$ source), the mass spectrum coincides with the mass spectrum with the real $j^0$ source (see \cref{eq:ScalarMass_UniversalReal,eq:PseudoScalarMass_UniversalReal}).

\subsection{Mixture \label{app:mixture}}

In the last of this section, we consider more generic source term.
In general, quark masses violate the isospin symmetry, and hence the source term in the effective theory should reflect the violation.
In particular, the isospin violation in the mass spectrum is important for the pion mass difference in our study.
The diagonalized quark mass matrix corresponds to the following source term in the LSM:
\eqs{
  H & = j^0 T^0 + j^3 T^3 \,, ~~~ (j^0, j^3 >0)\,.
}
In the presence of two sources $j^0$ and $j^3$, the vectorial flavor symmetry $SU(2)_V$ is also broken. 
The vacuum choice should reflect the breaking, and hence we assume the vacuum as follows:
\eqs{
  \langle \sigma^a \rangle = \bar \sigma^0 \delta_{a,0} + \bar \sigma^3 \delta_{a,3} \,, ~~~
  \langle \pi^a \rangle = \bar \pi^0 \delta_{a,0} + \bar \pi^3 \delta_{a,3} \,.
}
In the following, we find analytic formulae of the mass spectra.
We compute the spectrum by treating either $j^0$ or $j^3$ as a dominant source and by taking into account another as perturbation, and then we check the absence of tachyonic modes.

First, we consider the case with $j^0 \gg j^3$. 
From the dominant source $j^0$, the VEV $\bar \sigma^0$ is given by \cref{eq:VEV_UniversalMass}, and mass spectrum is given in \cref{eq:ScalarMass_UniversalReal,eq:PseudoScalarMass_UniversalReal}.
The potential for $\sigma^3$ (mass: $2c+\lambda_2 f^2$ with $f^2 = \lambda^{-1} (c+\mu^2)$) is tilted by the source term $j^3$, and therefore $\sigma^3$ gets its VEV as $\bar\sigma^3 \simeq j^3 (2c+\lambda_2 f^2)^{-1}$ as far as the potential is mainly lifted up by the mass term.
Thus, the VEVs in the case with $j^0 \gg j^3$ are given by
\begin{align}
  \bar \sigma^0 & = f + \frac{j^0}{2 \lambda f^2} \,, & 
  \bar \sigma^3 & \simeq \frac{j^3}{2c+\lambda_2 f^2} \,, 
  \label{eq:mixture1_sigmaVEV} \\
  \bar \pi^0 & = 0 \,, & 
  \bar \pi^3 & = 0 \,.
  \label{eq:mixture1_piVEV}
\end{align}
The presence of $\bar \sigma^3$ leads to the modification of the mass spectrum of the order $\mathcal{O}[(j^3)^2]$, in particular the mass difference between $\pi^3$ and $\pi^{1,2}$. 
$\pi^0$--$\pi^3$ mixing arises from $\bar \sigma^0$ and $\bar \sigma^3$, and then the mass eigenvalues are 
\begin{align}
  m_{\pi^0}^2 & = 2c + \frac{j^0}{f} - \frac{\lambda_2 (j^3)^2 (2c-\lambda_2 f^2)}{2c(2c+\lambda_2 f^2)^2} \,, &
  m_{\pi^1}^2 & = \frac{j^0}{f} \,, &
  m_{\pi^3}^2 & = \frac{j^0}{f} - \frac{\lambda_2 (j^3)^2}{2c(2c+\lambda_2 f^2)} \,.
  \label{eq:pionmass_mixture1}
\end{align}
Here, $\tilde \pi^0$ and $\tilde \pi^3$ are mass eigenvalues of the $\pi^0$--$\pi^3$ system.
The presence of $j^3$ leads to the mass difference among $\pi^a \, (a =1,2,3)$ though they form the isospin triplet.
We note that the isospin violation in the pion mass arises of order $\mathcal{O}[(j^3)^2]$, not $\mathcal{O}[j^3]$.
The condition that the lightest pion has a positive mass squared puts the upper bound on $j^3$: $j^3_\mathrm{max} \geq j^3$ where
\eqs{
  j^3_\mathrm{max} = \sqrt{\frac{j^0}{\lambda_2} \frac{2c}{f} \left( 2 c + \lambda_2 f^2 \right)} \,.
}

Next, we consider the case with $j^3 \gg j^0$, \textit{i.e.}, we treat $j^3$ as the dominant source.
\cref{eq:VEV_Isospin} gives the VEVs when the source term contains only $j^3$. 
Both $\sigma^3$ and $\pi^3$ get their VEVs in this case since there is the mass mixing among $\sigma$ and $\pi$.
Therefore, both $\sigma^0$ and $\pi^0$ get their VEVs via $j^0$ though $j^0$ makes the potential only for $\sigma^0$ tilted. 
The mass matrix for $\sigma^0$ and $\pi^0$ has components up to $\mathcal{O}[(j^3)^2]$ as 
\begin{align}
  m_\sigma^2 & = \frac{\lambda_2}{4 c^2} (j^3)^2 \,, &
  m_\pi^2 & = 2 c + \lambda_2 f^2 - \frac{\lambda_2}{4 c^2} (j^3)^2 \,, &
  m_{\pi \sigma}^2 & = \frac{\lambda_2}{2 c} f j^3 \,,
\end{align}
where we define $f^2 = \lambda^{-1} (c+\mu^2)$, again.
The mass basis and the mass eigenvalues are defined by 
\begin{align}
  \begin{pmatrix}
    s_- \\
    s_+
  \end{pmatrix}
  & = R^T
  \begin{pmatrix}
    \sigma^0 \\
    \pi^0
  \end{pmatrix} \,, &
  \mathrm{diag} (m_-^2 , m_+^2) & =
  R^T
  \begin{pmatrix}
    m_\sigma^2 & m_{\pi\sigma}^2 \\
    m_{\pi\sigma}^2 & m_\pi^2
  \end{pmatrix} 
  R^T \,,
\end{align}
where the mixing matrix $R$ and the mass eigenvalues $m^2_\pm$ are 
\begin{align}
  R(\theta) & \equiv 
  \begin{pmatrix}
    \cos \theta & - \sin\theta \\
    \sin \theta & \cos\theta
  \end{pmatrix} \,, & 
  \tan 2 \theta & = \frac{2 m^2_{\pi \sigma}}{m_\sigma^2 - m_\pi^2} \,, \\
  m^2_\pm & = \frac12 \left[ m_\sigma^2 + m_\pi^2 \pm \sqrt{(m_\sigma^2 - m_\pi^2)^2 + 4 m_{\pi\sigma}^2} \right] \,.
\end{align}
Both $s_-$ and $s_+$ get their VEVs, and hence we obtain the VEVs in the original basis as follows.
\eqs{
  \bar \sigma^0 = j^0 \left( \frac{1}{m_-^2} \cos^2\theta - \frac{1}{m_+^2} \sin^2\theta \right) \,, \qquad
  \bar \pi^0 = j^0 \cos\theta \sin\theta \left( \frac{1}{m_-^2} + \frac{1}{m_+^2} \right) \,. 
}
In the limit of small sources, we obtain the VEVs:
\begin{align}
  \bar \sigma^0 & \simeq j^0\left[ \frac{\lambda_2 f^2}{2c(2 c + \lambda_2f^2)} + \frac{2c}{(j^3)^2 \lambda_2} \left( 2 c + \lambda_2f^2 \right)\right] \,, & 
  \bar \sigma^3 & = \frac{j^3}{2c}  \,, 
  \label{eq:mixture2_sigmaVEV} \\
  \bar \pi^0& \simeq - \frac{j^0}{j^3} f\,, & 
  \bar \pi^3 & = f - \frac{(j^3)^2}{8 c^2 f} \,.
  \label{eq:mixture2_piVEV}
\end{align}

\subsection{Higgs-Scalar Meson Mixing}
In this subsection, we consider back-reaction from the VEV of the dark Higgs $\phi_D$.
The potential including the LSM field and $\phi_D$ is given by
\eqs{
  V(\Phi, \phi_D) = V_{\phi_D}(\phi_D) + V_\mathrm{LSM}(\Phi) + V_\mathrm{mix}(\Phi,\phi_D) \,,
}
where each part of the potential is
\begin{align}
  V_{\phi_D}(\phi_D) & = - \mu^2_\phi |\phi_D|^2 + \frac{\lambda_\phi}{4} |\phi_D|^4 \,, \\
  V_\mathrm{LSM}(\Phi) & = - \mu^2 \mathrm{tr}(\Phi^\dag \Phi) + \lambda_1 [\mathrm{tr}(\Phi^\dag \Phi) ]^2 + \lambda_2 \mathrm{tr}(\Phi^\dag \Phi)^2 \nonumber \\
  & \qquad - c \left[ \det \Phi + \det \Phi^\dag \right] - \mathrm{tr}(j^a T^a \Phi + j^{a\ast} T^a \Phi^\dag)\,, 
\end{align}
where $j^a T^a$ is a diagonalized source term that originates from the dark-quark mass matrix.
$V_\mathrm{mix}$ denotes the mixing term of the LSM field and $\phi_D$
\eqs{
  V_\mathrm{mix}(\Phi,\phi_D) & = \frac{1}{v} \mathrm{tr} (\Phi_D j_Y \Phi) + \mathrm{h.c.} \,.
}
Here, $v$ is the VEV of the dark Higgs that is determined only by the Higgs potential, $v = \sqrt{2 \mu_\phi^2/\lambda_\phi}$, and $j_Y$ and $\Phi_D$ are the matrix-form coupling and the dark Higgs field defined by
\eqs{
  j_Y = 
  \begin{pmatrix}
    0 & j_1 \\
    j_2^\ast & 0
  \end{pmatrix} \,, \qquad
  \Phi_D = 
  \begin{pmatrix}
    \phi_D & 0 \\
    0 & \phi_D^\ast 
  \end{pmatrix} \,.
}
$j_Y$ originates from the Yukawa coupling between the dark quarks and the dark Higgs.
$j_1$ and $j_2$ are, respectively, proportional to Yukawa couplings $y_1$ and $y_2$ in \cref{eq:QuarkLagrangian}, and $j_a \, (a =1, 2)$ have mass dimension of three.
A combination $\Phi_D j_Y$ has only off-diagonal entries, and gives a source term of the off-diagonal component of $\Phi$ after $\phi_D$ gets its VEV.
We define the Higgs-dependent source term $H$ as follows:
\eqs{
  \mathcal{L} \supset  \mathrm{tr}(j^a T^a \Phi + j^{a\ast} T^a \Phi^\dag) - \frac{1}{v} \mathrm{tr}(\Phi_D j_Y \Phi + j_Y^\dag \Phi_D^\dag \Phi^\dag) 
  \equiv \mathrm{tr}(H \Phi + H^\dag \Phi^\dag) \,,
}
where
\eqs{
  H(\phi_D) \equiv j^a T^a - \frac{1}{v} \Phi_D j_Y \,.
}
When the source term is hermitian, $H= H^\dag$ (\textit{i.e.}, $j_1 = j_2$), the source matrix $H$ is factored out $\mathrm{tr}H(\Phi + \Phi^\dag)$.
In this case, the dark Higgs potential is tilted if the charged scalar meson obtains its VEV in the charged basis.
The Lagrangian possesses the parity invariance, $\Phi \to \Phi^\dag$. 
As shown in previous subsection, the dominant VEVs of the LSM field are $\bar \sigma^0$ and $\bar \pi^3$ in mass basis when parity is conserved.
When the system possesses the $SU(2)_V$ flavor symmetry, no $U(1)_D$-charged sigma meson gets its VEV in charge basis from $\bar \sigma^0$.
Due to the parity invariance, again, no $U(1)_D$-charged sigma meson obtains its VEV in charge basis from $\bar \pi^3$ in mass basis.
Therefore, we conclude that we have no back-reaction from the LSM field to the dark Higgs VEV when we have $P$-invariance and $SU(2)_V$ global flavor symmetry.

We therefore consider the following cases that the source term is not hermitian. 
\begin{itemize}
  \item $H = \frac{J}{2} U_A^2$ with an $SU(2)$ matrix $U_A$
  \item $H = J U_R^\dag T^3 U_L$ with $SU(2)$ matrices $U_L, U_R$
  \item $H = U_R^\dag (J_0T^0 + J_3 T^3) U_L$ with $SU(2)$ matrices $U_L, U_R$
\end{itemize}
Since the quark mass matrix is diagonalized before the $U(1)_D$ breaking, the original source term $j^a T^a$ is diagonalized and we parametrize the source term as follows.
\eqs{
  H(\phi_D)
  = \frac12 
  \begin{pmatrix}
    j_0+j_3 & - \frac{2}{v} \phi_D j_1 \\
    - \frac{2}{v} \phi_D^\ast j_2^\ast & j_0-j_3
  \end{pmatrix} \,.
  \label{eq:parametrization_source}
}

\subsubsection*{$H = \frac{J}{2} U_A^2$}

We consider the first case that $H = \frac{J}{2} U_A^2$ with an $SU(2)$ matrix $U_A$ and a real-positive parameter $J>0$. 
The matrix $U_A$ is parametrize phase $\alpha_A$ and angle $\theta_A$ as follows:
\eqs{
  U_A = 
  \begin{pmatrix}
    \cos \theta_A & - e^{-i\alpha_A} \sin\theta_A \\
    e^{i \alpha_A} \sin\theta_A & \cos\theta_A
  \end{pmatrix} \,.
}
By comparing with \cref{eq:parametrization_source}, we determine parameters as follows. 
\eqs{
  j_0 & = J \cos 2 \theta_A \,, \qquad 
  j_3 = 0 \,, \\
  \frac{\phi_D}{v} j_1 & = \frac{J}{2} e^{-i \alpha_A} \sin 2 \theta_A \,, \qquad 
  \frac{\phi_D^\ast}{v} j_2^\ast = - \frac{J}{2} e^{i \alpha_A} \sin 2 \theta_A \,. 
}
The parameters depend on the Higgs field as follows.
\eqs{
  J = \sqrt{j_0^2 - \frac{4 |\phi_D|^2 j_1 j_2^\ast}{v^2}} \,, \quad
  \tan 2 \theta_A = \frac{2}{j_0} \sqrt{- \frac{|\phi_D|^2 j_1 j_2^\ast}{v^2}}\,, \quad 
  \phi_D j_1 e^{i \alpha_A} = - \phi_D^\ast j_2 e^{-i \alpha_A} \,.
  \label{eq:parameters_Asq}
}
By use of the $SU(2)_L \times SU(2)_R$ rotation, $U_L = U_R^\dag = U_A^\dag$, we can diagonalize the source term.
$\Phi' \equiv U_A \Phi U_A$ refers the LSM field in this basis. 
In this basis, the source term is proportional to unity, and therefore $\Phi'$ obtains its VEV proportional to unity: $\langle \Phi'\rangle = \bar \sigma^0 \mathbf{1}$, and hence the VEV of $\Phi$ has an off-diagonal entry that provides $U(1)_D$ violation.
\eqs{
  \langle \Phi \rangle = U_A^\dag \langle \Phi' \rangle U_A^\dag = \bar \sigma^0 
  \begin{pmatrix}
    \cos 2 \theta_A & e^{-i\alpha_A} \sin2 \theta_A \\
    - e^{i \alpha_A} \sin2\theta_A & \cos2 \theta_A
  \end{pmatrix}\,.
}
At the LSM potential minimum, we have a linear term of $J$ and this gives a back-reaction potential to the dark Higgs potential.
\eqs{
  V_\mathrm{LSM}(\langle\Phi'\rangle) + V_\mathrm{mix}(\langle\Phi'\rangle,\phi_D) 
  = - \frac{(c+\mu^2)^2}{4 \lambda} - J(\phi_D) \sqrt{\frac{c+\mu^2}{\lambda}} +O(J^2) \,.
}
We define the back-reaction potential as $V_\mathrm{BR}(\phi_D)$, and expand $\phi_D$ around the VEV $v$, $\phi_D = v + h$.
\eqs{
  V_\mathrm{BR}(\phi_D) = - f J(\phi_D) = - f \sqrt{j_0^2 - 4 j_1 j_2^\ast \left(1+\frac{h}{v}\right)^2}\,,
}
where $f = \sqrt{\lambda^{-1}(c+\mu^2)}$ denotes the decay constant.
Once we assume that the shift of the VEV is negligible, the pion mass is given by the source $J$ with $\phi_D = v$, $m_\pi^2 \simeq J(v)/f $.
This term tilts the Higgs potential, and then the true Higgs VEV is shifted from $v$, $v_D = v + \delta v$. 
We determine $\delta v$ in a perturbative manner of the source.
For simplicity, we take the sizes of $j_1$ and $j_2$ are the same, but the signs are  opposite to satisfy \cref{eq:parameters_Asq}: $2 j_1 = - 2 j_2 \equiv j > 0$.
When $j_0 \simeq j$, the potential is 
\eqs{
  V_{\phi_D}(\phi_D) + V_\mathrm{BR}(\phi_D) 
  = \frac12 m_h^2 h^2 - fj\sqrt{1 + \left(1+\frac{h}{v}\right)^2} + \cdots \,,
}
and then the shift of the Higgs VEV is given by
\eqs{
  \delta v = \frac{1}{\sqrt{2}} \frac{fj}{m_h^2 v} + O(j^2)
  \simeq \frac{1}{2} \frac{m_q \langle \bar q q \rangle}{m_h^2 v} \,.
}
We use the relation between the source and the pion mass, $m_\pi^2 \simeq \sqrt{j_0^2 +j^2}/f \simeq \sqrt2 j/f$, and use the Gell-Mann-Oaks-Renner (GOR) relation, $m_\pi^2 f^2 \simeq m_q \langle \bar q q \rangle$ in the second equation.

When $j_0 \gg j$, the potential is 
\eqs{
  V_{\phi_D}(\phi_D) + V_\mathrm{BR}(\phi_D) & = \frac12 m_h^2 h^2 - fj_0\sqrt{1 + \frac{j^2}{j_0^2} \left(1+\frac{h}{v}\right)^2} + \cdots \\
  & = \frac12 m_h^2 h^2 - fj_0 -\frac{f j^2}{2 j_0} \left(1+\frac{h}{v}\right)^2 + \cdots
}
and then the shift of the Higgs VEV is given by
\eqs{
  \delta v = \frac{f}{m_h^2 v}\frac{j^2}{j_0} + O(j^3) 
  \simeq \frac{1}{m_h^2 v}\frac{j^2}{m_\pi^2} \,, 
}
In the second equation, we use $m_\pi^2 \simeq \sqrt{j_0^2 +j^2}/f \simeq j_0/f$.

When $j_0 \ll j$, the potential is 
\eqs{
  V_{\phi_D}(\phi_D) + V_\mathrm{BR}(\phi_D) & = \frac12 m_h^2 h^2 - fj\sqrt{\frac{j_0^2}{j^2} + \left(1+\frac{h}{v}\right)^2} + \cdots \\
& \simeq \frac12 m_h^2 h^2 - fj \left(1+\frac{h}{v}\right) + \cdots
}
and then the shift of the Higgs VEV is given by
\eqs{
  \delta v = \frac{fj}{m_h^2 v} + O(j^2)
  \simeq \frac{m_q \langle \bar q q \rangle}{m_h^2 v} \,.
}
We use $m_\pi^2 \simeq \sqrt{j_0^2 +j^2}/f \simeq j/f$ and the GOR relation, again.

\subsubsection*{$H = J U_R^\dag T^3 U_L$}

Next, we consider $H = J U_R^\dag T^3 U_L$ with $SU(2)$ matrices $U_L$ and $U_R$, and a real-positive parameter $J > 0$,
\eqs{
  U_L = 
  \begin{pmatrix}
    \cos \theta_L & - e^{-i\alpha_L} \sin\theta_L \\
    e^{i \alpha_L} \sin\theta_L & \cos\theta_L
  \end{pmatrix} \,, \quad 
  U_R = 
  \begin{pmatrix}
    \cos \theta_R & - e^{-i\alpha_R} \sin\theta_R \\
    e^{i \alpha_R} \sin\theta_R & \cos\theta_R
  \end{pmatrix} \,, 
}

We take $\alpha_L = \alpha_R \equiv \alpha_V$ for simplicity, 
\eqs{
  J U_R^\dag T^3 U_L = \frac{J}{2}
  \begin{pmatrix}
    \cos 2 \theta_V & - e^{-i\alpha_V} \sin 2\theta_V \\
    - e^{i \alpha_V} \sin 2\theta_V & - \cos 2\theta_V
  \end{pmatrix}\,.
}
Here, we define $2 \theta_V \equiv \theta_L + \theta_R$.
By comparing with \cref{eq:parametrization_source}, we determine parameters as follows. 
\eqs{
  j_0 & = 0 \,, \qquad 
  j_3 = J \cos 2 \theta_V \,, \\
  \frac{\phi_D}{v} j_1 & = \frac{J}{2} e^{-i \alpha_V} \sin 2 \theta_V \,, \qquad 
  \frac{\phi_D^\ast}{v} j_2^\ast = \frac{J}{2} e^{i \alpha_V} \sin 2 \theta_V \,. 
}
The parameters depend on the Higgs field as follows.
\eqs{
  J = \sqrt{j_3^2 + \frac{4 |\phi_D|^2 j_1 j_2^\ast}{v^2}} \,, \quad
  \tan 2 \theta_V = \frac{2}{j_3} \sqrt{\frac{|\phi_D|^2 j_1 j_2^\ast}{v^2}}\,, \quad 
  \phi_D j_1 e^{i \alpha_V} = \phi_D^\ast j_2 e^{-i \alpha_V} \,.
  \label{eq:parameters_RdagT3L}
}
By use of the $SU(2)_L \times SU(2)_R$ rotation, $U_L$ and $U_R$, we can diagonalize the source term.
$\Phi' \equiv U_L \Phi U_R^\dag$ refers the LSM field in this basis. 
In this basis, the source term is proportional to $T^3$, and therefore $\Phi'$ obtains its VEV proportional to $T^3$: as shown in \cref{app:ISoV}, both $\sigma^3$ and $\pi^3$ get their VEVs, $\langle \Phi'\rangle = (\bar \sigma^3 + i \bar \pi^3) T^3$, and hence the VEV of $\Phi$ has an off-diagonal entry that provides $U(1)_D$ violation.
\eqs{
  \langle \Phi \rangle = U_L^\dag \langle \Phi' \rangle U_R = (\bar \sigma^3 + i \bar \pi^3) 
  \begin{pmatrix}
    \cos 2 \theta_V & - e^{-i\alpha_V} \sin2 \theta_V \\
    - e^{i \alpha_V} \sin2\theta_V & - \cos2 \theta_V
  \end{pmatrix}\,.
}
Since the VEV $\bar \pi^3 = f$ does not contain the linear term of $J$, the pion mass is proportional to $J^2$, 
\eqs{
  m_\pi^2 = \frac{\lambda_2}{2c (2c+\lambda_2 f^2)} J^2(v) \,,
}
where $f = \sqrt{\lambda^{-1}(c+\mu^2)}$ denotes the decay constant, again.
We assume that the shift of the VEV is negligible, and then the pion mass is given by the source $J$ with $\phi_D = v$.
At the LSM potential minimum, we have a quadratic term of $J$ and this gives a back-reaction potential to the dark Higgs potential.
\eqs{
  V_\mathrm{LSM}(\langle\Phi'\rangle)+V_\mathrm{mix}(\langle\Phi'\rangle,\phi_D) 
  = - \frac{(c+\mu^2)^2}{4 \lambda} - \frac{J^2(\phi_D)}{4c} +O(J^3)
}
Similar to the previous case, we define the back-reaction potential as $V_\mathrm{BR}(\phi_D)$, and expand $\phi_D$ around the VEV $v$, $\phi_D = v + h$.
\eqs{
  V_\mathrm{BR}(\phi_D) = - \frac{J^2(\phi_D)}{4c} 
  = - \frac{1}{4c} \left[j_3^2 + 4 j_1 j_2^\ast \left( 1 + \frac{h}{v} \right)^2 \right]\,,
}
This gives a shift of the Higgs field as follows up to the second order of sources.
\eqs{
  \delta v \simeq  \frac{2 j_1 j_2^\ast}{m_h^2 v c}\,.
}

\subsubsection*{$H = U_R^\dag (J_0T^0 + J_3 T^3) U_L$}

We consider more generic source term that is characterized by $T^0$ and $T^3$.
This is a combination of the aforementioned two cases.
We consider $H = U_R^\dag (J_0T^0 + J_3 T^3) U_L$ with $SU(2)$ matrices $U_L$ and $U_R$, which we use in the previous case, and a real-positive parameter $J_0, J_3 > 0$.
We take $\alpha_L = \alpha_R \equiv \alpha_V$ for simplicity, again.
\eqs{
  U_R^\dag (J_0T^0 + J_3 T^3) U_L = \frac{1}{2}
  \begin{pmatrix}
    J_0 \cos 2 \theta_A + J_3 \cos 2 \theta_V & - e^{-i\alpha_V} (J_0 \sin 2 \theta_A + J_3 \sin 2 \theta_V) \\
    - e^{i \alpha_V} (- J_0 \sin 2 \theta_A + J_3 \sin 2 \theta_V) & J_0 \cos 2 \theta_A - J_3 \cos 2 \theta_V
  \end{pmatrix}\,.
}
Here we define $2 \theta_V = \theta_L + \theta_R \,, 2 \theta_A = \theta_L - \theta_R$.
By comparing with \cref{eq:parametrization_source}, we determine parameters as follows. 
\eqs{
  j_0 & = J_0 \cos 2 \theta_V \,, \qquad 
  j_3 = J_3 \cos 2 \theta_V \,, \\
  \frac{\phi_D}{v} j_1 & = \frac{e^{-i \alpha_V}}{2} \left[ J_0 \sin 2 \theta_A + J_3 \sin 2 \theta_V \right] \,, \qquad 
  \frac{\phi_D^\ast}{v} j_2^\ast = \frac{e^{i \alpha_V}}{2} \left[ - J_0 \sin 2 \theta_A + J_3 \sin 2 \theta_V \right] \,. 
}
The parameters depend on the Higgs field as follows.
\begin{align}
  J_0^2 & = j_0^2 + \frac{4}{v^2} (\phi_D j_1 e^{i \alpha_V} - \phi_D^\ast j_2^\ast e^{-i\alpha_V})^2 \,, &
  J_3^2 & = j_3^2 + \frac{4}{v^2} (\phi_D j_1 e^{i \alpha_V} + \phi_D^\ast j_2^\ast e^{-i\alpha_V})^2 \nonumber \\
  \tan2\theta_A & = \frac{\phi_D j_1 e^{i \alpha_V} - \phi_D^\ast j_2^\ast e^{-i\alpha_V}}{2 j_0 v} \,, &
  \tan2\theta_V & = \frac{\phi_D j_1 e^{i \alpha_V} + \phi_D^\ast j_2^\ast e^{-i\alpha_V}}{2 j_3 v} \,, \nonumber \\
  e^{i \alpha_V} & = \frac{\phi_D^\ast j_2^\ast}{|\phi_D j_2|} = \frac{\phi_D^\ast j_1^\ast}{|\phi_D j_1|}\,.
  \label{eq:parameters_Generic}
\end{align}
By use of the $SU(2)_L \times SU(2)_R$ rotation, $U_L$ and $U_R$, we can diagonalize the source term.
$\Phi' \equiv U_L \Phi U_R^\dag$ refers the LSM field in this basis. 
As with the previous case, while the VEVs are diagonal in the mass basis, there are off-diagonal entries in $\langle \Phi \rangle$.

It is challenging to find an analytic formula that is valid whole parameter range. 
In \cref{app:mixture}, we find the approximate solutions where each of sources is larger than another and we treat the smaller one as the perturbation.
When $J_0$ dominates the source term, the VEVs are given by \cref{eq:mixture1_sigmaVEV,eq:mixture1_piVEV},
\begin{align}
  \bar \sigma^0 & = f + \frac{J_0}{2 \lambda f^2} \,, & 
  \bar \sigma^3 & \simeq \frac{J_3}{2c+\lambda_2 f^2} \,, \\
  \bar \pi^0 & = 0 \,, & 
  \bar \pi^3 & = 0 \,.
\end{align}
At the LSM potential minimum, we have a linear term of $J_0$ and this gives a back-reaction potential to the dark Higgs potential.
\eqs{
  V_\mathrm{LSM}(\langle\Phi'\rangle)+V_\mathrm{mix}(\langle\Phi'\rangle,\phi_D) 
  \simeq - \frac{(c+\mu^2)^2}{4 \lambda} - J_0(\phi_D) f \,, 
}
This back-reaction potential gives the same results with the case of $H = \frac{J}{2} U_A^2$.

On the other hand, when $J_3$ dominates the source term, the VEVs are given by \cref{eq:mixture2_sigmaVEV,eq:mixture2_piVEV},
\begin{align}
  \bar \sigma^0 & \simeq  \frac{J_0 2c}{(J_3)^2 \lambda_2} \left( 2 c + \lambda_2f^2 \right) \,, & 
  \bar \sigma^3 & = \frac{J_3}{2c}  \,, \\
  \bar \pi^0& \simeq - \frac{J_0}{J_3} f\,, & 
  \bar \pi^3 & = f - \frac{(J_3)^2}{8 c^2 f} \,.
\end{align}
As with the case of $H = J U_R^\dag T^3 U_L$, at the LSM potential minimum, we only have a quadratic term of $J_3$ and this gives a back-reaction potential to the dark Higgs potential.
\eqs{
  V_\mathrm{LSM}(\langle\Phi'\rangle)+V_\mathrm{mix}(\langle\Phi'\rangle,h) 
  \simeq - \frac{(c+\mu^2)^2}{4 \lambda} - \frac{J_3^2(\phi_D)}{4c} \,.
}
This back-reaction potential gives the same results with the case of $H = J R^\dag T^3 L$.

\section{Hadron Spectrum \label{sec:spectrum}}

In the previous appendix, we discuss the generic mass spectrum in the LSM without corrections from QED. 
In this appendix, we incorporate the QED correction into the mass spectrum.
While the QED corrections usually arise as the loop corrections, we treat them as effective interactions by regarding the charge matrices as spurion fields. 

\subsection{Pion Mass Matrix}
Let us consider meson spectrum that originates from the universal source discussed in \cref{app:UniV}.
The meson masses are 
\eqs{
  - \mathcal{L} & \supset \frac12 \sum_{a=1}^3 m_\pi^2 \pi^a \pi^a+ \frac12 m_\eta^2 (\pi^0)^2 \,, \\
  m_\eta^2 & = 2 c + \frac{j^0}{f} \,, \qquad m_\pi^2= \frac{j^0}{f} \,.
}
Here, we assume that the source $j^0$ is positive. 
When the quark mass basis differs from its charge basis, we obtain the charge basis by the flavor rotation.
It is expected that the charge basis of mesons is identified with the mass basis rotated by the same flavor rotation.
Let $\Phi$ and $\hat \Phi$ be the LSM fields in the mass basis and in the charge basis, respectively.
\eqs{
  \hat \Phi = U_L \Phi U_R^\dag \supset \hat \pi^a T^a = \frac12 
  \begin{pmatrix}
    \hat \eta+\hat \pi & \hat \pi^+ \\
    \hat \pi^- & \hat \eta-\hat \pi
  \end{pmatrix} \,,
}
where the flavor rotation matrices $U_L$ and $U_R$ that relate the quark mass basis to its charge bases.
In particular, when we take $U_L = U_R = U_V$ and their phases to be $\alpha_L = \alpha_R = 0$, the charge neutral pion $\hat \pi$ is a linear combination of $\pi^1$ and $\pi^3$,
\eqs{
  \begin{pmatrix}
    \hat \pi \\
    \mathrm{Re}(\hat \pi^-)
  \end{pmatrix}
  =
  \begin{pmatrix}
    \cos 2 \theta_V & - \sin 2 \theta_V \\
    \sin 2 \theta_V & \cos 2 \theta_V
  \end{pmatrix}
  \begin{pmatrix}
    \pi^3 \\
    \pi^1 
  \end{pmatrix} \,,
}
and $\hat \eta = \pi^0$ and $\mathrm{\hat \pi^-} = \pi^2$.
Since the charge neutral pion has a coupling to dark photons via chiral anomaly, pions can decay through the (off-shell) dark photons through the mixing.

Once we have $U(1)_D$ coupling, we can also add the charge masses for the dark pions.
In general, the masses in the mass basis are given by
\eqs{
  -\mathcal{L} 
  & \supset 4\pi \alpha' f_\pi^2 \left[ 
  c_1 \mathrm{tr} (\Pi^\dag q_L \Pi q_R)
  + c_2 \mathrm{tr} (\Pi^\dag q_L^2 \Pi) 
  + c_3 \mathrm{tr} (\Pi^\dag \Pi q_R^2) 
  +  \mathrm{h.c.} \right] \,,
}
Here, $q_L$ and $q_R$ are the charge matrices that are defined in \cref{eq:covariantD_Phi}, and $\Pi$ is the matrix-form pion fields in the mass basis.
\eqs{
  \Pi \equiv \frac12 
  \begin{pmatrix}
    \pi_3 & \pi_1 - i \pi_2 \\
    \pi_1 + i \pi_2 & -\pi_3
  \end{pmatrix} \,.
}
Compared to the chiral perturbation theory, $\Pi^\dag \Pi$ is not constant in the LSM, and then we can add the second and the third terms.
Indeed, these terms are important for giving the correct charge masses.
When $L$ and $R$ that relate the 
quark mass basis to its charge bases are unity, the mass basis coincides with the charge basis.
In this case, the charge matrices are the same $q_L = q_R = \hat q = \mathrm{diag}(\frac23,-\frac13)$ and $\pi_1$ and $\pi_2$ are the charged pions, then we obtain
\begin{align}
  -\mathcal{L} 
  \supset & \frac{c_1}{2} 4\pi \alpha' f_\pi^2 \pi_3^2 
  + \frac{1}{2} (c_2+c_3) 4\pi \alpha' f_\pi^2 (\pi_1^2 + \pi_2^2 + \pi_3^2)
  = \frac{c_Q^{\pi'}}{2} 4\pi \alpha' f_\pi^2 (\pi_1^2 + \pi_2^2) \,,
\end{align}
In the last line, we determine the coefficients in order to correctly give charge masses to the charged pions, $c_2 = c_3 = - 2 c_1 \equiv c_Q^{\pi'}/2$.

Next, we take into account isospin breaking terms.
For simplicity, we take the flavor rotation matrices $U_L = U_R \equiv U_V$, and then the charge matrices are the same $q_L = q_R = U_V \mathrm{diag}(\frac23,-\frac13) U_V^\dag$, where 
\eqs{
  U_V = 
  \begin{pmatrix}
    \cos \theta_V & - e^{-i\alpha_V} \sin\theta_V \\
    e^{i \alpha_V} \sin\theta_V & \cos\theta_V
  \end{pmatrix} \,.
}
In this case, the mass terms of pions are given by
\eqs{
  -\mathcal{L} 
  \supset & - \frac{m_D^2}{2} (\sin2 \theta_V \cos\alpha_V \pi_1 + \sin 2 \theta_V \sin\alpha_V \pi_2 - \cos2 \theta_V \pi_3)^2 \\
  & + \frac{m_D^2}{2} (\pi_1^2 + \pi_2^2 + \pi_3^2)
  + \frac{m_{\pi^3}^2}{2}\pi_3^2 
  + \frac{m_{\pi^1}^2}{2} (\pi_1^2 + \pi_2^2 ) \,,
}
where we define $m_D^2 = 4\pi \alpha' c_Q^{\pi'} f_\pi^2$.
$m_{\pi_3}^2$ and $m_{\pi_{1,2}}^2$ are defined in \cref{eq:pionmass_mixture1}, which include the isospin violation.
The mass eigenvalues are 
\eqs{
  \begin{pmatrix} \pi_1' \\ \pi_2' \\ \pi_3' \end{pmatrix}
  \equiv
  \begin{pmatrix} 
    c_{\beta_V} & & s_{\beta_V} \\
    & 1 & \\
    - s_{\beta_V} & & c_{\beta_V}
  \end{pmatrix}
  \begin{pmatrix} 
    c_{\alpha_V} & s_{\alpha_V} & \\
    - s_{\alpha_V} & c_{\alpha_V} & \\
    && 1
  \end{pmatrix}
  \begin{pmatrix} \pi_1 \\ \pi_2 \\ \pi_3 \end{pmatrix} \,,
}
where $c_\alpha = \cos \alpha$ and $s_\alpha = \sin \alpha$, and the mixing angle $\beta_V$ is defined by 
\eqs{
  \tan 2 \beta_V = \frac{m_D^2 \sin 4 \theta_V}{m_{\pi^3}^2-m_{\pi^1}^2 - m_D^2 \cos 4\theta_V}\,.
}
When there is no $U(1)_D$ correction, $m_D = 0$, $\beta_V$ is zero.
If there is no isospin violation in mass basis in advance, $m_{\pi^3}^2 = m_{\pi^1}^2$, the degeneracy of the dark pions is resolved only by the $U(1)_D$ correction. 
This means that the charge basis and the mass basis coincide with each other, and indeed $\beta_V = - 2 \theta_V$ as $m_{\pi^3}^2 = m_{\pi^1}^2$.
The mass eigenvalues are 
\begin{align}
  m_{\pi^1}^2+m_D^2 \,,
  \frac12 \left( m_{\pi^3}^2 + m_{\pi^1}^2+m_D^2 \pm \sqrt{(m_{\pi^3}^2 - m_{\pi^1}^2)^2 +  2 \cos4 \theta_V m_D^2 (m_{\pi^3}^2 - m_{\pi^1}^2)+ m_D^4} \right) \,.
\end{align}
In this mass basis, $U(1)_D$ gauge interaction of pions is given by 
\eqs{
  \mathcal{L}_\mathrm{LSM} & \supset 
  e' A'^\mu \left[ 
  \cos(2 \theta_V + \beta) (\pi_1' \partial_\mu \pi_2' - \pi_2' \partial_\mu \pi_1') 
  + \sin(2 \theta_V + \beta) (\pi_2' \partial_\mu \pi_3' - \pi_3' \partial_\mu \pi_2') 
  \right] \,.
}

\subsection{Nucleon Masses}

We now consider the mass spectrum of the light dark nucleons.
The dominant nucleon masses arise from the interaction with the LSM fields, while the global symmetry allows the vector-like mass term that originates from the quark mass term.
\eqs{
  \mathcal{L}_B & \supset - (\xi \overline N M_{Q'}^\mathrm{diag} N + g \overline N \Phi^\dag N + \mathrm{h.c.}) \,.
}
Here, we construct the hadron effective theory after we rotate the quark basis into their mass basis, and hence the vector-like mass term is proportional to the diagonalized quark mass matrix.
Even though the vector-like mass term is not dominant contribution, it is important for the nucleon mass difference.
The large-$N_C$ scaling of a coupling $g$ is $g = g_{\pi NN} \sim 4 \pi \sqrt{N_C}$.
We assume that the LSM field obtains its VEV to be diagonalized, $\Phi = f T^0 + \delta T^3$.
The nucleon masses are given by \cref{eq:nucleonmass}.
We can also include $U(1)_D$ correction to the nucleon mass,
\eqs{
  \mathcal{L}_B \supset - \frac{\alpha'}{4 \pi} \left[\overline N (q_R^N)^{\dag 2} \Phi^\dag N + \overline N \Phi^\dag (q_L^N)^2 N + \overline N (q_R^N)^{\dag}\Phi^\dag q_L^N N + \mathrm{h.c.} \right] \,.
}
The charge matrices are defined in the text, \cref{eq:charge_mat}.

We discuss the nucleon mass difference in the SM, proton-neutron mass difference, by use of the LSM.  
In the SM, $N_1$ and $N_2$ correspond to neutron and proton, respectively.
First, we consider the mass difference originates from isospin-violation by quark masses.
In the SM, the mass difference of quarks is smaller than their sum, hence we use VEVs given by \cref{eq:mixture1_sigmaVEV}.
\begin{align}
  m_n - m_p - \delta M^\gamma 
  & = - g \delta = \frac{f_\pi^2}{2 c + \lambda_2 f_\pi^2} \frac{m_n+m_p}{m_u+m_d} \frac{m_\pi^2}{f_\pi^2} (m_d-m_u) \nonumber \\
  & \simeq 3.68 (m_d-m_u) \,.
\end{align}
In the second equality, we rewrite the mass difference in terms of quark masses and hadron masses by use of the following relations.
\eqs{
  \frac{j^3}{j^0} = \frac{m_u - m_d}{m_u + m_d} \,, \qquad
  g = \frac{m_p + m_n}{f_\pi} \,, \qquad 
  j^0 = m_\pi^2 f _\pi \,.
}
Since the source terms originate from the quark mass terms, the ratio $j^3/j^0$ is determined by the quark mass difference and their sum.
The second relation is obtained from the sum of nucleon masses as quark masses are negligible.
Numerically, we take the SM values of parameters~\cite{Zyla:2020zbs}: $m_u = 2.16\,\mathrm{MeV}$, $m_d = 4.67\,\mathrm{MeV}$, $m_p = 938.27\,\mathrm{MeV}$, $m_n = 939.57\,\mathrm{MeV}$, $f_\pi = 92\,\mathrm{MeV}$, $\lambda_2 \simeq (4\pi)^2/3$, and $2c \simeq m_{\eta'}^2 \simeq (957.78\,\mathrm{MeV})^2$.
The mass difference in the LSM is not so far from the lattice QCD average of the difference is (See a review~\cite{Walker-Loud:2014iea})
\begin{align}
  m_n - m_p - \delta M^\gamma 
  & \simeq 0.95(8)(6) \times(m_d-m_u)~[\mu=2~\mathrm{GeV}] \,.
\end{align}
The electromagnetic correction is \cite{WalkerLoud:2012bg}
\begin{align}
  \delta M^\gamma 
  & \simeq 1.30(03)(47)~\mathrm{MeV}\,.
\end{align}

\section{Fitting \label{app:fitting}}

In this appendix, we demonstrate the validity of our approximation formulae \cref{eq:fitting_epsilon_max} and \cref{eq:fitting_epsilon_min} by applying the formulae to the specific models.
As we discussed in the text, as for the LHC lifetime frontier, FASER and MATHUSLA, we apply our formulae to the inelastic DM model \cite{Berlin:2018jbm}. 
On the other hand, as for the fixed-target experiments, SeaQuest and E137, we apply them to the dark vector meson search \cite{Berlin:2018tvf} for SeaQuest, and to the dark photon search \cite{Bjorken:2009mm}.

\subsection{MATHUSLA and FASER}

\begin{figure}
  \centering
  \includegraphics[width=7cm]{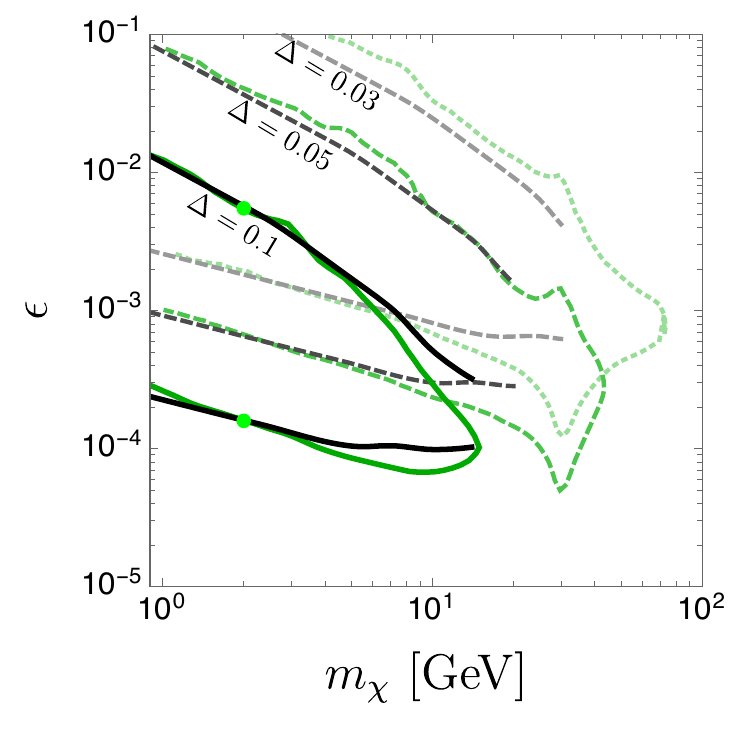}
  \includegraphics[width=7cm]{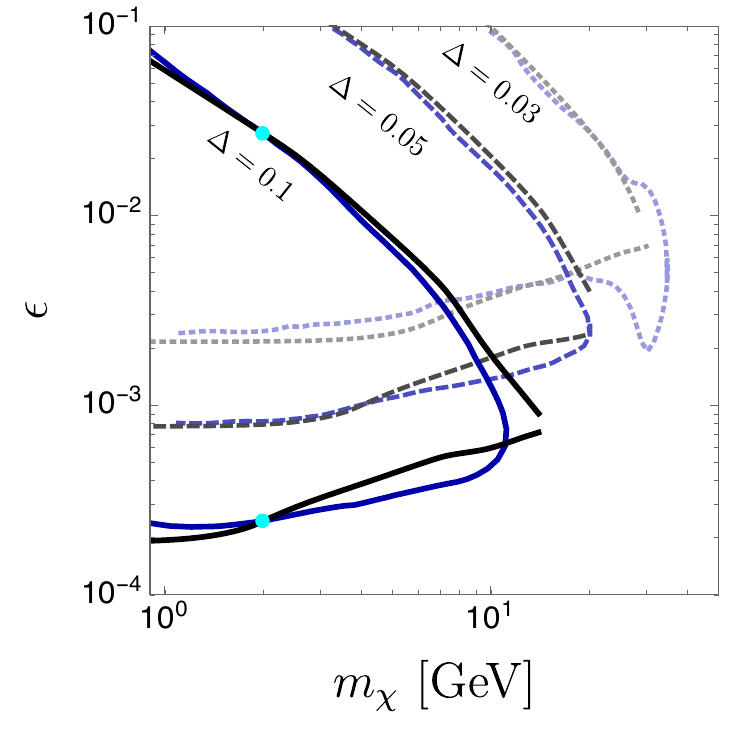}
  \caption{
    Our fitting with the results of Ref.~\cite{Berlin:2018jbm} for MATHUSLA (left) and FASER (right). 
    We take the parameters to be $m_{A'} = 3 m_{\chi}$ and $\alpha' = 0.1$, which are the same as those in Ref.~\cite{Berlin:2018jbm}. 
    The mass difference parameter, $\Delta$, is taken to be $\Delta = 0.1$ (solid), $\Delta = 0.05$ (dashed), and $\Delta = 0.03$ (dotted).
    The green (blue) lines show the future sensitivities at MATHUSLA (FASER) taken from Ref.~\cite{Berlin:2018jbm}. 
    The light green (cyan) points in the plot are the reference points of MATHUSLA (FASER) that we have used in the text. 
    The black lines show our approximate sensitivity given by \cref{eq:fitting_epsilon_max} and \cref{eq:fitting_epsilon_min}.
    }
  \label{fig:MATFAS_Fit}
\end{figure}

We show our fitting in \cref{fig:MATFAS_Fit} by use of our approximation formulae, \cref{eq:fitting_epsilon_max} and \cref{eq:fitting_epsilon_min}.
The MATHUSLA sensitivity plots are shown as green lines in the left panel, while the FASER plots are shown as blue lines in the right panel, which are read from Ref.~\cite{Berlin:2018jbm}.
We take the dark photon mass to be $m_{A'} = 3 m_{\chi}$ with the dark matter mass $m_\chi$ and the $U(1)_D$ coupling to be $\alpha' = 0.1$ that are same as the parameters taken in the literature. 
The on-shell $A'$ is produced via the SM meson decay, the Bremsstrahlung, and Drell-Yan at the LHC, and its decay $A' \to \chi_1 \chi_2$ produces the long-lived particle $\chi_2$.
The mass difference of the inelastic dark matter, defined by $\Delta \equiv (m_{\chi_2}-m_\chi)/m_\chi$ with the masses of the excited state $m_{\chi_2}$, is taken to be $\Delta = 0.1$ (solid), $\Delta = 0.05$ (dashed), and $\Delta = 0.03$ (dotted). 
We assume that the long-lived particles are produced only via Drell-Yan production at the LHC. 
We do not take into account $Z$-resonance for the production, and hence the peaky behavior around $m_\chi \simeq 30\,\mathrm{GeV}$ (corresponding to $m_{A'} \simeq m_Z$) are not reproduced in these figures. 
We take into account the difference of $m_{A'}$-dependence of $N_{A'}$ of $\chi_2$, which are scaled as $\epsilon^2 m_{A'}^{-4}$ in forward direction and $\epsilon^2 m_{A'}^{-2}$ in total, as we mentioned in the text. 

We take into account the efficiency due to the energy threshold as discussed in the text, $A_\mathrm{eff} = A^\mathrm{geo} A^\mathrm{prod}$.
$p_\mathrm{min}$ is the larger momentum between $p_\mathrm{geo}$ and $p_\mathrm{thr}$;
$p_\mathrm{thr} = 1\,\mathrm{TeV}(0.1/\Delta)$ for FASER and $p_\mathrm{thr} = 12\,\mathrm{GeV}(0.1/\Delta)$ for MATHUSLA, while $p_\mathrm{geo} = \sqrt{m_{A'}^2-4 m_\chi^2}/2 \theta$ with a typical angle $\theta$ where the detector is located: $\theta \simeq 0.5$ for MATHUSLA and $\theta \lesssim 2 \times 10^{-3}$ for FASER.
Meanwhile, we take $p_\mathrm{max}$ to be the same as what we have taken in the text: $p_\mathrm{max} = 100 \,\mathrm{GeV}$ for MATHUSLA and $p_\mathrm{max} = 7 \,\mathrm{TeV}$ for FASER.

To determine the reference values in \cref{eq:fitting_epsilon_max,eq:fitting_epsilon_min}, we take $m_{\chi} = 2.0\, \mathrm{GeV}$, and we read the boundary values of $\epsilon$ at $m_{\chi} = 2.0\, \mathrm{GeV}$ from the sensitivity plots with $\Delta = 0.1$, which are shown as green and cyan dots in the figure.
In the case of MATHUSLA, we obtain the reference values defined in \cref{eq:fitting_epsilon_max,eq:fitting_epsilon_min} from this calibration as follows.
\eqs{
  \frac{r_\mathrm{min}}{(c \tau_{\chi} p_\mathrm{max} m_{\chi}^{-1})_0} \simeq 16.2 \,, \qquad
  \frac{r_\mathrm{max}-r_\mathrm{min}}{(c \tau_{\chi} p_\mathrm{min} m_{\chi}^{-1})_0} \simeq 0.083 \,.
}
Here, $(c \tau_{\chi} p_\mathrm{max} m_{\chi}^{-1})_0 \,, (c \tau_{\chi} p_\mathrm{min} m_{\chi}^{-1})_0$ are the reference values $d_0(p_\mathrm{max})$ and $d_0(p_\mathrm{min})$ in the formulae, respectively. 
On the other hand, in the case of FASER, we obtain the reference values from this calibration as follows.
\eqs{
  \frac{r_\mathrm{min}}{(c \tau_{\chi} p_\mathrm{max} m_{\chi}^{-1})_0} \simeq 19.3 \,, \qquad
  \frac{r_\mathrm{max}-r_\mathrm{min}}{(c \tau_{\chi} p_\mathrm{min} m_{\chi}^{-1})_0} \simeq 0.0001 \,.
}

We show the boundaries with our approximation formulae as black lines in the figure.
In both panels, their line types are the same as the sensitivity plots: $\Delta = 0.1$ (solid), $\Delta = 0.05$ (dashed), and $\Delta = 0.03$ (dotted).
We just use the reference points on the sensitivity plot for $\Delta = 0.1$, the other shape of the sensitivity plots is almost reproduced by our formulae even for other choice of $\Delta$.
Since we assume that $\chi_2$ is produced only via the Drell-Yan process, the fitting of lower sensitivity curves gets worse for $m_{\chi} \simeq 1\,\mathrm{GeV}$, where the Bremsstrahlung process would be more important.

\subsection{SeaQuest}

\begin{figure}
  \centering
  \includegraphics[width=0.48\textwidth]{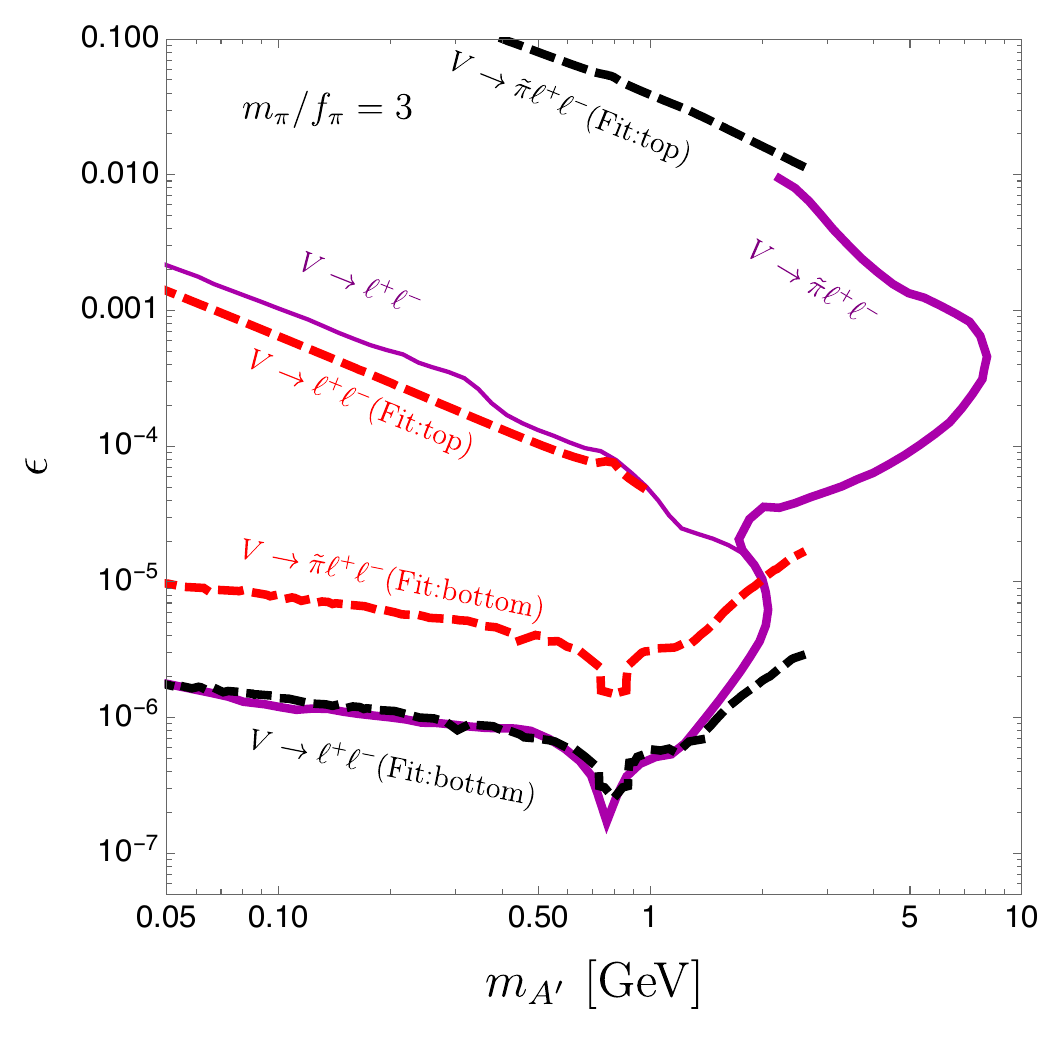}
  \includegraphics[width=0.48\textwidth]{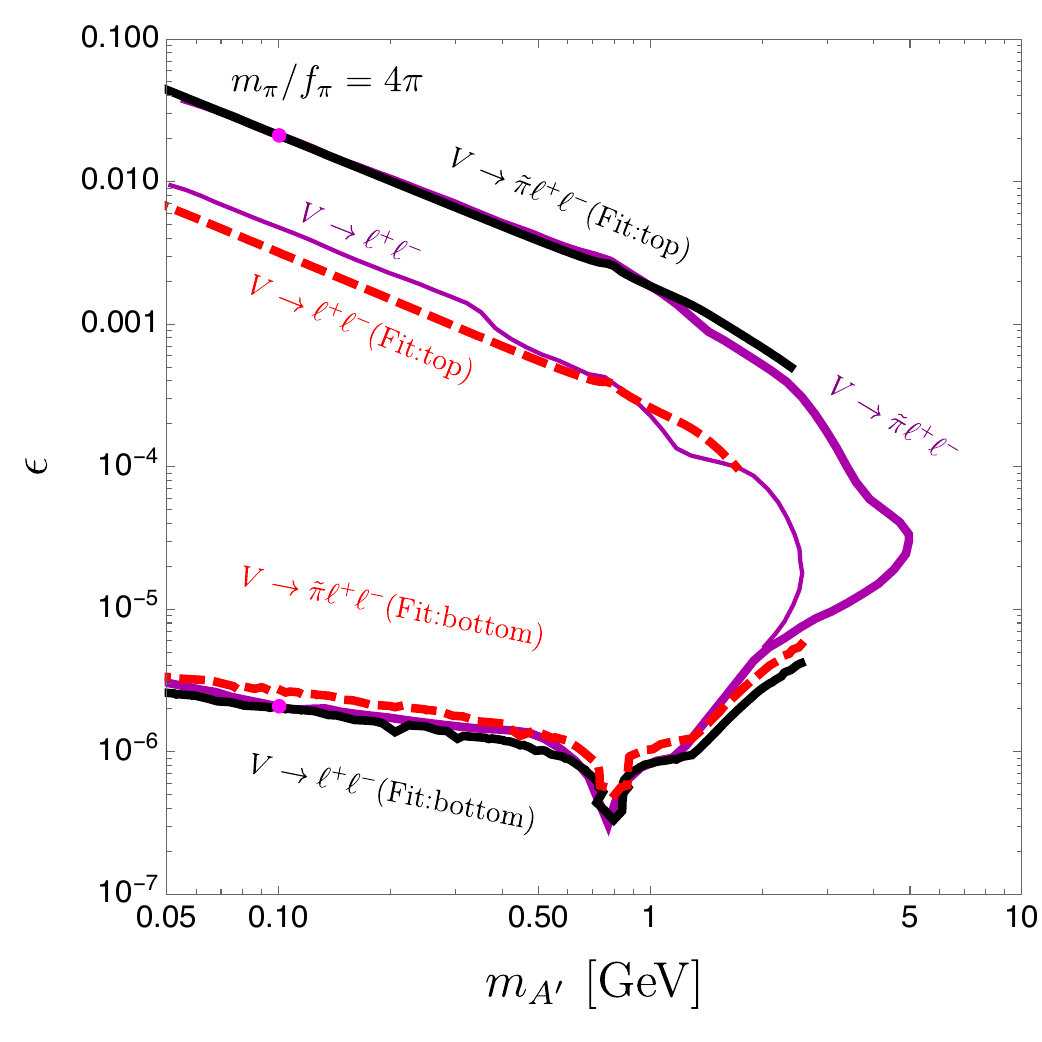}
  \caption{
    Our fitting with the results of Ref.~\cite{Berlin:2018tvf} for SeaQuest: for fixed $m_\pi/f_\pi = 3$ (left) $m_\pi/f_\pi = 4\pi$ (right). 
    The magenta lines show their results, and black and red lines show our fitting: magenta thick lines correspond to the sensitivity lines for visible decay of dark vector meson, 3-body decay ($V \to \tilde \pi \ell^+ \ell^-$) and 2-body decay ($V \to \ell^+ \ell^-$), while magenta thin lines correspond to the sensitivity lines only for 2-body decay.
    We take the dark pion mass to be $m_\pi = m_{A'}/3$ and the mass of the dark vector meson to be $m_V = 0.6 m_{A'}$.
    We take the reference points (magenta dots in the right panel) for our fitting.
  }
  \label{fig:SeaQFit}
\end{figure}

We show the validity of our fitting for the SeaQuest sensitivities on the decay of dark vector mesons in \cref{fig:SeaQFit} by applying our approximation formulae, \cref{eq:fit_epsilon_max_pion,eq:fit_epsilon_min_pion} to the SIMP model.
We use the sensitivity plots of the dark vector mesons decay discussed in Ref.~\cite{Berlin:2018tvf}.
In the literature, the dark photon is heavier than the dark hadrons. 
The on-shell $A'$ is produced through the SM meson decay, Bremsstrahlung, and Drell-Yan at SeaQuest. 
The produced dark photon mainly decays into the dark mesons, in particular $A' \to \tilde \pi \tilde \pi$ and $A' \to \tilde \pi V$ where $V$ and $\tilde \pi$ collectively denotes vector mesons and pseudo-scalar mesons, respectively. 

The $U(1)_D$-neutral dark vector mesons mix with dark photons as with the $\rho^0$-$\gamma$ mixing in the SM, and then the neutral mesons decay into the SM lepton pair, $\ell^+ \ell^-$, through the dark photon mixing.
On the other hand, the $U(1)_D$-charged dark vector mesons decay into the SM leptons through the off-shell dark photon by emitting $U(1)_D$-charged dark pions.
The decay length of the three-body decay tends to be longer than that of the two-body decay due to three-body phase space factor.

Magenta lines in \cref{fig:SeaQFit} show sensitivities of the dark vector meson $V$ decay at SeaQuest: thick lines correspond to 3-body decay ($V \to \tilde \pi \ell^+ \ell^-$) and 2-body decay ($V \to \ell^+ \ell^-$), while thin lines correspond to the sensitivity lines only for 2-body decay.
The thick sensitivity lines are composed of two parts: upper parts from the three-body decay $\Gamma(V \to \pi \ell^+ \ell)$, while lower parts from the two-body decay $\Gamma(V \to \ell^+ \ell)$.
It is assumed the ratio of the dark pion mass and the pion decay constant to be $m_\pi/f_\pi = 3$ ($= 4 \pi$) in the left (right) panel.
In both panels, the dark pion mass is $m_\pi = m_{A'}/3$ and the mass of the dark vector meson is $m_V = 0.6 m_{A'}$.
The produced numbers of $V$ depend on $m_\pi/f_\pi$, and increases as $m_\pi/f_\pi$ gets larger.
In particular, the branching fraction of $V$ is almost unity when we take $m_\pi/f_\pi = 4 \pi$. 

The reference values in \cref{eq:fit_epsilon_max_pion,eq:fit_epsilon_min_pion} are determined by the point at $m_{A'} = 0.1\,\mathrm{GeV}$ (magenta dots in the right panel).
We use the plot for $m_\pi/f_\pi = 4 \pi$ for the calibration, and hence the branching fraction to the vector meson is assumed to be unity. 
The lower boundary of the three-body decay is hidden since the boundary is overlapped with the sensitivity to the two-body decay.
Therefore, we take the reference points from the upper boundary of the three-body decay and the lower boundary of the two-body decay.
We obtain the reference values as follows.
\eqs{
  \frac{r_\mathrm{min}}{[c \tau_{V} (V \to \tilde \pi \ell^+\ell^-) p_{\mathrm{max}} m_{V}^{-1}]_0} \simeq 34.6 \,, \qquad
  \frac{r_\mathrm{max}-r_\mathrm{min}}{[c \tau_{V} (V \to \ell^+\ell^-) p_\mathrm{min} m_{V}^{-1}]_0} \simeq 0.00001 \,.
}
Here, $[c \tau_{V} (V \to \tilde \pi \ell^+\ell^-) p_{\mathrm{max}} m_{V}^{-1}]_0$ and $[c \tau_{V} (V \to \ell^+\ell^-) p_\mathrm{min} m_{V}^{-1}]_0$ correspond to the boosted decay lengths referred as to $d_0(p_\mathrm{max})$ and $d_0(p_\mathrm{min})$ in the text, respectively.
The black lines show the boundaries computed by our approximation formulae, and the boundaries well coincide with the sensitivity computed by Ref.~\cite{Berlin:2018tvf}.
Since we do not include the dark photon production through Drell-Yan, the boundaries are cut at a few GeV.
The black-dashed lines in the left panel also show the boundaries with our formulae.
Although the upper boundary of the sensitivity is not shown below 1~GeV in the literature, it is expected that our formulae well explain its extrapolation.
By use of our formulae, we also evaluate the upper-boundary of the two-body decay and lower boundary of the three-body decay, which as shown as red-dashed lines in \cref{fig:SeaQFit}.
The sensitivity of the upper boundaries is also computed by Ref.~\cite{Berlin:2018tvf}, which is shown as the thin-magenta lines, and our formula roughly reproduce their result.

\subsection{E137}
\begin{figure}
  \centering
  \includegraphics[width=0.48\textwidth]{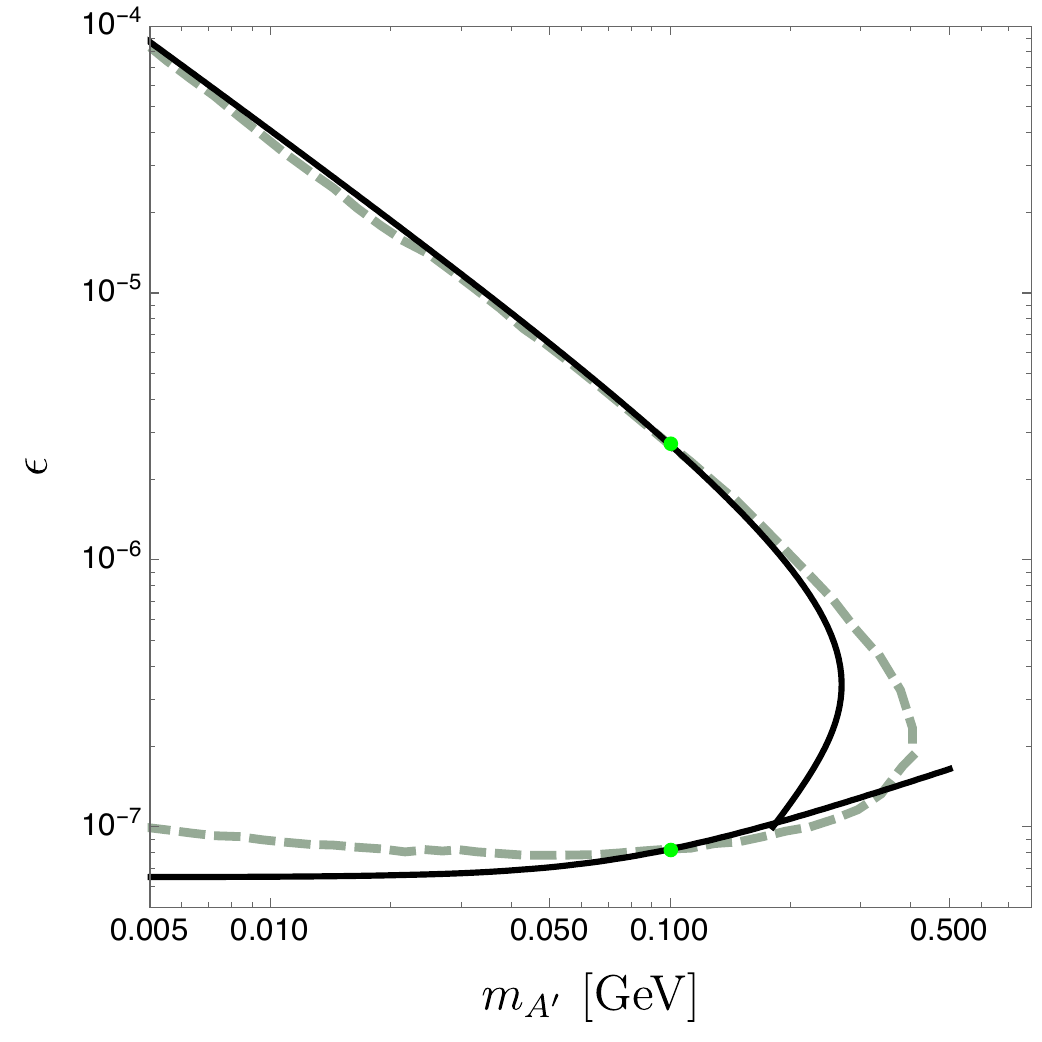}
  \caption{
  Our fitting with the results of Ref.~\cite{Bjorken:2009mm} for E137. 
  Dashed light green line shows the boundary of the existing constraint by E137 which is taken from Ref.~\cite{Bjorken:2009mm}, while the black solid lines show our fit.
  Green points on the boundaries are our reference points for the fitting.
  }
  \label{fig:E137Fit}
\end{figure}

As for the existing constraint on the dark pion decay, we use the constraint on the visible dark photon decay by E137. 
We take the reference point at $m_{A'} = 0.1 \, \mathrm{GeV}$, and read out the kinetic mixing $\epsilon$ on the upper and lower boundaries of the sensitivity area. 
\cref{fig:E137Fit} shows our fitting of dark photon visible decay with the constraint by Ref.~\cite{Bjorken:2009mm}.
The black lines show our fitting, dashed light green line shows the boundary of the existing constraint by Ref.~\cite{Bjorken:2009mm}, and the green points are our reference points. 
At these points, we get the reference values as follows.
\eqs{
  \frac{r_\mathrm{min}}{[c \tau_{A'} (A' \to \ell^+\ell^-) p_{\mathrm{max}} m_{A'}^{-1}]_0} \simeq 8.7 \,, \qquad
  \frac{r_\mathrm{max}-r_\mathrm{min}}{[c \tau_{A'} (A' \to \ell^+\ell^-) p_\mathrm{min} m_{A'}^{-1}]_0} \simeq 0.0093 \,.
}
Our fitting works well for the dark photon visible decay search at E137.

\section{Production and Geometric Efficiency \label{sec:production}}

\begin{figure}
  \centering
  \includegraphics[width=0.48\textwidth]{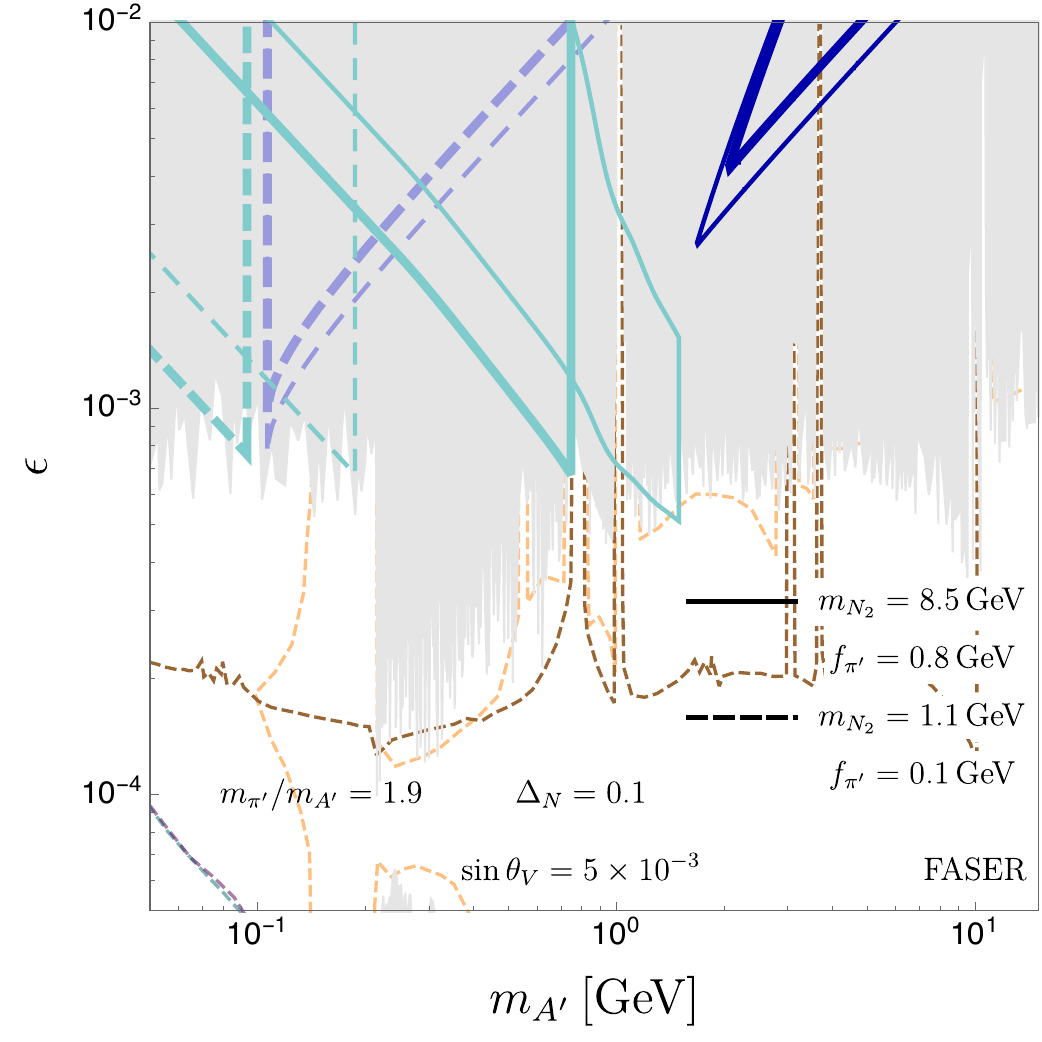}
  \includegraphics[width=0.48\textwidth]{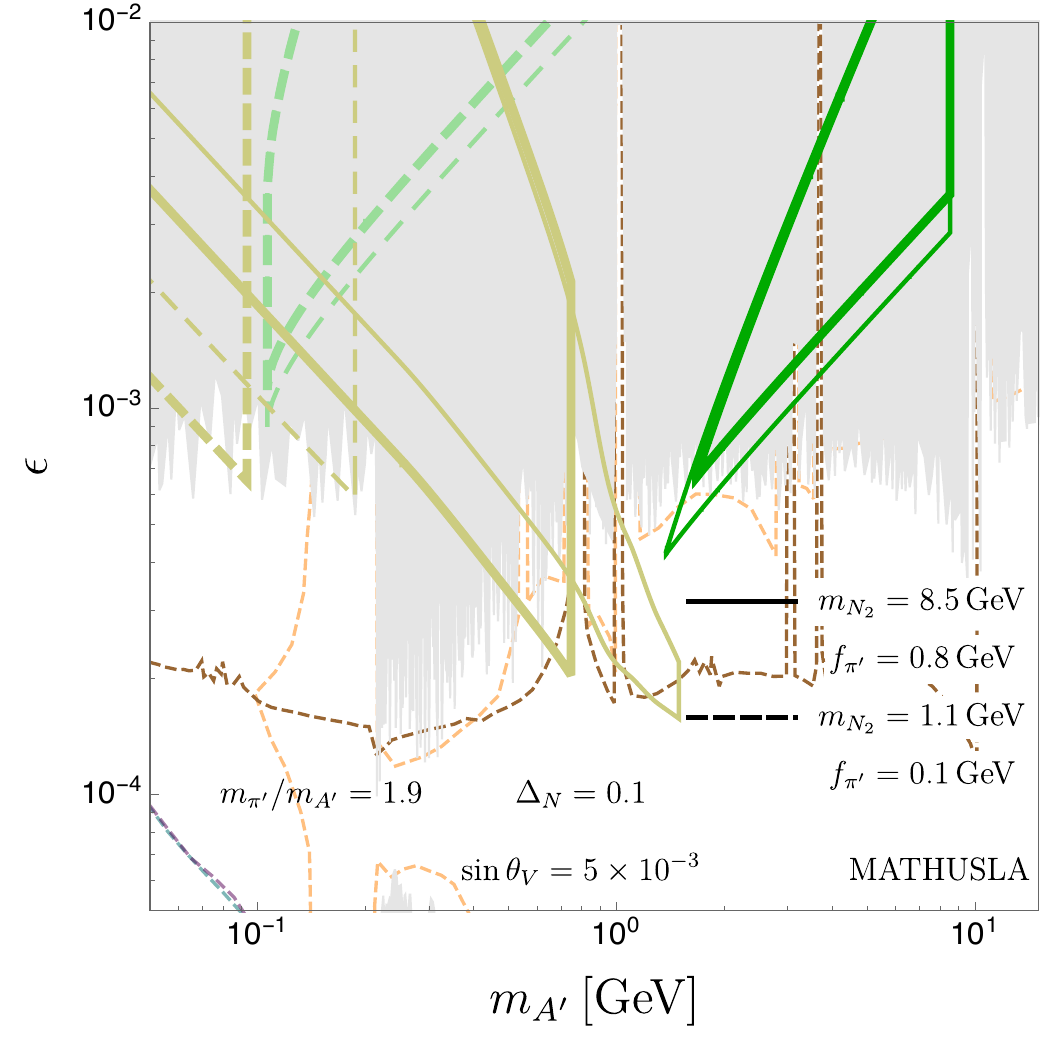}
  \caption{
    Sensitivity plots for the dark nucleon and dark pion searches at LHC lifetime frontier, including variation of multiplicity of long-lived particles: FASER (left) and MATHUSLA (right).
    The dark nucleon searches: FASER (blue solid-lines for $m_{N_2} = 8.5\,\mathrm{GeV}$, blue dashed-lines for $m_{N_2} = 1.1\,\mathrm{GeV}$) and MATHUSLA (green solid-lines for $m_{N_2} = 8.5\,\mathrm{GeV}$, green dashed-lines for $m_{N_2} = 1.1\,\mathrm{GeV}$). 
    We take different parameters for dark nucleons from \cref{fig:nucleon_epsilon-mA} in order that the FASER sensitivity lines appear even for $m_{N_2} = 8.5\,\mathrm{GeV}$: $\sin\theta_V = 5 \times 10^{-3}$ and $\Delta_N = 0.1$.
    The dark pion searches: FASER (cyan solid-lines for $f_{\pi'} = 0.8\,\mathrm{GeV}$, cyan dashed-lines for $f_{\pi'} = 0.1\,\mathrm{GeV}$) and MATHUSLA (yellow solid-lines for $f_{\pi'} = 0.8\,\mathrm{GeV}$, yellow dashed-lines for $f_{\pi'} = 0.1\,\mathrm{GeV}$).
    We take $m_{\pi'}/m_{A'} = 1.9$ for the visible decay of dark pions.
    The thick lines are the same as shown in \cref{fig:nucleon_epsilon-mA,fig:pion_epsilon-mA} with a different choice of $\sin\theta_V$ and $\Delta_N$, while the corresponding thin lines show the sensitivity curves when the multiplicities of hadrons are similar, in particular $n_{N_1} = 0.1$ and $n_{\pi'} = 0.2$.
    For the latter case, the dark pion sensitivities extend to $3 m_{\pi'} < m_{N_2}$ since the multiplicity is below unity.
    As in \cref{fig:nucleon_epsilon-mA,fig:pion_epsilon-mA}, we also show the existence constraints and the future sensitivities to the visible dark photon decay (the top and left-bottom shaded areas and the future sensitivity curves from Belle-II, LHCb, SHiP, and SeaQuest).
  }
  \label{fig:LHCProduction}
\end{figure}
We use the produced number of the long-lived particles given by \cref{eq:ProducedNumberN,eq:ProducedNumberPi_hadronization} for the visible decay search at the LHC lifetime frontier.
Here, we na\"ively assume that dark nucleons and dark pions are produced when off-shell dark photons with a fictitious mass of the dark dynamical scale are produced. 
This produced number depends on multiplicity of the produced particles, and we assume $n_p = 0.04$ and $n_{\pi'} = 2.0$ in the text, which are the analogy of the SM particle production at the $J/\psi$ threshold~\cite{DASP:1978ftr}.
Since pions are expected to be more produced than nucleons when they are much lighter than nucleons, the multiplicity of pions is larger than that of nucleons
In our study, the mass difference of dark nucleons and dark pions can be smaller than that of the SM hadrons. 
Therefore, we change the multiplicity of hadrons: in particular, to be the similar multiplicity, $n_{N_1} = 0.1$ and $n_{\pi'} = 0.2$. 
We show sensitivity plots in \cref{fig:LHCProduction} when the dark sector hadrons are produced on an almost equal footing.

In this figure, we take $\sin\theta_V = 5 \times 10^{-3}$ and $\Delta_N = 0.1$ for the visible decay of dark nucleons, while $m_{\pi'}/m_{A'} = 1.9$ for the visible decay of dark pions. 
We commonly use $U(1)_D$ coupling to be $\alpha' = 0.05$ ($\alpha' = 7 \times 10^{-3}$) for $n_{g'} = 1$ ($n_{g'} = 8$).
\cref{fig:LHCProduction} show the changes of the sensitivity plots of FASER (left panel) and MATHUSLA (right panel), and both panels include the visible decays from dark nucleons and dark pions.
The color codes are the same as the previous figures for FASER (MATHUSLA): 
the blue-solid (green-solid) lines for $m_{N_2} = 8.5\,\mathrm{GeV}$ and the cyan-solid (yellow-solid) lines for $f_{\pi'} = 0.8\,\mathrm{GeV}$, while the blue-dashed (green-dashed) lines for $m_{N_2} = 1.1\,\mathrm{GeV}$ and the cyan-dashed (yellow-dashed) lines for $f_{\pi'} = 0.1\,\mathrm{GeV}$.
Thick lines show the sensitivity curves with \cref{eq:ProducedNumberN,eq:ProducedNumberPi_hadronization}, while thin lines show the sensitivity curves with different choice of multiplicities.
In the text, the six dark pions are required to be produced through the dark hadronization due to the isospin symmetry and multiplicity. 
The pion multiplicity smaller than unity implies only three dark pions are required as the final state through the dark hadronization at one hadronic event.
Hence, the dark-pion sensitivity curves extends to $3 m_{\pi'} < m_{N_2}$ in the case of $n_{\pi'} = 0.2$.

\begin{figure}
  \centering
  \includegraphics[width=0.48\textwidth]{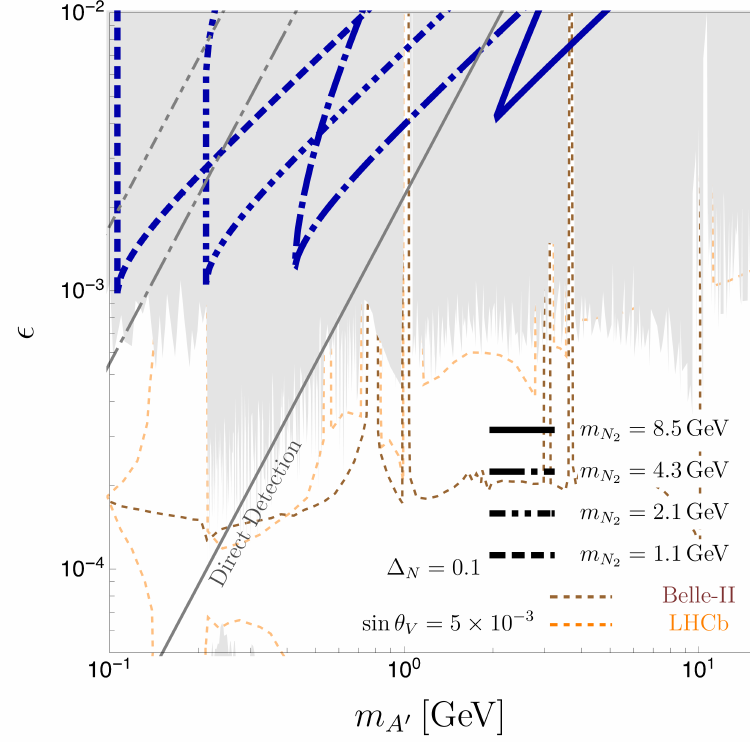}
  \includegraphics[width=0.48\textwidth]{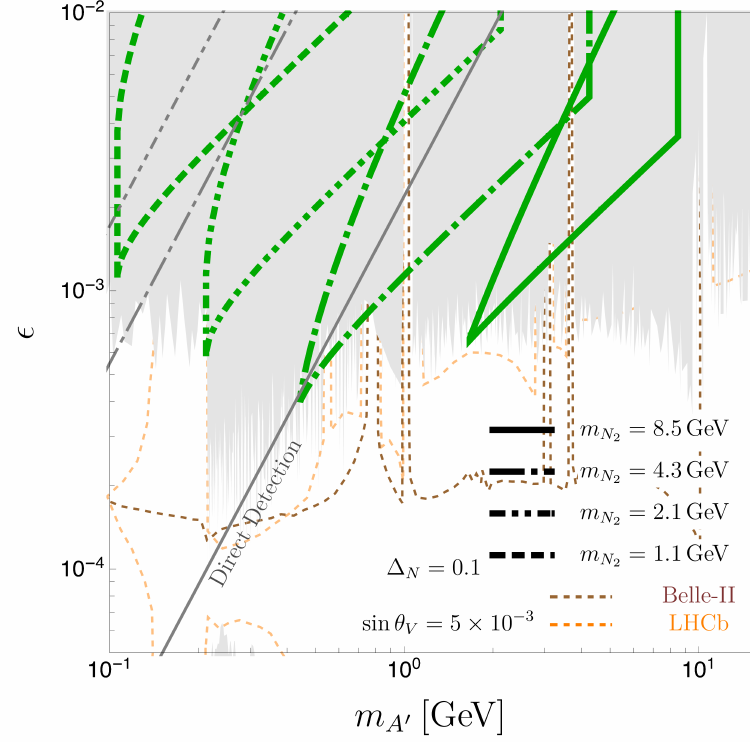}
  \includegraphics[width=0.48\textwidth]{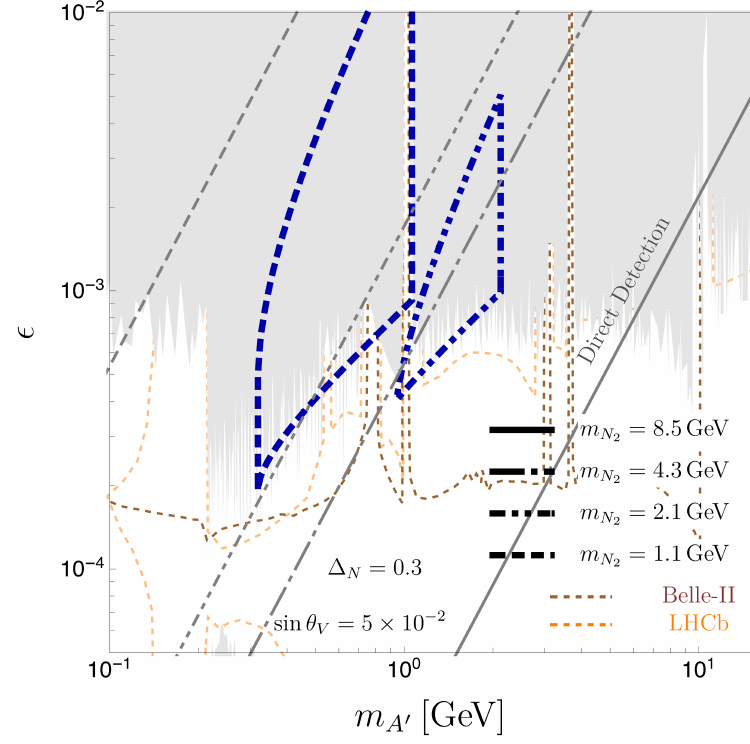}
  \includegraphics[width=0.48\textwidth]{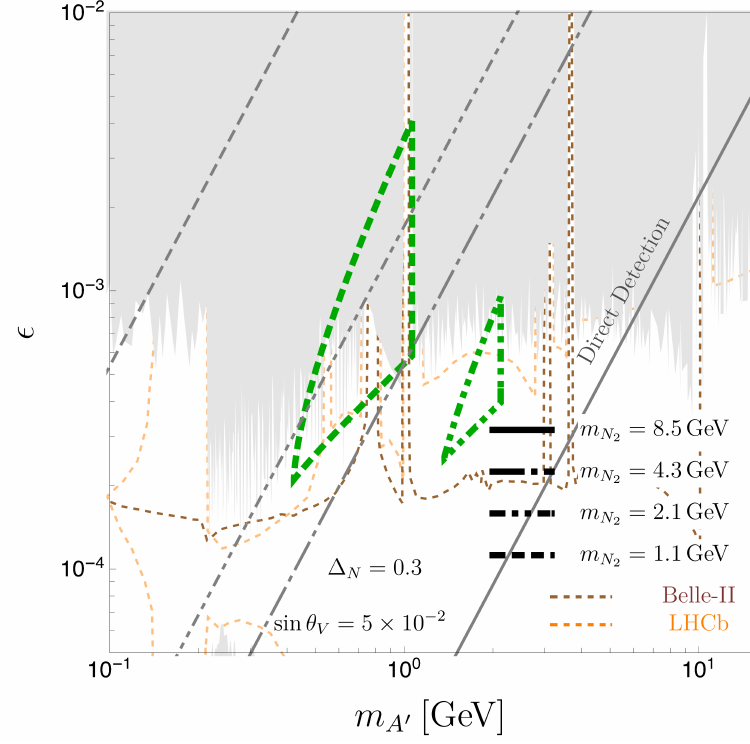}
  \caption{
    Dark nucleon searches at the LHC lifetime frontier on $\epsilon$-$m_{A'}$ plane: FASER (left columns) and MATHUSLA (right columns).
    We choose different values from the parameters taken in \cref{fig:nucleon_epsilon-mA}: we take the mass difference to be $\Delta_N = 0.1$ (top) and 0.3 (bottom), and the mixing angle to be $\sin\theta_V = 5 \times 10^{-3}$ and $\sin\theta_V = 5 \times 10^{-2}$.
    The color codes and the existing constraints are the same as in \cref{fig:nucleon_epsilon-mA}.
  }
  \label{fig:nucleon_epsilon-mA_Parameters}
\end{figure}

We show the parameter dependences of dark nucleon sensitivities at the LHC lifetime frontier in \cref{fig:nucleon_epsilon-mA_Parameters}.
Existing constraints by dark photon searches and the direct detection are the same as in \cref{fig:nucleon_epsilon-mA}. 
For the larger $\Delta_N$, the decay with on-shell dark photon easily opens, and therefore the sensitivity to the three-body decay of the nucleons gets narrow in $m_{A'}$ axis.
However, it is sensitive to the smaller $\epsilon$ since the lifetime gets shorter.
Due to the same reason as what we explained in the text, the sensitivity range gets shrink once we take tha larger couplings or the larger DM mass. 
\cref{fig:nucleon_epsilon-mA_Parameters} shows that some sensitivity curves disappear when $\sin\theta_V = 5 \times 10^{-2}$ and $\Delta_N = 0.3$.

\begin{figure}
  \centering
  \includegraphics[width=0.32\textwidth]{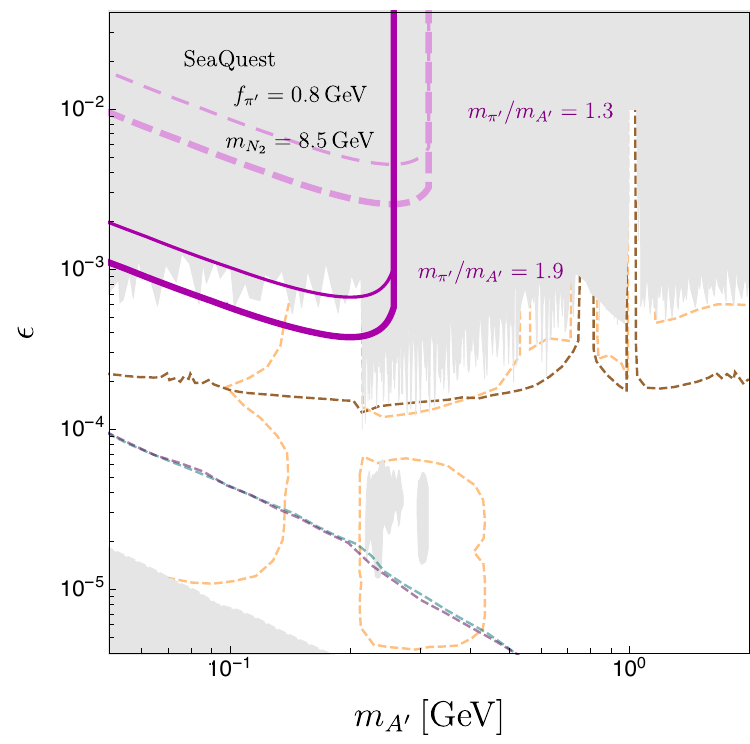}
  \includegraphics[width=0.32\textwidth]{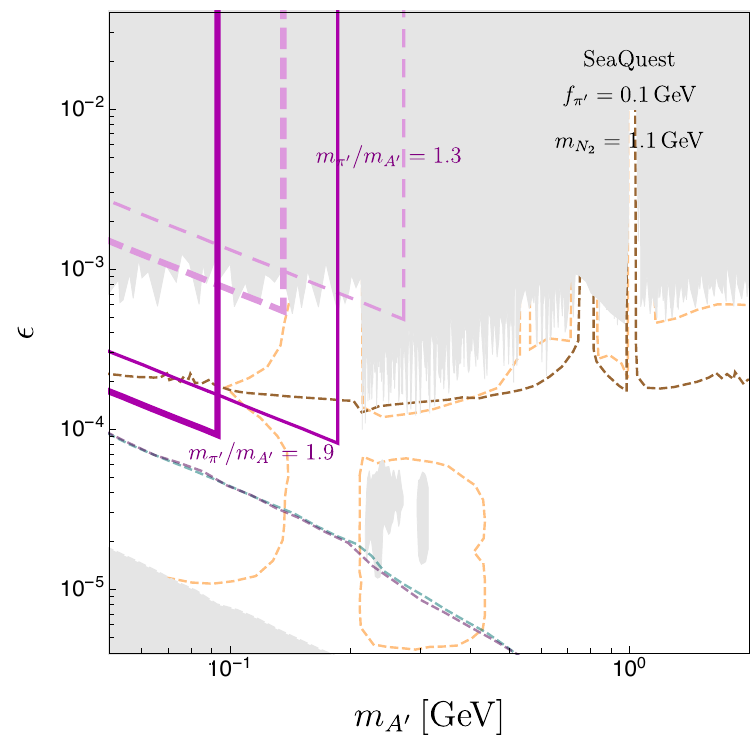}
  \includegraphics[width=0.32\textwidth]{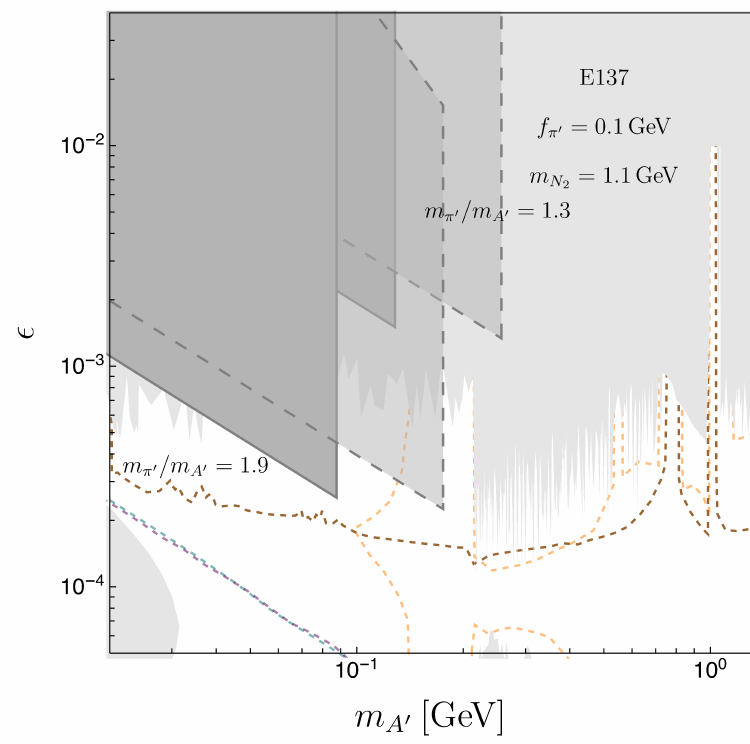}
  \caption{
    Sensitivity plots and exclusion plots for dark pion searches, including variation of productions at SeaQuest and E137.
    We take $f_{\pi'} = 0.8\,\mathrm{GeV}$ (left) and $f_{\pi'} = 0.1\,\mathrm{GeV}$ (middle) for the sensitivity of SeaQuest, while $f_{\pi'} = 0.1\,\mathrm{GeV}$ for the exclusion limit from E137 (right).
    We take the different values of $m_{\pi'}/m_{A'}$: $m_{\pi'}/m_{A'} = 1.9$ (solid boundaries) and $m_{\pi'}/m_{A'} = 1.3$ (dashed boundaries).
    For the SeaQuest plots, we take the production part to be ten times smaller than that used in the text, which is illustrated as thin lines in the figure.
    For the E137 plot, we again take the production part to be ten times smaller than that used in the text, which is illustrated as dashed-black lines in the figure.
  }
  \label{fig:BeamDumpProduction}
\end{figure}

As for the dark pions at SeaQuest, they are produced via a virtual dark photon with the fictitious mass of order of the SM $\rho$ mesons, and we use \cref{eq:ProducedNumberPi_FixedpBrems} to estimate their produced number. 
The proton Bremsstrahlung dominates their production, and the main contribution comes from the $\rho$ meson resonances due to the broad width of $\rho$ mesons.
As for the dark photon searches at the SeaQuest~\cite{Berlin:2018pwi}, as we discussed in the text, the dark photons from meson decays have less energetic and are produced with a larger angle off from beam-axis in comparison with the dark photon from proton bremsstrahlung.
Therefore, the geometric efficiency at SeaQuest decreases for the dark photons from meson decay.

Similarly to the dark photons, the searches for dark pions can be affected by the geometric efficiency.
Even though the production of dark pions is basically from proton Bremsstrahlung, the geometric efficiency for dark pions could change since $\rho$ meson dominates their production. 
In other words, the invariant mass of final states, $\sqrt{s_f}$, is close to the $\rho$ meson mass, $\sqrt{s_f} \sim m_\rho$, and hence the efficiency could be reduced in analogy to the efficiency of dark photons from meson decay.

\cref{fig:BeamDumpProduction} shows the sensitivities of SeaQuest to the dark pion searches, including variation of the geometric efficiency at SeaQuest. 
The different line-types correspond to the different values of $m_{\pi'}/m_{A'}$: $m_{\pi'}/m_{A'} = 1.9$ (solid) and $m_{\pi'}/m_{A'} = 1.3$ (dashed).
We take different decay constants, $f_{\pi'} = 0.8\,\mathrm{GeV}$ (left) and $f_{\pi'} = 0.1\,\mathrm{GeV}$ (right).
When the geometric efficiency is ten times smaller than the Bremsstrahlung, the sensitivity areas get smaller and the sensitivity lines are shown as thin lines in the figures.

\bibliographystyle{utphys}
\bibliography{ref}

\providecommand{\href}[2]{#2}\begingroup\raggedright\begin{thebibliography}{100}

\bibitem{Essig:2009nc}
R.~Essig, P.~Schuster, and N.~Toro, ``{Probing Dark Forces and Light Hidden
  Sectors at Low-Energy e+e- Colliders},''
  \href{http://dx.doi.org/10.1103/PhysRevD.80.015003}{{\em Phys. Rev. D}
  {\bfseries 80} (2009) 015003},
  \href{http://arxiv.org/abs/0903.3941}{{\ttfamily arXiv:0903.3941 [hep-ph]}}.

\bibitem{Reece:2009un}
M.~Reece and L.-T. Wang, ``{Searching for the light dark gauge boson in
  GeV-scale experiments},''
  \href{http://dx.doi.org/10.1088/1126-6708/2009/07/051}{{\em JHEP} {\bfseries
  07} (2009) 051}, \href{http://arxiv.org/abs/0904.1743}{{\ttfamily
  arXiv:0904.1743 [hep-ph]}}.

\bibitem{Aubert:2009cp}
{\bfseries BaBar} Collaboration, B.~Aubert {\em et~al.}, ``{Search for Dimuon
  Decays of a Light Scalar Boson in Radiative Transitions Upsilon $\to$ gamma
  A0},'' \href{http://dx.doi.org/10.1103/PhysRevLett.103.081803}{{\em Phys.
  Rev. Lett.} {\bfseries 103} (2009) 081803},
  \href{http://arxiv.org/abs/0905.4539}{{\ttfamily arXiv:0905.4539 [hep-ex]}}.

\bibitem{Bjorken:2009mm}
J.~D. Bjorken, R.~Essig, P.~Schuster, and N.~Toro, ``{New Fixed-Target
  Experiments to Search for Dark Gauge Forces},''
  \href{http://dx.doi.org/10.1103/PhysRevD.80.075018}{{\em Phys. Rev. D}
  {\bfseries 80} (2009) 075018},
  \href{http://arxiv.org/abs/0906.0580}{{\ttfamily arXiv:0906.0580 [hep-ph]}}.

\bibitem{Batell:2009di}
B.~Batell, M.~Pospelov, and A.~Ritz, ``{Exploring Portals to a Hidden Sector
  Through Fixed Targets},''
  \href{http://dx.doi.org/10.1103/PhysRevD.80.095024}{{\em Phys. Rev. D}
  {\bfseries 80} (2009) 095024},
  \href{http://arxiv.org/abs/0906.5614}{{\ttfamily arXiv:0906.5614 [hep-ph]}}.

\bibitem{Essig:2010gu}
R.~Essig, R.~Harnik, J.~Kaplan, and N.~Toro, ``{Discovering New Light States at
  Neutrino Experiments},''
  \href{http://dx.doi.org/10.1103/PhysRevD.82.113008}{{\em Phys. Rev. D}
  {\bfseries 82} (2010) 113008},
  \href{http://arxiv.org/abs/1008.0636}{{\ttfamily arXiv:1008.0636 [hep-ph]}}.

\bibitem{Andreas:2012mt}
S.~Andreas, C.~Niebuhr, and A.~Ringwald, ``{New Limits on Hidden Photons from
  Past Electron Beam Dumps},''
  \href{http://dx.doi.org/10.1103/PhysRevD.86.095019}{{\em Phys. Rev. D}
  {\bfseries 86} (2012) 095019},
  \href{http://arxiv.org/abs/1209.6083}{{\ttfamily arXiv:1209.6083 [hep-ph]}}.

\bibitem{Lees:2014xha}
{\bfseries BaBar} Collaboration, J.~Lees {\em et~al.}, ``{Search for a Dark
  Photon in $e^+e^-$ Collisions at BaBar},''
  \href{http://dx.doi.org/10.1103/PhysRevLett.113.201801}{{\em Phys. Rev.
  Lett.} {\bfseries 113} no.~20, (2014) 201801},
  \href{http://arxiv.org/abs/1406.2980}{{\ttfamily arXiv:1406.2980 [hep-ex]}}.

\bibitem{Anastasi:2015qla}
A.~Anastasi {\em et~al.}, ``{Limit on the production of a low-mass vector boson
  in $\mathrm{e}^{+}\mathrm{e}^{-} \to \mathrm{U}\gamma$, $\mathrm{U} \to
  \mathrm{e}^{+}\mathrm{e}^{-}$ with the KLOE experiment},''
  \href{http://dx.doi.org/10.1016/j.physletb.2015.10.003}{{\em Phys. Lett. B}
  {\bfseries 750} (2015) 633--637},
  \href{http://arxiv.org/abs/1509.00740}{{\ttfamily arXiv:1509.00740
  [hep-ex]}}.

\bibitem{Auerbach:2001wg}
{\bfseries LSND} Collaboration, L.~B. Auerbach {\em et~al.}, ``{Measurement of
  electron - neutrino - electron elastic scattering},''
  \href{http://dx.doi.org/10.1103/PhysRevD.63.112001}{{\em Phys. Rev. D}
  {\bfseries 63} (2001) 112001},
  \href{http://arxiv.org/abs/hep-ex/0101039}{{\ttfamily arXiv:hep-ex/0101039}}.

\bibitem{deNiverville:2011it}
P.~deNiverville, M.~Pospelov, and A.~Ritz, ``{Observing a light dark matter
  beam with neutrino experiments},''
  \href{http://dx.doi.org/10.1103/PhysRevD.84.075020}{{\em Phys. Rev. D}
  {\bfseries 84} (2011) 075020},
  \href{http://arxiv.org/abs/1107.4580}{{\ttfamily arXiv:1107.4580 [hep-ph]}}.

\bibitem{Bjorken:1988as}
J.~D. Bjorken, S.~Ecklund, W.~R. Nelson, A.~Abashian, C.~Church, B.~Lu, L.~W.
  Mo, T.~A. Nunamaker, and P.~Rassmann, ``{Search for Neutral Metastable
  Penetrating Particles Produced in the SLAC Beam Dump},''
  \href{http://dx.doi.org/10.1103/PhysRevD.38.3375}{{\em Phys. Rev. D}
  {\bfseries 38} (1988) 3375}.

\bibitem{Batell:2014mga}
B.~Batell, R.~Essig, and Z.~Surujon, ``{Strong Constraints on Sub-GeV Dark
  Sectors from SLAC Beam Dump E137},''
  \href{http://dx.doi.org/10.1103/PhysRevLett.113.171802}{{\em Phys. Rev.
  Lett.} {\bfseries 113} no.~17, (2014) 171802},
  \href{http://arxiv.org/abs/1406.2698}{{\ttfamily arXiv:1406.2698 [hep-ph]}}.

\bibitem{Aguilar-Arevalo:2017mqx}
{\bfseries MiniBooNE} Collaboration, A.~A. Aguilar-Arevalo {\em et~al.},
  ``{Dark Matter Search in a Proton Beam Dump with MiniBooNE},''
  \href{http://dx.doi.org/10.1103/PhysRevLett.118.221803}{{\em Phys. Rev.
  Lett.} {\bfseries 118} no.~22, (2017) 221803},
  \href{http://arxiv.org/abs/1702.02688}{{\ttfamily arXiv:1702.02688
  [hep-ex]}}.

\bibitem{Lees:2017lec}
{\bfseries BaBar} Collaboration, J.~P. Lees {\em et~al.}, ``{Search for
  Invisible Decays of a Dark Photon Produced in ${e}^{+}{e}^{-}$ Collisions at
  BaBar},'' \href{http://dx.doi.org/10.1103/PhysRevLett.119.131804}{{\em Phys.
  Rev. Lett.} {\bfseries 119} no.~13, (2017) 131804},
  \href{http://arxiv.org/abs/1702.03327}{{\ttfamily arXiv:1702.03327
  [hep-ex]}}.

\bibitem{Celentano:2014wya}
{\bfseries HPS} Collaboration, A.~Celentano, ``{The Heavy Photon Search
  experiment at Jefferson Laboratory},''
  \href{http://dx.doi.org/10.1088/1742-6596/556/1/012064}{{\em J. Phys. Conf.
  Ser.} {\bfseries 556} no.~1, (2014) 012064},
  \href{http://arxiv.org/abs/1505.02025}{{\ttfamily arXiv:1505.02025
  [physics.ins-det]}}.

\bibitem{Anelli:2015pba}
{\bfseries SHiP} Collaboration, M.~Anelli {\em et~al.}, ``{A facility to Search
  for Hidden Particles (SHiP) at the CERN SPS},''
  \href{http://arxiv.org/abs/1504.04956}{{\ttfamily arXiv:1504.04956
  [physics.ins-det]}}.

\bibitem{Alekhin:2015byh}
S.~Alekhin {\em et~al.}, ``{A facility to Search for Hidden Particles at the
  CERN SPS: the SHiP physics case},''
  \href{http://dx.doi.org/10.1088/0034-4885/79/12/124201}{{\em Rept. Prog.
  Phys.} {\bfseries 79} no.~12, (2016) 124201},
  \href{http://arxiv.org/abs/1504.04855}{{\ttfamily arXiv:1504.04855
  [hep-ph]}}.

\bibitem{Gardner:2015wea}
S.~Gardner, R.~J. Holt, and A.~S. Tadepalli, ``{New Prospects in Fixed Target
  Searches for Dark Forces with the SeaQuest Experiment at Fermilab},''
  \href{http://dx.doi.org/10.1103/PhysRevD.93.115015}{{\em Phys. Rev. D}
  {\bfseries 93} no.~11, (2016) 115015},
  \href{http://arxiv.org/abs/1509.00050}{{\ttfamily arXiv:1509.00050
  [hep-ph]}}.

\bibitem{Aidala:2017ofy}
{\bfseries SeaQuest} Collaboration, C.~A. Aidala {\em et~al.}, ``{The SeaQuest
  Spectrometer at Fermilab},''
  \href{http://dx.doi.org/10.1016/j.nima.2019.03.039}{{\em Nucl. Instrum. Meth.
  A} {\bfseries 930} (2019) 49--63},
  \href{http://arxiv.org/abs/1706.09990}{{\ttfamily arXiv:1706.09990
  [physics.ins-det]}}.

\bibitem{Berlin:2018pwi}
A.~Berlin, S.~Gori, P.~Schuster, and N.~Toro, ``{Dark Sectors at the Fermilab
  SeaQuest Experiment},''
  \href{http://dx.doi.org/10.1103/PhysRevD.98.035011}{{\em Phys. Rev. D}
  {\bfseries 98} no.~3, (2018) 035011},
  \href{http://arxiv.org/abs/1804.00661}{{\ttfamily arXiv:1804.00661
  [hep-ph]}}.

\bibitem{Izaguirre:2014bca}
E.~Izaguirre, G.~Krnjaic, P.~Schuster, and N.~Toro, ``{Testing GeV-Scale Dark
  Matter with Fixed-Target Missing Momentum Experiments},''
  \href{http://dx.doi.org/10.1103/PhysRevD.91.094026}{{\em Phys. Rev. D}
  {\bfseries 91} no.~9, (2015) 094026},
  \href{http://arxiv.org/abs/1411.1404}{{\ttfamily arXiv:1411.1404 [hep-ph]}}.

\bibitem{Akesson:2018vlm}
{\bfseries LDMX} Collaboration, T.~\r{A}kesson {\em et~al.}, ``{Light Dark
  Matter eXperiment (LDMX)},''
  \href{http://arxiv.org/abs/1808.05219}{{\ttfamily arXiv:1808.05219
  [hep-ex]}}.

\bibitem{Essig:2013vha}
R.~Essig, J.~Mardon, M.~Papucci, T.~Volansky, and Y.-M. Zhong, ``{Constraining
  Light Dark Matter with Low-Energy $e^+e^-$ Colliders},''
  \href{http://dx.doi.org/10.1007/JHEP11(2013)167}{{\em JHEP} {\bfseries 11}
  (2013) 167}, \href{http://arxiv.org/abs/1309.5084}{{\ttfamily arXiv:1309.5084
  [hep-ph]}}.

\bibitem{Kou:2018nap}
{\bfseries Belle-II} Collaboration, W.~Altmannshofer {\em et~al.}, ``{The Belle
  II Physics Book},'' \href{http://dx.doi.org/10.1093/ptep/ptz106}{{\em PTEP}
  {\bfseries 2019} no.~12, (2019) 123C01},
  \href{http://arxiv.org/abs/1808.10567}{{\ttfamily arXiv:1808.10567
  [hep-ex]}}. [Erratum: PTEP 2020, 029201 (2020)].

\bibitem{Izaguirre:2015zva}
E.~Izaguirre, G.~Krnjaic, and B.~Shuve, ``{Discovering Inelastic Thermal-Relic
  Dark Matter at Colliders},''
  \href{http://dx.doi.org/10.1103/PhysRevD.93.063523}{{\em Phys. Rev. D}
  {\bfseries 93} no.~6, (2016) 063523},
  \href{http://arxiv.org/abs/1508.03050}{{\ttfamily arXiv:1508.03050
  [hep-ph]}}.

\bibitem{Ilten:2016tkc}
P.~Ilten, Y.~Soreq, J.~Thaler, M.~Williams, and W.~Xue, ``{Proposed Inclusive
  Dark Photon Search at LHCb},''
  \href{http://dx.doi.org/10.1103/PhysRevLett.116.251803}{{\em Phys. Rev.
  Lett.} {\bfseries 116} no.~25, (2016) 251803},
  \href{http://arxiv.org/abs/1603.08926}{{\ttfamily arXiv:1603.08926
  [hep-ph]}}.

\bibitem{Liu:2018wte}
J.~Liu, Z.~Liu, and L.-T. Wang, ``{Enhancing Long-Lived Particles Searches at
  the LHC with Precision Timing Information},''
  \href{http://dx.doi.org/10.1103/PhysRevLett.122.131801}{{\em Phys. Rev.
  Lett.} {\bfseries 122} no.~13, (2019) 131801},
  \href{http://arxiv.org/abs/1805.05957}{{\ttfamily arXiv:1805.05957
  [hep-ph]}}.

\bibitem{Chou:2016lxi}
J.~P. Chou, D.~Curtin, and H.~J. Lubatti, ``{New Detectors to Explore the
  Lifetime Frontier},''
  \href{http://dx.doi.org/10.1016/j.physletb.2017.01.043}{{\em Phys. Lett. B}
  {\bfseries 767} (2017) 29--36},
  \href{http://arxiv.org/abs/1606.06298}{{\ttfamily arXiv:1606.06298
  [hep-ph]}}.

\bibitem{Feng:2017uoz}
J.~L. Feng, I.~Galon, F.~Kling, and S.~Trojanowski, ``{ForwArd Search
  ExpeRiment at the LHC},''
  \href{http://dx.doi.org/10.1103/PhysRevD.97.035001}{{\em Phys. Rev. D}
  {\bfseries 97} no.~3, (2018) 035001},
  \href{http://arxiv.org/abs/1708.09389}{{\ttfamily arXiv:1708.09389
  [hep-ph]}}.

\bibitem{Gligorov:2017nwh}
V.~V. Gligorov, S.~Knapen, M.~Papucci, and D.~J. Robinson, ``{Searching for
  Long-lived Particles: A Compact Detector for Exotics at LHCb},''
  \href{http://dx.doi.org/10.1103/PhysRevD.97.015023}{{\em Phys. Rev. D}
  {\bfseries 97} no.~1, (2018) 015023},
  \href{http://arxiv.org/abs/1708.09395}{{\ttfamily arXiv:1708.09395
  [hep-ph]}}.

\bibitem{Gudnason:2006yj}
S.~B. Gudnason, C.~Kouvaris, and F.~Sannino, ``{Dark Matter from new
  Technicolor Theories},''
  \href{http://dx.doi.org/10.1103/PhysRevD.74.095008}{{\em Phys. Rev.}
  {\bfseries D74} (2006) 095008},
\href{http://arxiv.org/abs/hep-ph/0608055}{{\ttfamily arXiv:hep-ph/0608055
  [hep-ph]}}.

\bibitem{Dietrich:2006cm}
D.~D. Dietrich and F.~Sannino, ``{Conformal window of SU(N) gauge theories with
  fermions in higher dimensional representations},''
  \href{http://dx.doi.org/10.1103/PhysRevD.75.085018}{{\em Phys. Rev.}
  {\bfseries D75} (2007) 085018},
\href{http://arxiv.org/abs/hep-ph/0611341}{{\ttfamily arXiv:hep-ph/0611341
  [hep-ph]}}.

\bibitem{Khlopov:2007ic}
M.~{\relax Yu}. Khlopov and C.~Kouvaris, ``{Strong Interactive Massive
  Particles from a Strong Coupled Theory},''
  \href{http://dx.doi.org/10.1103/PhysRevD.77.065002}{{\em Phys. Rev.}
  {\bfseries D77} (2008) 065002},
\href{http://arxiv.org/abs/0710.2189}{{\ttfamily arXiv:0710.2189 [astro-ph]}}.

\bibitem{Khlopov:2008ty}
M.~{\relax Yu}. Khlopov and C.~Kouvaris, ``{Composite dark matter from a model
  with composite Higgs boson},''
  \href{http://dx.doi.org/10.1103/PhysRevD.78.065040}{{\em Phys. Rev.}
  {\bfseries D78} (2008) 065040},
\href{http://arxiv.org/abs/0806.1191}{{\ttfamily arXiv:0806.1191 [astro-ph]}}.

\bibitem{Foadi:2008qv}
R.~Foadi, M.~T. Frandsen, and F.~Sannino, ``{Technicolor Dark Matter},''
  \href{http://dx.doi.org/10.1103/PhysRevD.80.037702}{{\em Phys. Rev.}
  {\bfseries D80} (2009) 037702},
\href{http://arxiv.org/abs/0812.3406}{{\ttfamily arXiv:0812.3406 [hep-ph]}}.

\bibitem{Mardon:2009gw}
J.~Mardon, Y.~Nomura, and J.~Thaler, ``{Cosmic Signals from the Hidden
  Sector},'' \href{http://dx.doi.org/10.1103/PhysRevD.80.035013}{{\em Phys.
  Rev.} {\bfseries D80} (2009) 035013},
\href{http://arxiv.org/abs/0905.3749}{{\ttfamily arXiv:0905.3749 [hep-ph]}}.

\bibitem{Kribs:2009fy}
G.~D. Kribs, T.~S. Roy, J.~Terning, and K.~M. Zurek, ``{Quirky Composite Dark
  Matter},'' \href{http://dx.doi.org/10.1103/PhysRevD.81.095001}{{\em Phys.
  Rev.} {\bfseries D81} (2010) 095001},
\href{http://arxiv.org/abs/0909.2034}{{\ttfamily arXiv:0909.2034 [hep-ph]}}.

\bibitem{Barbieri:2010mn}
R.~Barbieri, S.~Rychkov, and R.~Torre, ``{Signals of composite
  electroweak-neutral Dark Matter: LHC/Direct Detection interplay},''
  \href{http://dx.doi.org/10.1016/j.physletb.2010.04.010}{{\em Phys. Lett.}
  {\bfseries B688} (2010) 212--215},
\href{http://arxiv.org/abs/1001.3149}{{\ttfamily arXiv:1001.3149 [hep-ph]}}.

\bibitem{Blennow:2010qp}
M.~Blennow, B.~Dasgupta, E.~Fernandez-Martinez, and N.~Rius, ``{Aidnogenesis
  via Leptogenesis and Dark Sphalerons},''
  \href{http://dx.doi.org/10.1007/JHEP03(2011)014}{{\em JHEP} {\bfseries 03}
  (2011) 014},
\href{http://arxiv.org/abs/1009.3159}{{\ttfamily arXiv:1009.3159 [hep-ph]}}.

\bibitem{Lewis:2011zb}
R.~Lewis, C.~Pica, and F.~Sannino, ``{Light Asymmetric Dark Matter on the
  Lattice: SU(2) Technicolor with Two Fundamental Flavors},''
  \href{http://dx.doi.org/10.1103/PhysRevD.85.014504}{{\em Phys. Rev.}
  {\bfseries D85} (2012) 014504},
\href{http://arxiv.org/abs/1109.3513}{{\ttfamily arXiv:1109.3513 [hep-ph]}}.

\bibitem{Appelquist:2013ms}
{\bfseries Lattice Strong Dynamics (LSD)} Collaboration, T.~Appelquist {\em
  et~al.}, ``{Lattice calculation of composite dark matter form factors},''
  \href{http://dx.doi.org/10.1103/PhysRevD.88.014502}{{\em Phys. Rev.}
  {\bfseries D88} no.~1, (2013) 014502},
\href{http://arxiv.org/abs/1301.1693}{{\ttfamily arXiv:1301.1693 [hep-ph]}}.

\bibitem{Hietanen:2013fya}
A.~Hietanen, R.~Lewis, C.~Pica, and F.~Sannino, ``{Composite Goldstone Dark
  Matter: Experimental Predictions from the Lattice},''
  \href{http://dx.doi.org/10.1007/JHEP12(2014)130}{{\em JHEP} {\bfseries 12}
  (2014) 130},
\href{http://arxiv.org/abs/1308.4130}{{\ttfamily arXiv:1308.4130 [hep-ph]}}.

\bibitem{Cline:2013zca}
J.~M. Cline, Z.~Liu, G.~Moore, and W.~Xue, ``{Composite strongly interacting
  dark matter},'' \href{http://dx.doi.org/10.1103/PhysRevD.90.015023}{{\em
  Phys. Rev.} {\bfseries D90} no.~1, (2014) 015023},
\href{http://arxiv.org/abs/1312.3325}{{\ttfamily arXiv:1312.3325 [hep-ph]}}.

\bibitem{Appelquist:2014jch}
{\bfseries Lattice Strong Dynamics (LSD)} Collaboration, T.~Appelquist {\em
  et~al.}, ``{Composite bosonic baryon dark matter on the lattice: SU(4) baryon
  spectrum and the effective Higgs interaction},''
  \href{http://dx.doi.org/10.1103/PhysRevD.89.094508}{{\em Phys. Rev.}
  {\bfseries D89} no.~9, (2014) 094508},
\href{http://arxiv.org/abs/1402.6656}{{\ttfamily arXiv:1402.6656 [hep-lat]}}.

\bibitem{Hietanen:2014xca}
A.~Hietanen, R.~Lewis, C.~Pica, and F.~Sannino, ``{Fundamental Composite Higgs
  Dynamics on the Lattice: SU(2) with Two Flavors},''
  \href{http://dx.doi.org/10.1007/JHEP07(2014)116}{{\em JHEP} {\bfseries 07}
  (2014) 116},
\href{http://arxiv.org/abs/1404.2794}{{\ttfamily arXiv:1404.2794 [hep-lat]}}.

\bibitem{Krnjaic:2014xza}
G.~Krnjaic and K.~Sigurdson, ``{Big Bang Darkleosynthesis},''
  \href{http://dx.doi.org/10.1016/j.physletb.2015.11.001}{{\em Phys. Lett.}
  {\bfseries B751} (2015) 464--468},
\href{http://arxiv.org/abs/1406.1171}{{\ttfamily arXiv:1406.1171 [hep-ph]}}.

\bibitem{Detmold:2014qqa}
W.~Detmold, M.~McCullough, and A.~Pochinsky, ``{Dark Nuclei I: Cosmology and
  Indirect Detection},''
  \href{http://dx.doi.org/10.1103/PhysRevD.90.115013}{{\em Phys. Rev.}
  {\bfseries D90} no.~11, (2014) 115013},
\href{http://arxiv.org/abs/1406.2276}{{\ttfamily arXiv:1406.2276 [hep-ph]}}.

\bibitem{Detmold:2014kba}
W.~Detmold, M.~McCullough, and A.~Pochinsky, ``{Dark nuclei. II. Nuclear
  spectroscopy in two-color QCD},''
  \href{http://dx.doi.org/10.1103/PhysRevD.90.114506}{{\em Phys. Rev.}
  {\bfseries D90} no.~11, (2014) 114506},
\href{http://arxiv.org/abs/1406.4116}{{\ttfamily arXiv:1406.4116 [hep-lat]}}.

\bibitem{Asano:2014wra}
M.~Asano and R.~Kitano, ``{Partially Composite Dark Matter},''
  \href{http://dx.doi.org/10.1007/JHEP09(2014)171}{{\em JHEP} {\bfseries 09}
  (2014) 171},
\href{http://arxiv.org/abs/1406.6374}{{\ttfamily arXiv:1406.6374 [hep-ph]}}.

\bibitem{Brod:2014loa}
J.~Brod, J.~Drobnak, A.~L. Kagan, E.~Stamou, and J.~Zupan, ``{Stealth QCD-like
  strong interactions and the $t \bar {t}$ asymmetry},''
  \href{http://dx.doi.org/10.1103/PhysRevD.91.095009}{{\em Phys. Rev.}
  {\bfseries D91} no.~9, (2015) 095009},
\href{http://arxiv.org/abs/1407.8188}{{\ttfamily arXiv:1407.8188 [hep-ph]}}.

\bibitem{Antipin:2014qva}
O.~Antipin, M.~Redi, and A.~Strumia, ``{Dynamical generation of the weak and
  Dark Matter scales from strong interactions},''
  \href{http://dx.doi.org/10.1007/JHEP01(2015)157}{{\em JHEP} {\bfseries 01}
  (2015) 157},
\href{http://arxiv.org/abs/1410.1817}{{\ttfamily arXiv:1410.1817 [hep-ph]}}.

\bibitem{Hardy:2014mqa}
E.~Hardy, R.~Lasenby, J.~March-Russell, and S.~M. West, ``{Big Bang Synthesis
  of Nuclear Dark Matter},''
  \href{http://dx.doi.org/10.1007/JHEP06(2015)011}{{\em JHEP} {\bfseries 06}
  (2015) 011},
\href{http://arxiv.org/abs/1411.3739}{{\ttfamily arXiv:1411.3739 [hep-ph]}}.

\bibitem{Appelquist:2015yfa}
T.~Appelquist {\em et~al.}, ``{Stealth Dark Matter: Dark scalar baryons through
  the Higgs portal},'' \href{http://dx.doi.org/10.1103/PhysRevD.92.075030}{{\em
  Phys. Rev.} {\bfseries D92} no.~7, (2015) 075030},
\href{http://arxiv.org/abs/1503.04203}{{\ttfamily arXiv:1503.04203 [hep-ph]}}.

\bibitem{Appelquist:2015zfa}
T.~Appelquist {\em et~al.}, ``{Detecting Stealth Dark Matter Directly through
  Electromagnetic Polarizability},''
  \href{http://dx.doi.org/10.1103/PhysRevLett.115.171803}{{\em Phys. Rev.
  Lett.} {\bfseries 115} no.~17, (2015) 171803},
\href{http://arxiv.org/abs/1503.04205}{{\ttfamily arXiv:1503.04205 [hep-ph]}}.

\bibitem{Antipin:2015xia}
O.~Antipin, M.~Redi, A.~Strumia, and E.~Vigiani, ``{Accidental Composite Dark
  Matter},'' \href{http://dx.doi.org/10.1007/JHEP07(2015)039}{{\em JHEP}
  {\bfseries 07} (2015) 039},
\href{http://arxiv.org/abs/1503.08749}{{\ttfamily arXiv:1503.08749 [hep-ph]}}.

\bibitem{Hardy:2015boa}
E.~Hardy, R.~Lasenby, J.~March-Russell, and S.~M. West, ``{Signatures of Large
  Composite Dark Matter States},''
  \href{http://dx.doi.org/10.1007/JHEP07(2015)133}{{\em JHEP} {\bfseries 07}
  (2015) 133},
\href{http://arxiv.org/abs/1504.05419}{{\ttfamily arXiv:1504.05419 [hep-ph]}}.

\bibitem{Co:2016akw}
R.~T. Co, K.~Harigaya, and Y.~Nomura, ``{Chiral Dark Sector},''
  \href{http://dx.doi.org/10.1103/PhysRevLett.118.101801}{{\em Phys. Rev.
  Lett.} {\bfseries 118} no.~10, (2017) 101801},
\href{http://arxiv.org/abs/1610.03848}{{\ttfamily arXiv:1610.03848 [hep-ph]}}.

\bibitem{Dienes:2016vei}
K.~R. Dienes, F.~Huang, S.~Su, and B.~Thomas, ``{Dynamical Dark Matter from
  Strongly-Coupled Dark Sectors},''
  \href{http://dx.doi.org/10.1103/PhysRevD.95.043526}{{\em Phys. Rev.}
  {\bfseries D95} no.~4, (2017) 043526},
\href{http://arxiv.org/abs/1610.04112}{{\ttfamily arXiv:1610.04112 [hep-ph]}}.

\bibitem{Ishida:2016fbp}
H.~Ishida, S.~Matsuzaki, and Y.~Yamaguchi, ``{Bosonic-Seesaw Portal Dark
  Matter},'' \href{http://dx.doi.org/10.1093/ptep/ptx132}{{\em PTEP} {\bfseries
  2017} no.~10, (2017) 103B01},
\href{http://arxiv.org/abs/1610.07137}{{\ttfamily arXiv:1610.07137 [hep-ph]}}.

\bibitem{Lonsdale:2017mzg}
S.~J. Lonsdale, M.~Schroor, and R.~R. Volkas, ``{Asymmetric Dark Matter and the
  hadronic spectra of hidden QCD},''
  \href{http://dx.doi.org/10.1103/PhysRevD.96.055027}{{\em Phys. Rev.}
  {\bfseries D96} no.~5, (2017) 055027},
\href{http://arxiv.org/abs/1704.05213}{{\ttfamily arXiv:1704.05213 [hep-ph]}}.

\bibitem{Berryman:2017twh}
J.~M. Berryman, A.~de~Gouv{\^e}a, K.~J. Kelly, and Y.~Zhang, ``{Dark Matter and
  Neutrino Mass from the Smallest Non-Abelian Chiral Dark Sector},''
  \href{http://dx.doi.org/10.1103/PhysRevD.96.075010}{{\em Phys. Rev.}
  {\bfseries D96} no.~7, (2017) 075010},
\href{http://arxiv.org/abs/1706.02722}{{\ttfamily arXiv:1706.02722 [hep-ph]}}.

\bibitem{Gresham:2017zqi}
M.~I. Gresham, H.~K. Lou, and K.~M. Zurek, ``{Nuclear Structure of Bound States
  of Asymmetric Dark Matter},''
  \href{http://dx.doi.org/10.1103/PhysRevD.96.096012}{{\em Phys. Rev.}
  {\bfseries D96} no.~9, (2017) 096012},
\href{http://arxiv.org/abs/1707.02313}{{\ttfamily arXiv:1707.02313 [hep-ph]}}.

\bibitem{Gresham:2017cvl}
M.~I. Gresham, H.~K. Lou, and K.~M. Zurek, ``{Early Universe synthesis of
  asymmetric dark matter nuggets},''
  \href{http://dx.doi.org/10.1103/PhysRevD.97.036003}{{\em Phys. Rev.}
  {\bfseries D97} no.~3, (2018) 036003},
\href{http://arxiv.org/abs/1707.02316}{{\ttfamily arXiv:1707.02316 [hep-ph]}}.

\bibitem{Mitridate:2017oky}
A.~Mitridate, M.~Redi, J.~Smirnov, and A.~Strumia, ``{Dark Matter as a weakly
  coupled Dark Baryon},'' \href{http://dx.doi.org/10.1007/JHEP10(2017)210}{{\em
  JHEP} {\bfseries 10} (2017) 210},
\href{http://arxiv.org/abs/1707.05380}{{\ttfamily arXiv:1707.05380 [hep-ph]}}.

\bibitem{Gresham:2018anj}
M.~I. Gresham, H.~K. Lou, and K.~M. Zurek, ``{Astrophysical Signatures of
  Asymmetric Dark Matter Bound States},''
  \href{http://dx.doi.org/10.1103/PhysRevD.98.096001}{{\em Phys. Rev.}
  {\bfseries D98} no.~9, (2018) 096001},
\href{http://arxiv.org/abs/1805.04512}{{\ttfamily arXiv:1805.04512 [hep-ph]}}.

\bibitem{Ibe:2018juk}
M.~Ibe, A.~Kamada, S.~Kobayashi, and W.~Nakano, ``{Composite Asymmetric Dark
  Matter with a Dark Photon Portal},''
  \href{http://dx.doi.org/10.1007/JHEP11(2018)203}{{\em JHEP} {\bfseries 11}
  (2018) 203}, \href{http://arxiv.org/abs/1805.06876}{{\ttfamily
  arXiv:1805.06876 [hep-ph]}}.

\bibitem{Braaten:2018xuw}
E.~Braaten, D.~Kang, and R.~Laha, ``{Production of dark-matter bound states in
  the early universe by three-body recombination},''
  \href{http://dx.doi.org/10.1007/JHEP11(2018)084}{{\em JHEP} {\bfseries 11}
  (2018) 084},
\href{http://arxiv.org/abs/1806.00609}{{\ttfamily arXiv:1806.00609 [hep-ph]}}.

\bibitem{Francis:2018xjd}
A.~Francis, R.~J. Hudspith, R.~Lewis, and S.~Tulin, ``{Dark Matter from Strong
  Dynamics: The Minimal Theory of Dark Baryons},''
\href{http://arxiv.org/abs/1809.09117}{{\ttfamily arXiv:1809.09117 [hep-ph]}}.

\bibitem{Bai:2018dxf}
Y.~Bai, A.~J. Long, and S.~Lu, ``{Dark Quark Nuggets},''
  \href{http://dx.doi.org/10.1103/PhysRevD.99.055047}{{\em Phys. Rev. D}
  {\bfseries 99} no.~5, (2019) 055047},
  \href{http://arxiv.org/abs/1810.04360}{{\ttfamily arXiv:1810.04360
  [hep-ph]}}.

\bibitem{Chu:2018faw}
X.~Chu, C.~Garcia-Cely, and H.~Murayama, ``{Finite-size dark matter and its
  effect on small-scale structure},''
  \href{http://dx.doi.org/10.1103/PhysRevLett.124.041101}{{\em Phys. Rev.
  Lett.} {\bfseries 124} no.~4, (2020) 041101},
  \href{http://arxiv.org/abs/1901.00075}{{\ttfamily arXiv:1901.00075
  [hep-ph]}}.

\bibitem{Hall:2019rld}
E.~Hall, T.~Konstandin, R.~McGehee, and H.~Murayama, ``{Asymmetric Matters from
  a Dark First-Order Phase Transition},''
  \href{http://arxiv.org/abs/1911.12342}{{\ttfamily arXiv:1911.12342
  [hep-ph]}}.

\bibitem{Tsai:2020vpi}
Y.-D. Tsai, R.~McGehee, and H.~Murayama, ``{Resonant Self-Interacting Dark
  Matter from Dark QCD},'' \href{http://arxiv.org/abs/2008.08608}{{\ttfamily
  arXiv:2008.08608 [hep-ph]}}.

\bibitem{Asadi:2021yml}
P.~Asadi, E.~D. Kramer, E.~Kuflik, G.~W. Ridgway, T.~R. Slatyer, and
  J.~Smirnov, ``{Accidentally Asymmetric Dark Matter},''
  \href{http://dx.doi.org/10.1103/PhysRevLett.127.211101}{{\em Phys. Rev.
  Lett.} {\bfseries 127} no.~21, (2021) 211101},
  \href{http://arxiv.org/abs/2103.09822}{{\ttfamily arXiv:2103.09822
  [hep-ph]}}.

\bibitem{Zhang:2021orr}
M.~Zhang, ``{Leptophilic composite asymmetric dark matter and its detection},''
  \href{http://dx.doi.org/10.1103/PhysRevD.104.055008}{{\em Phys. Rev. D}
  {\bfseries 104} no.~5, (2021) 055008},
  \href{http://arxiv.org/abs/2104.06988}{{\ttfamily arXiv:2104.06988
  [hep-ph]}}.

\bibitem{Bottaro:2021aal}
S.~Bottaro, M.~Costa, and O.~Popov, ``{Asymmetric accidental composite dark
  matter},'' \href{http://dx.doi.org/10.1007/JHEP11(2021)055}{{\em JHEP}
  {\bfseries 11} (2021) 055}, \href{http://arxiv.org/abs/2104.14244}{{\ttfamily
  arXiv:2104.14244 [hep-ph]}}.

\bibitem{Hall:2021zsk}
E.~Hall, R.~McGehee, H.~Murayama, and B.~Suter, ``{Asymmetric Dark Matter May
  Not Be Light},'' \href{http://arxiv.org/abs/2107.03398}{{\ttfamily
  arXiv:2107.03398 [hep-ph]}}.

\bibitem{Kribs:2016cew}
G.~D. Kribs and E.~T. Neil, ``{Review of strongly-coupled composite dark matter
  models and lattice simulations},''
  \href{http://dx.doi.org/10.1142/S0217751X16430041}{{\em Int. J. Mod. Phys.}
  {\bfseries A31} no.~22, (2016) 1643004},
\href{http://arxiv.org/abs/1604.04627}{{\ttfamily arXiv:1604.04627 [hep-ph]}}.

\bibitem{Holdom:1985ag}
B.~Holdom, ``{Two U(1)'s and Epsilon Charge Shifts},''
  \href{http://dx.doi.org/10.1016/0370-2693(86)91377-8}{{\em Phys. Lett. B}
  {\bfseries 166} (1986) 196--198}.

\bibitem{Archilli:2011zc}
{\bfseries KLOE-2} Collaboration, F.~Archilli {\em et~al.}, ``{Search for a
  vector gauge boson in $\phi$ meson decays with the KLOE detector},''
  \href{http://dx.doi.org/10.1016/j.physletb.2011.11.033}{{\em Phys. Lett. B}
  {\bfseries 706} (2012) 251--255},
  \href{http://arxiv.org/abs/1110.0411}{{\ttfamily arXiv:1110.0411 [hep-ex]}}.

\bibitem{Babusci:2012cr}
{\bfseries KLOE-2} Collaboration, D.~Babusci {\em et~al.}, ``{Limit on the
  production of a light vector gauge boson in phi meson decays with the KLOE
  detector},'' \href{http://dx.doi.org/10.1016/j.physletb.2013.01.067}{{\em
  Phys. Lett. B} {\bfseries 720} (2013) 111--115},
  \href{http://arxiv.org/abs/1210.3927}{{\ttfamily arXiv:1210.3927 [hep-ex]}}.

\bibitem{Anastasi:2016ktq}
{\bfseries KLOE-2} Collaboration, A.~Anastasi {\em et~al.}, ``{Limit on the
  production of a new vector boson in $\mathrm{e^+ e^-}\rightarrow {\rm
  U}\gamma$, U$\rightarrow \pi^+\pi^-$ with the KLOE experiment},''
  \href{http://dx.doi.org/10.1016/j.physletb.2016.04.019}{{\em Phys. Lett. B}
  {\bfseries 757} (2016) 356--361},
  \href{http://arxiv.org/abs/1603.06086}{{\ttfamily arXiv:1603.06086
  [hep-ex]}}.

\bibitem{Bauer:2018onh}
M.~Bauer, P.~Foldenauer, and J.~Jaeckel, ``{Hunting All the Hidden Photons},''
  \href{http://dx.doi.org/10.1007/JHEP07(2018)094}{{\em JHEP} {\bfseries 07}
  (2018) 094}, \href{http://arxiv.org/abs/1803.05466}{{\ttfamily
  arXiv:1803.05466 [hep-ph]}}.

\bibitem{Ibe:2018tex}
M.~Ibe, A.~Kamada, S.~Kobayashi, T.~Kuwahara, and W.~Nakano, ``{Ultraviolet
  Completion of a Composite Asymmetric Dark Matter Model with a Dark Photon
  Portal},'' \href{http://dx.doi.org/10.1007/JHEP03(2019)173}{{\em JHEP}
  {\bfseries 03} (2019) 173}, \href{http://arxiv.org/abs/1811.10232}{{\ttfamily
  arXiv:1811.10232 [hep-ph]}}.

\bibitem{Fukuda:2014xqa}
H.~Fukuda, S.~Matsumoto, and S.~Mukhopadhyay, ``{Asymmetric dark matter in
  early Universe chemical equilibrium always leads to an antineutrino
  signal},'' \href{http://dx.doi.org/10.1103/PhysRevD.92.013008}{{\em Phys.
  Rev. D} {\bfseries 92} no.~1, (2015) 013008},
  \href{http://arxiv.org/abs/1411.4014}{{\ttfamily arXiv:1411.4014 [hep-ph]}}.

\bibitem{Desai:2004pq}
{\bfseries Super-Kamiokande} Collaboration, S.~Desai {\em et~al.}, ``{Search
  for dark matter WIMPs using upward through-going muons in
  Super-Kamiokande},'' \href{http://dx.doi.org/10.1103/PhysRevD.70.083523}{{\em
  Phys. Rev. D} {\bfseries 70} (2004) 083523},
  \href{http://arxiv.org/abs/hep-ex/0404025}{{\ttfamily arXiv:hep-ex/0404025}}.
  [Erratum: Phys.Rev.D 70, 109901 (2004)].

\bibitem{Covi:2009xn}
L.~Covi, M.~Grefe, A.~Ibarra, and D.~Tran, ``{Neutrino Signals from Dark Matter
  Decay},'' \href{http://dx.doi.org/10.1088/1475-7516/2010/04/017}{{\em JCAP}
  {\bfseries 04} (2010) 017}, \href{http://arxiv.org/abs/0912.3521}{{\ttfamily
  arXiv:0912.3521 [hep-ph]}}.

\bibitem{Ibe:2019yra}
M.~Ibe, S.~Kobayashi, R.~Nagai, and W.~Nakano, ``{Oscillating Composite
  Asymmetric Dark Matter},''
  \href{http://dx.doi.org/10.1007/JHEP01(2020)027}{{\em JHEP} {\bfseries 01}
  (2020) 027}, \href{http://arxiv.org/abs/1907.11464}{{\ttfamily
  arXiv:1907.11464 [hep-ph]}}.

\bibitem{Ilten:2018crw}
P.~Ilten, Y.~Soreq, M.~Williams, and W.~Xue, ``{Serendipity in dark photon
  searches},'' \href{http://dx.doi.org/10.1007/JHEP06(2018)004}{{\em JHEP}
  {\bfseries 06} (2018) 004}, \href{http://arxiv.org/abs/1801.04847}{{\ttfamily
  arXiv:1801.04847 [hep-ph]}}.

\bibitem{Zyla:2020zbs}
{\bfseries Particle Data Group} Collaboration, P.~Zyla {\em et~al.}, ``{Review
  of Particle Physics},'' \href{http://dx.doi.org/10.1093/ptep/ptaa104}{{\em
  PTEP} {\bfseries 2020} no.~8, (2020) 083C01}.

\bibitem{Manohar:1983md}
A.~Manohar and H.~Georgi, ``{Chiral Quarks and the Nonrelativistic Quark
  Model},'' \href{http://dx.doi.org/10.1016/0550-3213(84)90231-1}{{\em Nucl.
  Phys. B} {\bfseries 234} (1984) 189--212}.

\bibitem{Georgi:1992dw}
H.~Georgi, ``{Generalized dimensional analysis},''
  \href{http://dx.doi.org/10.1016/0370-2693(93)91728-6}{{\em Phys. Lett. B}
  {\bfseries 298} (1993) 187--189},
  \href{http://arxiv.org/abs/hep-ph/9207278}{{\ttfamily arXiv:hep-ph/9207278}}.

\bibitem{tHooft:1973alw}
G.~'t~Hooft, ``{A Planar Diagram Theory for Strong Interactions},''
  \href{http://dx.doi.org/10.1016/0550-3213(74)90154-0}{{\em Nucl. Phys. B}
  {\bfseries 72} (1974) 461}.

\bibitem{tHooft:1974pnl}
G.~'t~Hooft, ``{A Two-Dimensional Model for Mesons},''
  \href{http://dx.doi.org/10.1016/0550-3213(74)90088-1}{{\em Nucl. Phys. B}
  {\bfseries 75} (1974) 461--470}.

\bibitem{Witten:1979vv}
E.~Witten, ``{Current Algebra Theorems for the U(1) Goldstone Boson},''
  \href{http://dx.doi.org/10.1016/0550-3213(79)90031-2}{{\em Nucl. Phys. B}
  {\bfseries 156} (1979) 269--283}.

\bibitem{Witten:1979kh}
E.~Witten, ``{Baryons in the 1/n Expansion},''
  \href{http://dx.doi.org/10.1016/0550-3213(79)90232-3}{{\em Nucl. Phys. B}
  {\bfseries 160} (1979) 57--115}.

\bibitem{Coleman:1980mx}
S.~R. Coleman and E.~Witten, ``{Chiral Symmetry Breakdown in Large N
  Chromodynamics},'' \href{http://dx.doi.org/10.1103/PhysRevLett.45.100}{{\em
  Phys. Rev. Lett.} {\bfseries 45} (1980) 100}.

\bibitem{Witten:1980sp}
E.~Witten, ``{Large N Chiral Dynamics},''
  \href{http://dx.doi.org/10.1016/0003-4916(80)90325-5}{{\em Annals Phys.}
  {\bfseries 128} (1980) 363}.

\bibitem{Kobayashi:1970ji}
M.~Kobayashi and T.~Maskawa, ``{Chiral symmetry and eta-x mixing},''
  \href{http://dx.doi.org/10.1143/PTP.44.1422}{{\em Prog. Theor. Phys.}
  {\bfseries 44} (1970) 1422--1424}.

\bibitem{tHooft:1976rip}
G.~'t~Hooft, ``{Symmetry Breaking Through Bell-Jackiw Anomalies},''
  \href{http://dx.doi.org/10.1103/PhysRevLett.37.8}{{\em Phys. Rev. Lett.}
  {\bfseries 37} (1976) 8--11}.

\bibitem{Gasser:1982ap}
J.~Gasser and H.~Leutwyler, ``{Quark Masses},''
  \href{http://dx.doi.org/10.1016/0370-1573(82)90035-7}{{\em Phys. Rept.}
  {\bfseries 87} (1982) 77--169}.

\bibitem{Donoghue:1996zn}
J.~F. Donoghue and A.~F. Perez, ``{The Electromagnetic mass differences of
  pions and kaons},'' \href{http://dx.doi.org/10.1103/PhysRevD.55.7075}{{\em
  Phys. Rev. D} {\bfseries 55} (1997) 7075--7092},
  \href{http://arxiv.org/abs/hep-ph/9611331}{{\ttfamily arXiv:hep-ph/9611331}}.

\bibitem{Colangelo:2010et}
G.~Colangelo {\em et~al.}, ``{Review of lattice results concerning low energy
  particle physics},''
  \href{http://dx.doi.org/10.1140/epjc/s10052-011-1695-1}{{\em Eur. Phys. J. C}
  {\bfseries 71} (2011) 1695}, \href{http://arxiv.org/abs/1011.4408}{{\ttfamily
  arXiv:1011.4408 [hep-lat]}}.

\bibitem{Ibe:2011hq}
M.~Ibe, S.~Matsumoto, and T.~T. Yanagida, ``{The GeV-scale dark matter with
  B--L asymmetry},''
  \href{http://dx.doi.org/10.1016/j.physletb.2012.01.032}{{\em Phys. Lett. B}
  {\bfseries 708} (2012) 112--118},
  \href{http://arxiv.org/abs/1110.5452}{{\ttfamily arXiv:1110.5452 [hep-ph]}}.

\bibitem{Ibe:2019ena}
M.~Ibe, A.~Kamada, S.~Kobayashi, T.~Kuwahara, and W.~Nakano, ``{Baryon-Dark
  Matter Coincidence in Mirrored Unification},''
  \href{http://dx.doi.org/10.1103/PhysRevD.100.075022}{{\em Phys. Rev. D}
  {\bfseries 100} no.~7, (2019) 075022},
  \href{http://arxiv.org/abs/1907.03404}{{\ttfamily arXiv:1907.03404
  [hep-ph]}}.

\bibitem{Harigaya:2016rwr}
K.~Harigaya and Y.~Nomura, ``{Light Chiral Dark Sector},''
  \href{http://dx.doi.org/10.1103/PhysRevD.94.035013}{{\em Phys. Rev. D}
  {\bfseries 94} no.~3, (2016) 035013},
  \href{http://arxiv.org/abs/1603.03430}{{\ttfamily arXiv:1603.03430
  [hep-ph]}}.

\bibitem{Ibe:2021gil}
M.~Ibe, S.~Kobayashi, and K.~Watanabe, ``{Chiral Composite Asymmetric Dark
  Matter},'' \href{http://arxiv.org/abs/2105.07642}{{\ttfamily arXiv:2105.07642
  [hep-ph]}}.

\bibitem{Izaguirre:2013uxa}
E.~Izaguirre, G.~Krnjaic, P.~Schuster, and N.~Toro, ``{New Electron Beam-Dump
  Experiments to Search for MeV to few-GeV Dark Matter},''
  \href{http://dx.doi.org/10.1103/PhysRevD.88.114015}{{\em Phys. Rev. D}
  {\bfseries 88} (2013) 114015},
  \href{http://arxiv.org/abs/1307.6554}{{\ttfamily arXiv:1307.6554 [hep-ph]}}.

\bibitem{Banerjee:2016tad}
{\bfseries NA64} Collaboration, D.~Banerjee {\em et~al.}, ``{Search for
  invisible decays of sub-GeV dark photons in missing-energy events at the CERN
  SPS},'' \href{http://dx.doi.org/10.1103/PhysRevLett.118.011802}{{\em Phys.
  Rev. Lett.} {\bfseries 118} no.~1, (2017) 011802},
  \href{http://arxiv.org/abs/1610.02988}{{\ttfamily arXiv:1610.02988
  [hep-ex]}}.

\bibitem{Banerjee:2017hhz}
{\bfseries NA64} Collaboration, D.~Banerjee {\em et~al.}, ``{Search for vector
  mediator of Dark Matter production in invisible decay mode},''
  \href{http://dx.doi.org/10.1103/PhysRevD.97.072002}{{\em Phys. Rev. D}
  {\bfseries 97} no.~7, (2018) 072002},
  \href{http://arxiv.org/abs/1710.00971}{{\ttfamily arXiv:1710.00971
  [hep-ex]}}.

\bibitem{Poulin:2016anj}
V.~Poulin, J.~Lesgourgues, and P.~D. Serpico, ``{Cosmological constraints on
  exotic injection of electromagnetic energy},''
  \href{http://dx.doi.org/10.1088/1475-7516/2017/03/043}{{\em JCAP} {\bfseries
  03} (2017) 043}, \href{http://arxiv.org/abs/1610.10051}{{\ttfamily
  arXiv:1610.10051 [astro-ph.CO]}}.

\bibitem{Katz:2020ywn}
A.~Katz, E.~Salvioni, and B.~Shakya, ``{Split SIMPs with Decays},''
  \href{http://dx.doi.org/10.1007/JHEP10(2020)049}{{\em JHEP} {\bfseries 10}
  (2020) 049}, \href{http://arxiv.org/abs/2006.15148}{{\ttfamily
  arXiv:2006.15148 [hep-ph]}}.

\bibitem{Ren:2018gyx}
{\bfseries PandaX-II} Collaboration, X.~Ren {\em et~al.}, ``{Constraining Dark
  Matter Models with a Light Mediator at the PandaX-II Experiment},''
  \href{http://dx.doi.org/10.1103/PhysRevLett.121.021304}{{\em Phys. Rev.
  Lett.} {\bfseries 121} no.~2, (2018) 021304},
  \href{http://arxiv.org/abs/1802.06912}{{\ttfamily arXiv:1802.06912
  [hep-ph]}}.

\bibitem{PandaX-II:2021lap}
{\bfseries PandaX-II} Collaboration, J.~Yang {\em et~al.}, ``{Constraining
  self-interacting dark matter with the full dataset of PandaX-II},''
  \href{http://dx.doi.org/10.1007/s11433-021-1740-2}{{\em Sci. China Phys.
  Mech. Astron.} {\bfseries 64} no.~11, (2021) 111062},
  \href{http://arxiv.org/abs/2104.14724}{{\ttfamily arXiv:2104.14724
  [hep-ex]}}.

\bibitem{XENON:2019gfn}
{\bfseries XENON} Collaboration, E.~Aprile {\em et~al.}, ``{Light Dark Matter
  Search with Ionization Signals in XENON1T},''
  \href{http://dx.doi.org/10.1103/PhysRevLett.123.251801}{{\em Phys. Rev.
  Lett.} {\bfseries 123} no.~25, (2019) 251801},
  \href{http://arxiv.org/abs/1907.11485}{{\ttfamily arXiv:1907.11485
  [hep-ex]}}.

\bibitem{LUX:2016ggv}
{\bfseries LUX} Collaboration, D.~S. Akerib {\em et~al.}, ``{Results from a
  search for dark matter in the complete LUX exposure},''
  \href{http://dx.doi.org/10.1103/PhysRevLett.118.021303}{{\em Phys. Rev.
  Lett.} {\bfseries 118} no.~2, (2017) 021303},
  \href{http://arxiv.org/abs/1608.07648}{{\ttfamily arXiv:1608.07648
  [astro-ph.CO]}}.

\bibitem{XENON:2018voc}
{\bfseries XENON} Collaboration, E.~Aprile {\em et~al.}, ``{Dark Matter Search
  Results from a One Ton-Year Exposure of XENON1T},''
  \href{http://dx.doi.org/10.1103/PhysRevLett.121.111302}{{\em Phys. Rev.
  Lett.} {\bfseries 121} no.~11, (2018) 111302},
  \href{http://arxiv.org/abs/1805.12562}{{\ttfamily arXiv:1805.12562
  [astro-ph.CO]}}.

\bibitem{DarkSide:2018bpj}
{\bfseries DarkSide} Collaboration, P.~Agnes {\em et~al.}, ``{Low-Mass Dark
  Matter Search with the DarkSide-50 Experiment},''
  \href{http://dx.doi.org/10.1103/PhysRevLett.121.081307}{{\em Phys. Rev.
  Lett.} {\bfseries 121} no.~8, (2018) 081307},
  \href{http://arxiv.org/abs/1802.06994}{{\ttfamily arXiv:1802.06994
  [astro-ph.HE]}}.

\bibitem{CRESST:2019jnq}
{\bfseries CRESST} Collaboration, A.~H. Abdelhameed {\em et~al.}, ``{First
  results from the CRESST-III low-mass dark matter program},''
  \href{http://dx.doi.org/10.1103/PhysRevD.100.102002}{{\em Phys. Rev. D}
  {\bfseries 100} no.~10, (2019) 102002},
  \href{http://arxiv.org/abs/1904.00498}{{\ttfamily arXiv:1904.00498
  [astro-ph.CO]}}.

\bibitem{Migdal}
A.~Migdal, ``Ionization of atoms accompanying $\alpha$- and $\beta$-decay,''
  {\em J.Phys.(USSR)} {\bfseries 4} (1941) 449.

\bibitem{Baur:1983}
G.~Baur, F.~Rosel, and D.~Trautmann, ``Ionisation induced by neutrons,''
  \href{http://dx.doi.org/10.1088/0022-3700/16/14/006}{{\em Journal of Physics
  B: Atomic and Molecular Physics} {\bfseries 16} no.~14, (Jul, 1983)
  L419--L423}. \url{https://doi.org/10.1088/0022-3700/16/14/006}.

\bibitem{Ibe:2017yqa}
M.~Ibe, W.~Nakano, Y.~Shoji, and K.~Suzuki, ``{Migdal Effect in Dark Matter
  Direct Detection Experiments},''
  \href{http://dx.doi.org/10.1007/JHEP03(2018)194}{{\em JHEP} {\bfseries 03}
  (2018) 194}, \href{http://arxiv.org/abs/1707.07258}{{\ttfamily
  arXiv:1707.07258 [hep-ph]}}.

\bibitem{XENON:2019zpr}
{\bfseries XENON} Collaboration, E.~Aprile {\em et~al.}, ``{Search for Light
  Dark Matter Interactions Enhanced by the Migdal Effect or Bremsstrahlung in
  XENON1T},'' \href{http://dx.doi.org/10.1103/PhysRevLett.123.241803}{{\em
  Phys. Rev. Lett.} {\bfseries 123} no.~24, (2019) 241803},
  \href{http://arxiv.org/abs/1907.12771}{{\ttfamily arXiv:1907.12771
  [hep-ex]}}.

\bibitem{Kamada:2020buc}
A.~Kamada, H.~J. Kim, and T.~Kuwahara, ``{Maximally self-interacting dark
  matter: models and predictions},''
  \href{http://dx.doi.org/10.1007/JHEP12(2020)202}{{\em JHEP} {\bfseries 12}
  (2020) 202}, \href{http://arxiv.org/abs/2007.15522}{{\ttfamily
  arXiv:2007.15522 [hep-ph]}}.

\bibitem{Ilten:2015hya}
P.~Ilten, J.~Thaler, M.~Williams, and W.~Xue, ``{Dark photons from charm mesons
  at LHCb},'' \href{http://dx.doi.org/10.1103/PhysRevD.92.115017}{{\em Phys.
  Rev. D} {\bfseries 92} no.~11, (2015) 115017},
  \href{http://arxiv.org/abs/1509.06765}{{\ttfamily arXiv:1509.06765
  [hep-ph]}}.

\bibitem{Aaij:2017rft}
{\bfseries LHCb} Collaboration, R.~Aaij {\em et~al.}, ``{Search for Dark
  Photons Produced in 13 TeV $pp$ Collisions},''
  \href{http://dx.doi.org/10.1103/PhysRevLett.120.061801}{{\em Phys. Rev.
  Lett.} {\bfseries 120} no.~6, (2018) 061801},
  \href{http://arxiv.org/abs/1710.02867}{{\ttfamily arXiv:1710.02867
  [hep-ex]}}.

\bibitem{Rrapaj:2015wgs}
E.~Rrapaj and S.~Reddy, ``{Nucleon-nucleon bremsstrahlung of dark gauge bosons
  and revised supernova constraints},''
  \href{http://dx.doi.org/10.1103/PhysRevC.94.045805}{{\em Phys. Rev. C}
  {\bfseries 94} no.~4, (2016) 045805},
  \href{http://arxiv.org/abs/1511.09136}{{\ttfamily arXiv:1511.09136
  [nucl-th]}}.

\bibitem{Chang:2016ntp}
J.~H. Chang, R.~Essig, and S.~D. McDermott, ``{Revisiting Supernova 1987A
  Constraints on Dark Photons},''
  \href{http://dx.doi.org/10.1007/JHEP01(2017)107}{{\em JHEP} {\bfseries 01}
  (2017) 107}, \href{http://arxiv.org/abs/1611.03864}{{\ttfamily
  arXiv:1611.03864 [hep-ph]}}.

\bibitem{Hardy:2016kme}
E.~Hardy and R.~Lasenby, ``{Stellar cooling bounds on new light particles:
  plasma mixing effects},''
  \href{http://dx.doi.org/10.1007/JHEP02(2017)033}{{\em JHEP} {\bfseries 02}
  (2017) 033}, \href{http://arxiv.org/abs/1611.05852}{{\ttfamily
  arXiv:1611.05852 [hep-ph]}}.

\bibitem{Mahoney:2017jqk}
C.~Mahoney, A.~K. Leibovich, and A.~R. Zentner, ``{Updated Constraints on
  Self-Interacting Dark Matter from Supernova 1987A},''
  \href{http://dx.doi.org/10.1103/PhysRevD.96.043018}{{\em Phys. Rev. D}
  {\bfseries 96} no.~4, (2017) 043018},
  \href{http://arxiv.org/abs/1706.08871}{{\ttfamily arXiv:1706.08871
  [hep-ph]}}.

\bibitem{Chang:2018rso}
J.~H. Chang, R.~Essig, and S.~D. McDermott, ``{Supernova 1987A Constraints on
  Sub-GeV Dark Sectors, Millicharged Particles, the QCD Axion, and an
  Axion-like Particle},'' \href{http://dx.doi.org/10.1007/JHEP09(2018)051}{{\em
  JHEP} {\bfseries 09} (2018) 051},
  \href{http://arxiv.org/abs/1803.00993}{{\ttfamily arXiv:1803.00993
  [hep-ph]}}.

\bibitem{Riordan:1987aw}
E.~M. Riordan {\em et~al.}, ``{A Search for Short Lived Axions in an Electron
  Beam Dump Experiment},''
  \href{http://dx.doi.org/10.1103/PhysRevLett.59.755}{{\em Phys. Rev. Lett.}
  {\bfseries 59} (1987) 755}.

\bibitem{Bross:1989mp}
A.~Bross, M.~Crisler, S.~H. Pordes, J.~Volk, S.~Errede, and J.~Wrbanek, ``{A
  Search for Shortlived Particles Produced in an Electron Beam Dump},''
  \href{http://dx.doi.org/10.1103/PhysRevLett.67.2942}{{\em Phys. Rev. Lett.}
  {\bfseries 67} (1991) 2942--2945}.

\bibitem{Davier:1989wz}
M.~Davier and H.~Nguyen~Ngoc, ``{An Unambiguous Search for a Light Higgs
  Boson},'' \href{http://dx.doi.org/10.1016/0370-2693(89)90174-3}{{\em Phys.
  Lett. B} {\bfseries 229} (1989) 150--155}.

\bibitem{Konaka:1986cb}
A.~Konaka {\em et~al.}, ``{Search for Neutral Particles in Electron Beam Dump
  Experiment},'' \href{http://dx.doi.org/10.1103/PhysRevLett.57.659}{{\em Phys.
  Rev. Lett.} {\bfseries 57} (1986) 659}.

\bibitem{Merkel:2014avp}
H.~Merkel {\em et~al.}, ``{Search at the Mainz Microtron for Light Massive
  Gauge Bosons Relevant for the Muon g-2 Anomaly},''
  \href{http://dx.doi.org/10.1103/PhysRevLett.112.221802}{{\em Phys. Rev.
  Lett.} {\bfseries 112} no.~22, (2014) 221802},
  \href{http://arxiv.org/abs/1404.5502}{{\ttfamily arXiv:1404.5502 [hep-ex]}}.

\bibitem{Essig:2010xa}
R.~Essig, P.~Schuster, N.~Toro, and B.~Wojtsekhowski, ``{An Electron Fixed
  Target Experiment to Search for a New Vector Boson A' Decaying to e+e-},''
  \href{http://dx.doi.org/10.1007/JHEP02(2011)009}{{\em JHEP} {\bfseries 02}
  (2011) 009}, \href{http://arxiv.org/abs/1001.2557}{{\ttfamily arXiv:1001.2557
  [hep-ph]}}.

\bibitem{APEX:2011dww}
{\bfseries APEX} Collaboration, S.~Abrahamyan {\em et~al.}, ``{Search for a New
  Gauge Boson in Electron-Nucleus Fixed-Target Scattering by the APEX
  Experiment},'' \href{http://dx.doi.org/10.1103/PhysRevLett.107.191804}{{\em
  Phys. Rev. Lett.} {\bfseries 107} (2011) 191804},
  \href{http://arxiv.org/abs/1108.2750}{{\ttfamily arXiv:1108.2750 [hep-ex]}}.

\bibitem{Battaglieri:2014hga}
M.~Battaglieri {\em et~al.}, ``{The Heavy Photon Search Test Detector},''
  \href{http://dx.doi.org/10.1016/j.nima.2014.12.017}{{\em Nucl. Instrum. Meth.
  A} {\bfseries 777} (2015) 91--101},
  \href{http://arxiv.org/abs/1406.6115}{{\ttfamily arXiv:1406.6115
  [physics.ins-det]}}.

\bibitem{HPS:2018xkw}
{\bfseries HPS} Collaboration, P.~H. Adrian {\em et~al.}, ``{Search for a dark
  photon in electroproduced $e^{+}e^{-}$ pairs with the Heavy Photon Search
  experiment at JLab},''
  \href{http://dx.doi.org/10.1103/PhysRevD.98.091101}{{\em Phys. Rev. D}
  {\bfseries 98} no.~9, (2018) 091101},
  \href{http://arxiv.org/abs/1807.11530}{{\ttfamily arXiv:1807.11530
  [hep-ex]}}.

\bibitem{Bergsma:1985is}
{\bfseries CHARM} Collaboration, F.~Bergsma {\em et~al.}, ``{A Search for
  Decays of Heavy Neutrinos in the Mass Range 0.5-{GeV} to 2.8-{GeV}},''
  \href{http://dx.doi.org/10.1016/0370-2693(86)91601-1}{{\em Phys. Lett. B}
  {\bfseries 166} (1986) 473--478}.

\bibitem{Gninenko:2012eq}
S.~N. Gninenko, ``{Constraints on sub-GeV hidden sector gauge bosons from a
  search for heavy neutrino decays},''
  \href{http://dx.doi.org/10.1016/j.physletb.2012.06.002}{{\em Phys. Lett. B}
  {\bfseries 713} (2012) 244--248},
  \href{http://arxiv.org/abs/1204.3583}{{\ttfamily arXiv:1204.3583 [hep-ph]}}.

\bibitem{Athanassopoulos:1997er}
{\bfseries LSND} Collaboration, C.~Athanassopoulos {\em et~al.}, ``{Evidence
  for muon-neutrino $\to$ electron-neutrino oscillations from pion decay in
  flight neutrinos},'' \href{http://dx.doi.org/10.1103/PhysRevC.58.2489}{{\em
  Phys. Rev. C} {\bfseries 58} (1998) 2489--2511},
  \href{http://arxiv.org/abs/nucl-ex/9706006}{{\ttfamily
  arXiv:nucl-ex/9706006}}.

\bibitem{Blumlein:2011mv}
J.~Blumlein and J.~Brunner, ``{New Exclusion Limits for Dark Gauge Forces from
  Beam-Dump Data},''
  \href{http://dx.doi.org/10.1016/j.physletb.2011.05.046}{{\em Phys. Lett. B}
  {\bfseries 701} (2011) 155--159},
  \href{http://arxiv.org/abs/1104.2747}{{\ttfamily arXiv:1104.2747 [hep-ex]}}.

\bibitem{Blumlein:2013cua}
J.~Bl\"umlein and J.~Brunner, ``{New Exclusion Limits on Dark Gauge Forces from
  Proton Bremsstrahlung in Beam-Dump Data},''
  \href{http://dx.doi.org/10.1016/j.physletb.2014.02.029}{{\em Phys. Lett. B}
  {\bfseries 731} (2014) 320--326},
  \href{http://arxiv.org/abs/1311.3870}{{\ttfamily arXiv:1311.3870 [hep-ph]}}.

\bibitem{Berlin:2018jbm}
A.~Berlin and F.~Kling, ``{Inelastic Dark Matter at the LHC Lifetime Frontier:
  ATLAS, CMS, LHCb, CODEX-b, FASER, and MATHUSLA},''
  \href{http://dx.doi.org/10.1103/PhysRevD.99.015021}{{\em Phys. Rev. D}
  {\bfseries 99} no.~1, (2019) 015021},
  \href{http://arxiv.org/abs/1810.01879}{{\ttfamily arXiv:1810.01879
  [hep-ph]}}.

\bibitem{deNiverville:2016rqh}
P.~deNiverville, C.-Y. Chen, M.~Pospelov, and A.~Ritz, ``{Light dark matter in
  neutrino beams: production modelling and scattering signatures at MiniBooNE,
  T2K and SHiP},'' \href{http://dx.doi.org/10.1103/PhysRevD.95.035006}{{\em
  Phys. Rev. D} {\bfseries 95} no.~3, (2017) 035006},
  \href{http://arxiv.org/abs/1609.01770}{{\ttfamily arXiv:1609.01770
  [hep-ph]}}.

\bibitem{Foroughi-Abari:2021zbm}
S.~Foroughi-Abari and A.~Ritz, ``{Dark Sector Production via Proton
  Bremsstrahlung},'' \href{http://arxiv.org/abs/2108.05900}{{\ttfamily
  arXiv:2108.05900 [hep-ph]}}.

\bibitem{Alpigiani:2018fgd}
{\bfseries MATHUSLA} Collaboration, C.~Alpigiani {\em et~al.}, ``{A Letter of
  Intent for MATHUSLA: A Dedicated Displaced Vertex Detector above ATLAS or
  CMS.},'' \href{http://arxiv.org/abs/1811.00927}{{\ttfamily arXiv:1811.00927
  [physics.ins-det]}}.

\bibitem{Lubatti:2019vkf}
{\bfseries MATHUSLA} Collaboration, H.~Lubatti {\em et~al.}, ``{Explore the
  lifetime frontier with MATHUSLA},''
  \href{http://dx.doi.org/10.1088/1748-0221/15/06/C06026}{{\em JINST}
  {\bfseries 15} no.~06, (2020) C06026},
  \href{http://arxiv.org/abs/1901.04040}{{\ttfamily arXiv:1901.04040
  [hep-ex]}}.

\bibitem{Ariga:2018zuc}
{\bfseries FASER} Collaboration, A.~Ariga {\em et~al.}, ``{Letter of Intent for
  FASER: ForwArd Search ExpeRiment at the LHC},''
  \href{http://arxiv.org/abs/1811.10243}{{\ttfamily arXiv:1811.10243
  [physics.ins-det]}}.

\bibitem{Ariga:2018pin}
{\bfseries FASER} Collaboration, A.~Ariga {\em et~al.}, ``{Technical Proposal
  for FASER: ForwArd Search ExpeRiment at the LHC},''
  \href{http://arxiv.org/abs/1812.09139}{{\ttfamily arXiv:1812.09139
  [physics.ins-det]}}.

\bibitem{Ariga:2018uku}
{\bfseries FASER} Collaboration, A.~Ariga {\em et~al.},
  ``{FASER\textquoteright{}s physics reach for long-lived particles},''
  \href{http://dx.doi.org/10.1103/PhysRevD.99.095011}{{\em Phys. Rev. D}
  {\bfseries 99} no.~9, (2019) 095011},
  \href{http://arxiv.org/abs/1811.12522}{{\ttfamily arXiv:1811.12522
  [hep-ph]}}.

\bibitem{Aielli:2019ivi}
G.~Aielli {\em et~al.}, ``{Expression of interest for the CODEX-b detector},''
  \href{http://dx.doi.org/10.1140/epjc/s10052-020-08711-3}{{\em Eur. Phys. J.
  C} {\bfseries 80} no.~12, (2020) 1177},
  \href{http://arxiv.org/abs/1911.00481}{{\ttfamily arXiv:1911.00481
  [hep-ex]}}.

\bibitem{DASP:1978ftr}
{\bfseries DASP} Collaboration, R.~Brandelik {\em et~al.}, ``{Charged Pion,
  Kaon and Nucleon Production by e+ e- Annihilation for C.M. Energies Between
  3.6-GeV and 5.2-GeV},''
  \href{http://dx.doi.org/10.1016/0550-3213(79)90134-2}{{\em Nucl. Phys. B}
  {\bfseries 148} (1979) 189--227}.

\bibitem{Basso:2015lua}
E.~Basso, C.~Bourrely, R.~Pasechnik, and J.~Soffer, ``{The
  Drell\textendash{}Yan process as a testing ground for parton distributions up
  to LHC},'' \href{http://dx.doi.org/10.1016/j.nuclphysa.2016.02.001}{{\em
  Nucl. Phys. A} {\bfseries 948} (2016) 63--77},
  \href{http://arxiv.org/abs/1509.07988}{{\ttfamily arXiv:1509.07988
  [hep-ph]}}.

\bibitem{Curtin:2018mvb}
D.~Curtin {\em et~al.}, ``{Long-Lived Particles at the Energy Frontier: The
  MATHUSLA Physics Case},''
  \href{http://dx.doi.org/10.1088/1361-6633/ab28d6}{{\em Rept. Prog. Phys.}
  {\bfseries 82} no.~11, (2019) 116201},
  \href{http://arxiv.org/abs/1806.07396}{{\ttfamily arXiv:1806.07396
  [hep-ph]}}.

\bibitem{Wess:1971yu}
J.~Wess and B.~Zumino, ``{Consequences of anomalous Ward identities},''
  \href{http://dx.doi.org/10.1016/0370-2693(71)90582-X}{{\em Phys. Lett. B}
  {\bfseries 37} (1971) 95--97}.

\bibitem{Witten:1983tw}
E.~Witten, ``{Global Aspects of Current Algebra},''
  \href{http://dx.doi.org/10.1016/0550-3213(83)90063-9}{{\em Nucl. Phys. B}
  {\bfseries 223} (1983) 422--432}.

\bibitem{Achasov:2002ud}
M.~N. Achasov {\em et~al.}, ``{Study of the process e+ e- ---\ensuremath{>} pi+
  pi- pi0 in the energy region s**(1/2) from 0.98-GeV to 1.38-GeV},''
  \href{http://dx.doi.org/10.1103/PhysRevD.66.032001}{{\em Phys. Rev. D}
  {\bfseries 66} (2002) 032001},
  \href{http://arxiv.org/abs/hep-ex/0201040}{{\ttfamily arXiv:hep-ex/0201040}}.

\bibitem{Achasov:2003ir}
M.~N. Achasov {\em et~al.}, ``{Study of the process e+ e- ---\ensuremath{>} pi+
  pi- pi0 in the energy region s**(1/2) below 0.98-GeV},''
  \href{http://dx.doi.org/10.1103/PhysRevD.68.052006}{{\em Phys. Rev. D}
  {\bfseries 68} (2003) 052006},
  \href{http://arxiv.org/abs/hep-ex/0305049}{{\ttfamily arXiv:hep-ex/0305049}}.

\bibitem{Hochberg:2014dra}
Y.~Hochberg, E.~Kuflik, T.~Volansky, and J.~G. Wacker, ``{Mechanism for Thermal
  Relic Dark Matter of Strongly Interacting Massive Particles},''
  \href{http://dx.doi.org/10.1103/PhysRevLett.113.171301}{{\em Phys. Rev.
  Lett.} {\bfseries 113} (2014) 171301},
  \href{http://arxiv.org/abs/1402.5143}{{\ttfamily arXiv:1402.5143 [hep-ph]}}.

\bibitem{Hochberg:2014kqa}
Y.~Hochberg, E.~Kuflik, H.~Murayama, T.~Volansky, and J.~G. Wacker, ``{Model
  for Thermal Relic Dark Matter of Strongly Interacting Massive Particles},''
  \href{http://dx.doi.org/10.1103/PhysRevLett.115.021301}{{\em Phys. Rev.
  Lett.} {\bfseries 115} no.~2, (2015) 021301},
  \href{http://arxiv.org/abs/1411.3727}{{\ttfamily arXiv:1411.3727 [hep-ph]}}.

\bibitem{Berlin:2018tvf}
A.~Berlin, N.~Blinov, S.~Gori, P.~Schuster, and N.~Toro, ``{Cosmology and
  Accelerator Tests of Strongly Interacting Dark Matter},''
  \href{http://dx.doi.org/10.1103/PhysRevD.97.055033}{{\em Phys. Rev. D}
  {\bfseries 97} no.~5, (2018) 055033},
  \href{http://arxiv.org/abs/1801.05805}{{\ttfamily arXiv:1801.05805
  [hep-ph]}}.

\bibitem{LHCb:2019vmc}
{\bfseries LHCb} Collaboration, R.~Aaij {\em et~al.}, ``{Search for
  $A'\to\mu^+\mu^-$ Decays},''
  \href{http://dx.doi.org/10.1103/PhysRevLett.124.041801}{{\em Phys. Rev.
  Lett.} {\bfseries 124} no.~4, (2020) 041801},
  \href{http://arxiv.org/abs/1910.06926}{{\ttfamily arXiv:1910.06926
  [hep-ex]}}.

\bibitem{Walker-Loud:2014iea}
A.~Walker-Loud, ``{Nuclear Physics Review},''
  \href{http://dx.doi.org/10.22323/1.187.0013}{{\em PoS} {\bfseries
  LATTICE2013} (2014) 013}, \href{http://arxiv.org/abs/1401.8259}{{\ttfamily
  arXiv:1401.8259 [hep-lat]}}.

\bibitem{WalkerLoud:2012bg}
A.~Walker-Loud, C.~E. Carlson, and G.~A. Miller, ``{The Electromagnetic
  Self-Energy Contribution to $M_p - M_n$ and the Isovector Nucleon
  MagneticPolarizability},''
  \href{http://dx.doi.org/10.1103/PhysRevLett.108.232301}{{\em Phys. Rev.
  Lett.} {\bfseries 108} (2012) 232301},
  \href{http://arxiv.org/abs/1203.0254}{{\ttfamily arXiv:1203.0254 [nucl-th]}}.

\end{thebibliography}\endgroup

\end{document}